\documentclass[12pt]{article}
\usepackage{latexsym}
\usepackage{epsf}

\input def2e.frm

\def\a{\alpha}
\def\b{\beta}
\def\c{\chi}
\def\d{\delta}
\def\e{\epsilon}                % Also, \varepsilon
\def\f{\phi}                    %       \varphi
\def\g{\gamma}
\def\h{\eta}

\def\j{\psi}
\def\k{\kappa}
\def\l{\lambda}
\def\m{\mu}
\def\n{\nu}
\def\o{\omega}
\def\p{\pi}                     % Also, \varpi
\def\r{\rho}                    %       \varrho
\def\s{\sigma}                  %       \varsigma
\def\t{\tau}

\def\x{\xi}
\def\z{\zeta}
\def\D{\Delta}
\def\F{\Phi}
\def\G{\Gamma}
\def\J{\Psi}
\def\L{\Lambda}
\def\O{\Omega}
\def\P{\Pi}

\def\cc{{\cal C}}
\def\cd{{\cal D}}

\def\cf{{\cal F}}

\def\cl{{\cal L}}

\def\co{{\cal O}}
\def\cp{{\cal P}}

\def\cs{{\cal S}}
\def\ct{{\cal T}}
\def\cu{{\cal U}}

\def\cbo{{\,\raise-.15ex\Sc [\,}}                       % curly "
\def\Sl#1{\rlap{\hbox{$\mskip 3 mu /$}}#1}      % " upper
\def\vev#1{\Big\langle #1 \Big\rangle}           % < >
\def\svev#1{\left\langle #1\right\rangle}       % variable < >
\def\lvec#1{\raisebox{0.0ex}{$\stackrel{\leftarrow}{#1}$} }  % <--  accent
\def\dt#1{{\buildrel {\raisebox{-0.5mm}{\large .}} \over {#1}}}
\def\ddt#1{{\buildrel {\hbox{\LARGE .\kern-2pt.}} \over {#1}}}% double dot-over
\def\beq{\begin{equation}}
\def\eeq{\end{equation}}
\def\bqry{\begin{eqnarray}}
\def\eqry{\end{eqnarray}}

\def\secteq#1{ \setcounter{equation}{0}
               \renewcommand{\theequation}{#1.\arabic{equation}} }

\def\beqn#1{ \renewcommand{\theequation}{#1}
             \begin{eqnarray} }
\def\eeqn{ \renewcommand{\theequation}{\arabic{equation}}
           \end{eqnarray} }

\def\beqr#1{ \setcounter{equation}{#1}
             \begin{eqnarray} }

\def\eeqr{\end{eqnarray}}
\def\NON{\nonumber\\}
\def\beqrabc#1{ \setcounter{equation}{0}
                \renewcommand{\theequation}{#1\alph{equation}}
                \begin{eqnarray} }
\def\beqrn#1#2{ \setcounter{equation}{#2}
                \renewcommand{\theequation}{#1.\arabic{equation}}
                \begin{eqnarray} }
\def\seeq#1{eq.~(\ref{#1})}
\def\seEq#1{Eq.~(\ref{#1})}
\def\seeqs#1{eqs.~(\ref{#1})}
\def\seEqs#1{Eqs.~(\ref{#1})}
\def\seneq#1{~(\ref{#1})}

\def\IJMP#1{Int. J. Mod. Phys. {\bf #1}}
\def\JMP#1{Jour. Math. Phys. {\bf #1}}
\def\NPB#1{Nucl. Phys. {\bf B#1}}

\def\PLB#1{Phys. Lett. {\bf B#1}}
\def\PRD#1{Phys. Rev. {\bf D#1}}

\def\PRP#1{Phys. Rep. {\bf #1}}

\def\MPL#1{Mod. Phys. Lett. {\bf A #1}}

\def\sstyle{\scriptstyle}

\def\rhs{\mbox{r.h.s.} }
\def\lhs{\mbox{l.h.s.} }
\def\ie{\mbox{\it i.e.} }
\def\eg{\mbox{e.g.} }

\def\geqx{\,\raisebox{-1.0ex}{$\stackrel{\textstyle >}{\sim}$}\,}
\def\frac#1#2{ {\sstyle {#1\over #2} } }
\def\det#1{{\rm det}\left(#1\right)}

\def\tr{{\rm tr}\,}

\def\half{{1\over 2}}

\def\det{{\rm det\,}}
\def\Det{{\rm Det}}

\def\bmath#1{ \mbox{\large\boldmath $#1$} }
\def\Bmath#1{\raisebox{-.0mm}{\Large\boldmath $#1$}}

\def\hmath#1{\raisebox{-1mm}{\huge\boldmath $#1$}}
\def\hmatn#1{\raisebox{-1.1mm}{\huge $#1$}}

\def\vev#1{\big\langle #1 \big\rangle}

\def\lvec#1{{\buildrel \leftarrow\over {#1}}}
\def\rvec#1{{\buildrel \rightarrow\over {#1}}}
\def\nord{\raisebox{-.0mm}{$ \buildrel
            {\hbox{\huge .}}\over{\hbox{\huge .}} $} }
\def\eqv{\raisebox{0.0mm}{$\stackrel{\circ}{=}$}}

\def\hq{\hat{q}}
\def\hJ{\hat{\J}}
\def\hb{\hat\beta}
\def\sst{\scriptscriptstyle}
\def\st{\scriptstyle}
\def\hG{\hat{G}}
\def\cff{F}
\def\tl#1{\tilde{#1}}
\def\tL{\tilde{\Lambda}}
\def\sumint#1{\sum_{#1}\hspace{-5.5mm}\int\hspace{1.5mm}}

\def\ds{\delta}
\def\dgen{\mbox{\boldmath $\delta$}}
\def\bards{\bar\delta}
\def\subv#1{_{\! #1}}
\def\scl{S_{\rm cl}}
\def\LYM{\L_{1}}

\def\sbsect#1#2{
  \vspace{3ex}
  \noindent {\bf {#1}.~{#2}}
  \vspace{2ex}
}

\def\beqabc#1{
  \setcounter{equation}{0}
  \renewcommand{\theequation}{#1\alph{equation}}
  \begin{eqnarray}
}

\def\eeqabc#1{
  \end{eqnarray}
  \setcounter{equation}{#1}
}

\begin{document}
\hyphenation{fer-mio-nic per-tur-ba-tive pa-ra-met-rized}

\noindent August 1999 \hfill TAUP--2582--99  \\

\begin{center}
% DRAFT -- DRAFT -- DRAFT -- DRAFT -- DRAFT -- DRAFT\\

\vspace{15mm}
{\large\bf Feynman rules for non-perturbative sectors \\[1mm]
and anomalous supersymmetry Ward identities}
\\[15mm]
Aharon Casher$\,^a$ and Yigal Shamir$\,^{b*}$
\\[5mm]
{\it School of Physics and Astronomy\\
Beverly and Raymond Sackler Faculty of Exact Sciences\\
Tel-Aviv University, Ramat~Aviv 69978, Israel}
\\[2mm]
$^a\!$ email: ronyc@post.tau.ac.il\\
$^b\!$ email: shamir@post.tau.ac.il
\\[5mm]
ABSTRACT
\\[2mm]
\end{center}

\begin{quotation}
We show that supersymmetry Ward identities contain an anomalous term which
takes the form of a surface term in Hilbert space.
In the one-instanton sector the anomalous term is
the integral of a total $\r$-derivative
where $\r$ is the instanton's size.
There are cases where the anomalous term is non-zero, and
cannot be modified by subtractions. This constitutes a
supersymmetry anomaly.
The derivation is based on Feynman rules suitable for
any non-perturbative sector of a weakly-coupled, renormalizable gauge theory.
\end{quotation}

%%%%%%%%%%%%
\newpage
\noindent {\large\bf 1.~~Introduction}
\vspace{3ex}
\secteq{1}

  Following 't~Hooft's pioneering work~\cite{thooft},
instanton physics has played an important role in understanding the dynamics
of asymptotically-free theories. Mostly, explicit instanton calculations were
limited to the semi-classical approximation. A two-loop calculation
which is particularly relevant for supersymmetry (SUSY)
can be found in ref.~\cite{MRS}.

  There is no consistent regularization method which preserve SUSY~\cite{RDR}.
In this sense, SUSY  is similar to a chiral symmetry.
Whether SUSY  is a true symmetry at the quantum level, or not,
can be decided only by non-perturbative studies.
This question was previously addressed in the semi-classical
approximation, where it was found that SUSY  is preserved.
Important works are based on the {\it assumption}
that this remains true beyond the semi-classical approximation,
and that SUSY  is an exact symmetry in the continuum limit~[4-10].

  We begin with the investigation of Ward identities
in the most general path integral setting. A Ward identity reads
\beq
  \svev{\dgen\co} = \svev{\dgen S\; \co} 
  + \svev{\dgen\m\; \co}
  + \svev{\,\mbox{\rm Hilbert-space surface terms}\,}\,,
\label{WIabs}
\eeq
where the bold letter $\dgen$ denotes any infinitesimal transformation.
Besides the familiar terms containing the variations of the action
($\dgen S$) and the measure ($\dgen\m$), \seeq{WIabs} contains
also (the expectation value of) {\it Hilbert-space surface terms},
which arise from {\it integration by parts}
in the functional integral.

  In more detail, in each non-perturbative sector the path integration
is over a given set of collective coordinates $\z_n$
that characterize the classical background,
as well as over the amplitudes of the quantum-field modes.
The expectation value of a (gauge invariant) operator $\co$ is
\beq
  \svev{\co}  = \int d^n\z\; e^{-\scl-W_1-W_2}\,\svev{\co}\subv{\z} \,.
\label{vevz}
\eeq
Here $\scl$ is the classical action,
and the semi-classical measure, $\exp(-W_1)$, is the product of functional
determinants and a tree-level jacobian~\cite{thooft}. 
$W_2$ is the sum of all connected bubble diagrams,
and $\svev{\co}\subv{\z}$
is the sum of all diagrams with external legs defined by the operator $\co$.
All diagrams are obtained here by integrating over the quantum fields,
keeping the collective coordinates fixed. \seEq{vevz} applies to
both bare and renormalized quantities.

\seEq{WIabs} is derived in Sec.~2.
Restricting our attention to the SUSY variation,
which we denote $\ds$, we write
$\d S$ and $\d\m$ as spectral sums, and show that they both vanish
mode by mode. Thus, $\vev{\ds\co}$ may fail to be zero only
due to the Hilbert-space surface term. Explicitly
\beq
  \svev{\ds\co} = \int d^n\z\; {\partial\over \partial\z_n}
  \left(e^{-\scl-W_1-W_2}\,\svev{\ds\z_n \, \co}\subv{\z}\right) \,,
\label{dOint}
\eeq
where $\ds\z_n$ is the SUSY variation of the $n$-th collective coordinate.

The tools of functional differentiation and integration used in Sec.~2
are very useful in clarifying the algebraic structure that underlies
\seeq{dOint}. However, manipulations of the (formal)
path integral should be supported by a detailed diagrammatic calculation.
In Sec.~3 we turn to the instanton sector of
four-dimensional SUSY gauge theories.
Only the integral of the total $\r$-derivative
is not automatically zero ($\r$ is the instanton size).
The anomalous term may thus be
{\it represented} as an amplitude involving a zero-size instanton.
An anomaly due to a topological singularity was previously found
in supersymmetric quantum mechanics in the presence of
fluxons with a fractional magnetic flux~\cite{ACY}.

We consider the SUSY Ward identity\seneq{dOint} at the leading
non-trivial order for a specific bi-local operator in
super Yang-Mills theory.
Our main result is the anomalous Ward identity \seeq{anml}, 
and its local version \seeq{local}.
Both equations are valid in any renormalization prescription
where perturbation theory is supersymmetric.
While the leading contribution to the  $\r \to 0$ surface term on the \rhs
of \seeq{anml} is a tree diagram, a straightforward
calculation of the expectation value of $\ds\co$
involves loop diagrams.
The proof of \seeq{anml} thus requires careful separation
of short- and long-distance contributions.
We also generalize the result to supersymmetric QCD.
An important question is how \seeq{anml} is modified by
higher-order corrections.
A renormalization-group argument indicates that the anomalous term
should remain finite (and non-zero) after the inclusion of next-order
logarithmic corrections.

The limit of a vanishing instanton size,  $\r \to 0$,
is studied in more detail in Sec.~4.
In correlation functions involving operators of sufficiently high
dimension, the lower limit of the $\r$-integral may diverge.
Under these circumstances one should perform {\it non-perturbative}
subtractions. Our main result is that no non-perturbative subtraction
can possibly modify \seeq{anml}. Therefore \seeq{anml}
constitutes a {\it supersymmetry anomaly}.
We also discuss how the anomaly should arise on the lattice.

There are important non-perturbative sectors where one is not
expanding around an exact solution of the classical field equations.
This is true in the one-instanton sector of theories with
a scalar (Higgs) VEV, as well as in the instanton-antiinstanton sector.
Aiming to cover these cases too,
we have developed Landau-gauge Feynman rules
for any non-perturbative sector of a renormalizable gauge theory (Appendix~B).
The Feynman rules define a systematic expansion provided
the considered correlation function is infra-red convergent,
and provided the background field is ``almost'' a classical solution.

Affirming the validity of the perturbative expansion must be done
on a case by case basis. (A more detailed discussion of
the one-instanton sector of SUSY-Higgs theories will be given elsewhere.)
However, the basic Ward identity\seneq{dOint} can be derived in a completely
general way using the Feynman rules of Appendix~B. This is done in Appendix~C.
We give a fully detailed diagrammatic derivation
in the (physically less interesting)
case of a theory without a gauge symmetry.
Because of its technical complexity, we outline the generalization
to gauge theories but omit the arithmetical details.

Much of the technical material contained in this paper is devoted
to various detailed proofs. A first acquaintance with
the algebraic structure may be obtained by reading Sec.~2
up to subsection~2.2. In order to understand the content of
\seeq{anml} it is enough to read Sec.~3 up to subsection~3.1.
The local form of \seeq{anml} is discussed in subsection~3.5,
and the renormalization-group argument is given in subsection~3.6.
The proof that \seeq{anml} cannot be modified by subtractions
is given in subsection~4.1.

The notation used throughout the paper is defined in
Appendix~A. In Appendix~D we explain how the general formalism
of Sec.~2 works in the case of translation invariance.
Our conclusions are given in Sec.~5.

%%%%%%%%%%%%%%%%%%%%%%%%%%%%%%%%%%%%%%%%%%%%%%%%%%%%
\vspace{5ex}
%\newpage
\noindent {\large\bf 2.~~Path-integral derivation of SUSY Ward identities}
\vspace{3ex}
\secteq{2}

In this section we derive the anomalous
Ward identity\seneq{dOint} using the tools of functional
integration and differentiation. While the (unregularized) path integral
is a formal construct, the derivation serves to clarify both the underlying
algebraic structure, and the role of physical boundary conditions.
The results of this section are supported by the
detailed diagrammatic calculations of Sec.~3 and Appendix~C.

The notation of Appendix~A is used below.
The general and the SUSY Ward identities
(\seeqs{WIabs} and\seneq{dOint} respectively)
will be derived using the transverse-field path integral
defined in the first four subsections of Appendix~B.
(The rest of Appendix~B, which is devoted to the construction of the
tree-level propagators etc., is not needed for this section.)

%%%%%%%%%%%%
\sbsect{2.1}{The field variation}

The independent variables of the path integral
include the collective coordinates $\z_n$,
the gauge degrees of freedom $\o^a(x)$,
and the amplitudes of the quantum modes $\hb_p$ and $\hJ_p$ where
\beq
  \hb(x) \equiv \sumint{p} \c^B_p(x) \hb_p \,, \qquad
  \hJ(x) \equiv \sumint{p} \c^F_p(x) \hJ_p \,,
\label{qp}
\eeq
and $\c^B_p(x)$ ($\c^F_p(x))$ is the eigenfunction corresponding to 
$\hb_p$ ($\hJ_p$).
In this paper the quantum bosonic field,
$\hb(x)$, is assumed to be {\it transverse}.
Namely, it obeys the background gauge condition
$\O^{\dagger a} \hb(x) = 0$ (\seeq{backgg})
in addition to the orthogonality
conditions $\bmath( b_{;m} \bmath| \hb \bmath) = 0$ (\seeq{ortho})
where $b_{;m}$ is the covariant derivative of the classical field
with respect to $\z_m$.
The above constraints may be summarized as
$\bmath( b_{;{\sst M}} \bmath| \hb \bmath) = 0$,
where the capital index ${\st M}$ runs over the ordinary
collective coordinates {\it and}
over the infinitely-many collective coordinates
associated with local gauge transformations, see Appendix~B.4.

We begin by expanding an arbitrary variation
of an elementary field in terms of variations of
the independent path-integral variables.
For the variation of a boson field, $\dgen B(x)$, one has
\beq
\dgen B(x) = \dgen \hb(x) +  \left(b_{;n}(x) + \hb_{;n}(x)\right) \dgen\z_n
          + \left(\O^a(b(x)) -ig T^a \hb(x) \right) \dgen\o^a(x)   \,.
\label{var}
\eeq
Covariant $\z_n$-derivatives of the quantum field 
are obtained by differentiating the eigenfunctions (see also Appendix~B.3)
\beq
  \hb_{;n}(x) = \sumint{p}  \c^B_{p;n}(x)\, \hb_p \,,
\label{dbetadzeta}
\eeq
while the quantum-field variation is, by definition
\beq
  \dgen \hb(x) \equiv \sumint{p}  \c^B_p(x)\, \dgen \hb_p \,.
\label{dqbeta}
\eeq
The field variation acts on the quantum amplitudes $\hb_p$
and {\it not} on the eigenfunctions.
The differential operator $\O^a$ generates infinitesimal
gauge transformations of the classical bosonic field (Appendix~B.1).
The variation of a fermion field, $\dgen \J(x)$, is expanded as
\beq
\dgen \J(x) = \dgen \hJ(x) + \dgen\z_n \, \hJ_{;n}(x)
          -ig \, \dgen\o^a(x)\, T^a \,\hJ(x)   \,,
\label{varF}
\eeq
where 
\beq
  \hJ_{;n}(x) = \sumint{p}  \c^F_{p;n}(x)\, \hJ_p \,, \qquad
  \dgen \hJ(x) \equiv \sumint{p}  \c^F_p(x)\, \dgen \hJ_p \,,
\label{dqpsi}
\eeq
in analogy with the bosonic case.
(The reader should note that $\dgen \J(x)$ and $\dgen \hJ(x)$
have different meanings.)
For the SUSY variation, ordering matters in \seeq{varF}.
It follows from \seeq{dqbeta} that $\dgen\hb(x)$
obeys the same orthogonality constraints as does $\hb(x)$.
This allows us to extract
the variations $\dgen\z_n$ and $\dgen\o^a(x)$.
Considering \seeq{var} and taking
the inner product\seneq{inner} with $b_{;n}$ leads to
\beq
  \ds\z_m = C^{-1}_{mn} \left\{ \Bmath( b_{;n} \Bmath| \G\hJ \Bmath)
  + ig \Bmath( b_{;n}^\dagger T^a \hb \Bmath| \d\o^a \Bmath) \right\} \,,
\label{dzeta}
\eeq
where $C_{mn}$ is defined in eq.~(B.14) and we have used \seeq{covdergauge}.
We have specialized to the SUSY
case, where the bosons' variation is linear in the fermions, see \seeq{susvar}.
(For any other variation, simply replace $\ds \to \dgen$ 
and $\G\hJ \to \dgen B$.)
Note that $b_{;n}$ is $O(1/g)$, and $\ds\z_m$ is $O(g)$.
We next apply $\O^\dagger$ on both sides of \seeq{var}.
Using the gauge-fixing constraint\seneq{backgg}
and eliminating $\ds\z_m$ we find
\bqry
  \ds\o^a & = & (\cf^{-1})^{ab} \Bmath| \O^{b\dagger}\, \G\hJ \Bmath) \NON
  & & + ig (\cf^{-1})^{ab} \Bmath| b_{;m}^\dagger T^b \hb \Bmath)
  \, C^{-1}_{mn} \, \Bmath( b_{;n} \Bmath| \G\hJ \Bmath) \,.
\label{domega}
\eqry
Here
$\cf^{-1} = \cf_0^{-1} - \cf_0^{-1} \cf_{\rm int} \cf_0^{-1} + \cdots$,
where $\cf_0$ and $\cf_{\rm int}$ occur respectively in the tree-level and
interaction ghost lagrangians (\seeq{fp0int}). \seEqs{dzeta}
and\seneq{domega} are summarized by the compact formula (cf.\ \seeq{cc})
\beq
  \ds\z_{\sst M}
  = \cc^{-1}_{\sst MN} \Bmath( b_{;{\sst N}} \Bmath| \G\hJ \Bmath) \,.
\label{dcoll}
\eeq

The variation of a given quantum amplitude is obtained
by taking the inner product of \seeq{var} or\seneq{varF} with the corresponding
eigenmode. Explicitly,
\beq
  \ds \hb_p =  \Bmath( \c^B_p \Bmath| \G\hJ \Bmath)
  - \ds\z_{\sst M} \Bmath( \c^B_p \Bmath| \hb_{; {\sst M}} \Bmath)\,,
\label{bosvar}
\eeq
for bosons, and
\beq
  \ds \hJ_p =  \Bmath( \c^F_p \Bmath| \ds\J \Bmath)
  - \ds\z_{\sst M} \Bmath( \c^F_p \Bmath| \hJ_{; {\sst M}} \Bmath)  \,.
\label{fervar}
\eeq
For fermions. The last term is these equations
compensates for the $\z_{\sst M}$-dependence of the eigenmodes.

%%%%%%%%%%%%
\sbsect{2.2}{Ward identities}

The variation of any local operator can be expressed in terms of
the variations of the independent variables. One has
\beq
  \ds\co=\ct\co \,,
\label{TO}
\eeq
where $\ct$ is the functional differentiation operator
\beq
 \ct =  \sumint{p} \ds \hq_p {\partial\over \partial\hq_p}
  + \sumint{\sst N} \ds \z_{\sst N}\, {\partial\over \partial\z_{\sst N}} \,,
\label{trans}
\eeq
and $\hq_p$ stands for any quantum-field amplitude, cf.\ Appendix~B.3.
In practice we will be interested in gauge invariant operators,
which effectively restricts the last sum to ${\st N} = n$ only.
The expectation value of $\ds\co$ is given by the
functional integral (\seeq{Zperp})
\beq
  \svev{\ds\co} = \int d^n\z\; \cd \hq \,\,
  {\Det\, \cc \over {\rm Det}^\half \, \cc_0} \,\, e^{-S} \,\, \ct\co \,.
\label{vevTO}
\eeq
Integrating by parts we obtain the generic Ward identity
for gauge invariant operators
\beq
  \svev{\ds\co}
  = \svev{\ds S \; \co} + \svev{\ds\m \; \co}
  + \int d^n\z\; {\partial\over \partial\z_n}
  \left\{ \int \cd \hq \,\,
  {\Det\, \cc \over {\rm Det}^\half \, \cc_0} \; e^{-S} \; \ds\z_n \, \co
  \right\} \,,
\label{WI}
\eeq
where $\ds S$ and $\ds\m$ are respectively the variations of the action
and the measure. Explicitly $\ds S = \ct S$ and
\beq
  -\ds\m \equiv
  \ct \log \left\{ {\Det\, \cc \over {\rm Det}^\half \, \cc_0} \right\}
  + \sumint{p}
    {\partial(\ds\hq_p) \over \partial\hq_p}
  + {\partial(\ds\z_n) \over \partial\z_n} \,.
\label{dm}
\eeq
The last term on the \rhs of \seeq{WI} is the Hilbert-space surface term
of \seeq{WIabs}. The corresponding surface term for the
quantum amplitudes $\hq_p$ is zero because of the Gaussian integration.

In the rest of this section we consider SUSY Ward identities.
We compute $\ds S$ and $\ds\m$,
and prove that they both vanish mode by mode.
This leaves us with the last term in \seeq{WI}.
When the functional integration over the quantum fields
is done, this term is recognized as the \rhs of \seeq{dOint}.

The SUSY transformation is off-diagonal in that it maps bosons into fermions
and vice versa. In view of this, the invariance of the measure, $\ds\m=0$,
is an expected result. Nevertheless, the below derivation gives
us valuable information on how the cancelation of the various contributions
to $\ds\m$ works in practice.
We also find that the physical boundary conditions play a direct role.
The same elements reappear in the diagrammatic
proof of Appendix~C which is more rigorous and, at the same time,
technically more involved.

The presence of a Hilbert-space surface term in Ward identities
is not an unfamiliar situation. In appendix~D we apply the generic
Ward identity\seneq{WI} to the case of translation invariance.
We show that momentum is conversed because the Hilbert-space surface term
coincides in this case with a spacetime surface term.

%%%%%%%%%%%%
%\newpage
\sbsect{2.3}{Variation of the action}

The functional variation of the action consists of a volume integral
and a surface integral. The latter may in general depend on the choice of
functional variables, and it is our objective to determine it.
We start with the action principle
\beq
  \int d^dx\, \left(\d\cl - \partial_\m(\d Q \, \P_\m) \right)
  = \int d^dx\, \d Q \left( -\P_{\m;\m} + \partial\cl/\partial Q \right) \,.
\label{eom}
\eeq
As usual $\cl=\cl(Q,Q_{,\m})$ and $\P_\m = \partial\cl/\partial{Q_{,\m}}$.
Substituting \seeqs{var} and\seneq{varF} into the r.h.s. of \seeq{eom}
and integrating by parts leads to
\setcounter{equation}{0}
\renewcommand{\theequation}{2.18\alph{equation}}
\bqry
\hspace{-20mm}
  \int d^dx\,
  \d Q \left( -\P_{\m;\m} + \partial\cl/\partial Q \right)
  & =  & \int d^dx\, \d \hat{q}
  \left( -\P_{\m;\m} +\partial\cl/\partial Q \right)
\label{tq} \\
  & & + \d\z_n \int d^dx\,  \Big( Q_{;n;\m} \P_\m +
   Q_{;n} (\partial\cl/\partial Q) \Big)
\label{tz} \\
  & & -  \d\z_n \oint d\s_\m\, J_\m^n \,.
\label{intj}
\eqry
\renewcommand{\theequation}{2.\arabic{equation}}
\setcounter{equation}{18}
\hspace{-2.5mm}Gauge invariance of the action was used.
(Note that $\partial\cl/\partial Q = \partial\cl/\partial \hat{q}$.)
Here
\beq
  J_\m^n =  Q_{;n}\, \P_\m\,.
\label{Jrho}
\eeq
Let us write $\ct = \ct_q + \ct_\z$ in correspondence with
the two terms on the \rhs of \seeq{trans}.
Since $\z_n$-derivatives act on the classical field or on
the wave-functions (\seeq{dqdz}),
expression\seneq{tz} is equal to $\ct_\z\, S$.
Now, in terms of the mode amplitudes, the bilinear part of the action reads
\beq
  S^{(2)} = {1\over 2} \sumint{p} \l_p\, \hat{q}_p^2 \,,
\label{S2modes}
\eeq
where $\l_p$ is an eigenvalue of the small fluctuations operator,
$L \c_p = \l_p \c_p$. Consequently
\beq
  \ct_q\, S^{(2)} = \sumint{p}  \l_p\, \d\hat{q}_p \, \hat{q}_p
  = \Bmath( \d\hat{q} \Bmath| L \Bmath| \hat{q} \Bmath) \,.
\label{ds2}
\eeq
In infinite volume $L$ must act on $\hat{q}(x)$, and not on $\d\hat{q}(x)$.
The expression on the \rhs of row\seneq{tq} is therefore equal to $\ct_q\, S$.
Putting everything together we arrive at the following result
\beq
  \ct S  =  \int  d^dx\, \d\cl -  \oint d\s_\m\, (\d Q \,\P_\m)
        + \d\z_n \oint d\s_\m\, J_\m^n \,.
\label{TS}
\eeq
\seEq{TS} is completely general.
In the SUSY case, $\ds\cl$ is a total derivative and
\beq
  \ct S  =  \oint d\s_\m\, \hat{S}_\m
        +  \d\z_n \oint d\s_\m\, J_\m^n \,,
\label{TSUS}
\eeq
where $\hat{S}_\m$ is the SUSY current.
In order to understand the last surface term we employ
a {\it finite volume} cutoff. This term is then completely determined by
the boundary conditions and, for any boundary conditions which ensure
the hermiticity of the small fluctuation operators,
\beq
  \oint d\s_\m\, \c_{p,n}\, \P_\m = 0 \,.
\label{bndry}
\eeq
We note that \seeq{bndry} holds for each eigenmode $\c_p$ separately.
This implies the operator statement $\oint d\s_\m\, J_\m^n = 0$.

%%%%%%%%%%%%
%\newpage
\sbsect{2.4}{Variation of the measure}

Computing the variation of the measure is a straightforward
algebraic task. The first term on the \rhs of \seeq{dm} is
\beq
  \ct \log \left\{ {\Det\, \cc \over {\rm Det}^\half \, \cc_0} \right\}
  =
  \cc^{-1}_{\sst LK}\, \ct\, \cc_{\sst KL}
   - \half (\cc_0)^{-1}_{\sst LK}\, \ct\, (\cc_0)_{\sst KL} \,.
\label{varJ}
\eeq
The classical-valued matrix $\cc_0$ depends only on the collective
coordinates, and
\beq
  \ct\, (\cc_0)_{\sst KL} = \ds\z_{\sst M}\, (\cc_0)_{{\sst KL};{\sst M}}
  =  \ds\z_{n}\, (\cc_0)_{{\sst KL};n} \,.
\label{varC0}
\eeq
The last equality is true since
$(\cc_0)_{{\sst KL};{\sst M}}=0$ for any collective coordinate $\z_{\sst M}$
which corresponds to an exact symmetry. The latter always include
(local) gauge transformations and (global) translations.
Hence the non-zero terms always correspond to a subset
of the ordinary collective coordinates $\z_n$.

In the calculation of the field-dependent part $\ct\cc_1$
we use the bosonic closure relation
\beq
  1 = \sumint{p} \bmath| \c_p^B \bmath) \bmath( \c_p^B \bmath|
  + \sumint{\sst M}
  \bmath| \tl{b}_{{\sst M}} \bmath) \bmath(  \tl{b}_{{\sst M}} \bmath| \,,
\label{clsr}
\eeq
where $\tl{b}_{{\sst M}} = (\cc_0^{-1/2})_{\sst MN}\, b_{;{\sst N}}$
are normalized. Including the contribution from $\cc_0$ one has
\beq
%\hspace{0mm}
  \ct\, \cc_{\sst KL} = \ds\z_{n}\, \cc_{{\sst KL};n}
  + \ds\z_{\sst M}\, \cc_{{\sst KM};{\sst L}}
  - \Bmath( b_{;{\sst K};{\sst L}} \Bmath| \G\hJ \Bmath)
  - \ds\z_{\sst M}\,
  \Bmath( b_{;{\sst K}} \Bmath| B_{;{\sst M};{\sst L}} \Bmath) \,.
\label{varC}
\eeq
Next, using \seeq{dcoll} we have
\beq
  {\partial(\ds\z_n) \over \partial\z_n}
  = - \cc^{-1}_{n{\sst K}}\, \cc_{{\sst KL};n} \, \ds\z_{\sst L}
  + \cc^{-1}_{n{\sst K}} \Bmath( b_{;{\sst K}} \Bmath| \G\hJ \Bmath)_{\!\! ,n}
\,.
\label{ddrho}
\eeq
What remains is the contribution from the functional trace.
For the fermions
\beq
  \sumint{p} {\partial(\ds\hJ_p)\over\partial\hJ_p}
  = \ds\z_n \, \sumint{p} \Bmath( \c_p^F \Bmath| \c_{p;n}^F \Bmath)
  - \cc^{-1}_{{\sst LK}} \,
  \Bmath( b_{;{\sst K}} \Bmath| \G\hJ_{;{\sst L}} \Bmath) \,,
\label{ddF}
\eeq
where \seeq{fervar} and the closure relation for fermions were used.
For the bosons, starting from \seeq{bosvar} one has
\beq
  \sumint{p} {\partial(\ds\hb_p)\over\partial\hb_p}
  = - \ds\z_n\, \sumint{p} \Bmath( \c_p^B \Bmath| \c_{p;n}^B \Bmath)
  -  \ds\z_{\sst M} \, \sumint{p}
  \Bmath( \c_p^B \Bmath| \hb_{;{\sst L}} \Bmath)
  \, \cc^{-1}_{\sst LK} \,
  \Bmath( b_{;{\sst K};{\sst M}} \Bmath| \c_p^B \Bmath)
\label{ddbtmp}
\eeq
Using the bosonic closure
relation\seneq{clsr} we find after a few algebraic steps
\bqry
  \sumint{p} {\partial(\ds\hb_p)\over\partial\hb_p}
  & = & - \ds\z_n\, \sumint{p} \Bmath( \c_p^B \Bmath| \c_{p;n}^B \Bmath)
  - \ds\z_n\, \sumint{\sst M}
  \Bmath( \tl{b}_{\sst M} \Bmath| \tl{b}_{{\sst M};n}\Bmath)
  - \ds\z_{n}\, \cc^{-1}_{\sst LK} \, \cc_{{\sst KL};n}
\NON
  & &  + \half \ds\z_{n}\, (\cc_0^{-1})_{\sst LK} (\cc_0)_{{\sst KL};n}
  + \ds\z_{\sst M}\, \cc^{-1}_{\sst LK} \,
  \Bmath( b_{;{\sst K}} \Bmath| B_{;{\sst L};{\sst M}} \Bmath) \,.
\label{ddb}
\eqry
Collecting all terms we obtain
\bqry
  \ds\m & = & \ds\z_n\, \sumint{p} (-)^F \Bmath( \c_p \Bmath| \c_{p;n} \Bmath)
 + \ds\z_n\, \sumint{\sst M}
  \Bmath( \tl{b}_{\sst M} \Bmath| \tl{b}_{{\sst M};n}\Bmath)
\NON
  & & + \ds\z_{\sst M}\, \cc^{-1}_{\sst LK} \,
  \Bmath( b_{;{\sst K}} \Bmath| B_{;{\sst M};{\sst L}}
                                - B_{;{\sst L};{\sst M}} \Bmath) \,.
\label{dmcom}
\eqry
The first term on the \rhs is a spectral trace over
(transverse) bosons and fermions.
As with the variation of the action we now employ a finite volume cutoff.
The modes are normalized, namely,
$\bmath( \c_p \bmath| \c_p \bmath) = 1$.
Hence $\bmath( \c_p \bmath| \c_{p;n} \bmath) = 0$.
Similarly,
$\bmath( \tl{b}_{\sst M} \Bmath| \tl{b}_{{\sst M};n}\bmath) = 0$.
Finally, as we show below, the commutator (the last term in \seeq{dmcom})
is in fact a spectral trace over the ghosts field, which is zero
for the same reason.
Thus, $\ds\m$ vanishes mode by mode.

%%%%%%%%%%%%
%\newpage
\sbsect{2.5}{Ghosts contribution}

We last consider the commutator in \seeq{dmcom}.
Using Appendices~B.3 and~B.4 and inserting a complete set of
ghosts eigenstates, the commutator can be rewritten as
\beq
  \Bmath( b_{;{\sst K}} \Bmath| B_{;{\sst M};{\sst L}}
                                - B_{;{\sst L};{\sst M}} \Bmath)
  =  \Bmath( b_{;{\sst K}} \Bmath|
  (\O^a - ig T^a \hb) F^a_{\sst ML} \Bmath)
  = \sumint{p} \cc_{{\sst K}p} \, F_{p{\sst ML}} \,,
\label{com}
\eeq
where $F^a_{\sst ML}$ is a generalized field strength and
\beq
  F_{p{\sst ML}} = \Bmath( c^a_p \Bmath| F^a_{\sst ML} \Bmath) \,.
\label{FpML}
\eeq
Hence
\beq
  \cc^{-1}_{\sst LK} \,
  \Bmath( b_{;{\sst K}} \Bmath| B_{;{\sst M};{\sst L}}
                                - B_{;{\sst L};{\sst M}} \Bmath)
  = \sumint{p} \cc^{-1}_{\sst LK} \, \cc_{{\sst K}p} \, F_{p{\sst ML}}
  = \sumint{p}  F_{p{\sst M}p} \,,
\label{FpMp}
\eeq
where explicitly
\beq
   F_{p{\sst M}p} = - \Bmath( c^a_p \Bmath| c^a_{p;{\sst M}} \Bmath) \,.
\label{Feq0}
\eeq
We see that the commutator in \seeq{dmcom} is a spectral
trace over the ghost field, which vanishes in finite volume too.

In summary, using the (formal) tools of
functional differentiation and integration we showed
that the only anomalous term in SUSY Ward identities
is a surface term in Hilbert space, cf.\ \seeq{dOint}.
The contributions of $\ds S$ and $\ds\m$ (cf.\ \seeq{WI})
vanish mode by mode. We note that the classical equation of motion
was nowhere used. Consequently, \seeq{dOint} should hold
in any non-perturbative sector, regardless of whether or not one is
expanding around an exact classical solution.
A detailed diagrammatic proof of this statement is
given in Appendix~C.

%%%%%%%%%%%%%%%%%%%%%%%%%%%%%%%%%%%%%%%%%%%%%%%%%%%%
%%%%%%%%%%%%%%%%%%%%%%%%%%%%%%%%%%%%%%%%%%%%%%%%%%%%
%\newpage
\vspace{5ex}
\noindent {\large\bf 3.~~Anomalous supersymmetry Ward identities}
\vspace{3ex}
\secteq{3}

In this section we turn to the one-instanton sector of super Yang-Mills
(SYM) theory. Our main result is
a SUSY Ward identity whose anomalous term is non-zero and unambiguous.
The same anomalous term is obtained using any
renormalization scheme where perturbation theory is supersymmetric.
Moreover, as shown in Sec.~4,
no non-perturbative subtraction can modify this result.
Similar statements apply to supersymmetric QCD (SQCD).

We begin (subsection~3.1) with a leading-order calculation of $\svev{\ds\co}$
for a specific gauge-invariant bi-local operator in SYM (\seeq{anml}).
\seEq{dOint} is used in the calculation.
As already mentioned in the introduction, the result arises
from (the $\r$-integral of) a total $\r$-derivative.

In the next two subsections we confirm \seeq{anml} by a detailed
diagrammatic calculation.
Working at fixed $\r$ we show (subsection~3.2)
that every diagram which contributes to $\svev{\ds\co}_{\!\r}$
is related to a diagram
which contributes to the $\r$-derivative of $\svev{\ds\r\,\co}_{\!\r}$.
While $\svev{\ds\r\,\co}_{\!\r}$ is a tree diagram to leading order,
one encounters {\it loop} diagrams in a direct same-order
calculation of $\svev{\ds\co}$.
In subsection~3.3 we prove that the (tree-diagram) \rhs
is equal to the (one-loop renormalized) \lhs of \seeq{anml}.
A key role is played by the recursion relations of Appendix~B.6.
(While this discussion is given in terms of a concrete example,
it actually proves the validity after renormalization of \seeq{dOint},
whenever its \rhs consists of tree diagrams only.)

In subsection~3.4 we generalize the result to SQCD,
and in subsection~3.5 we discuss the local form of the anomaly.
In subsection~3.6 we give a renormalization-group argument
that the anomalous term should remain finite, and non-zero,
after the inclusion of higher-order logarithmic corrections.

We now give a quick review of (super) Yang-Mills
instantons taking the gauge group to be SU(2).
In SYM one expands around an exact, scale-invariant classical
solution, and the Feynman rules are considerably simpler
than those of Appendix~B.
The (regular gauge) instanton field is
\beq
  a^{c}_\m = {2\over g}\,{\bar\h_{c\m\n}(x-x_0)_\n \over (x-x_0)^2+\r^2 } \,,
\label{ainst}
\eeq
where $\bar\h_{c\m\n}$ is the 't~Hooft symbol~\cite{thooft}.
($\bar\h_{c\m\n}$ is antisymmetric in the last two indices;
$\bar\h_{c\m\n}=\e_{c\m\n}$ if $\m,\n=1,2,3,$ 
and $\bar\h_{c4\n}=\d_{c\n}$.)
There are eight gauge-field zero modes ($b_{;n}$ in our generic notation).
Their explicit (unnormalized) form is $a^{c}_{\m;n}$.
As explained in Appendix~B.1, each zero mode constitutes of
an ordinary derivative of the classical field with respect to a
collective coordinate, plus a compensating (proper) gauge transformation
to enforce the background gauge $D_\m\, a_{\m;n} =0$.
We now list the various zero modes.
First, for the dilatation zero mode, covariant and ordinary differentiations
coincide \ie $a^{c}_{\m;\r} = a^{c}_{\m,\r}$.
For the zero modes associated with the translation
collective coordinates $x_0^\n$, the compensating
gauge transformation is generated by $\o^{c}_\n = a^{c}_\n$.
Explicitly
$a_{\m;\n} = a_{\m,\n} - D_\m\, a_\n = f_{\m\n}$.

Last we consider the three isospin zero modes $a^{c}_{\m;a'}$.
In a {\it singular} gauge, one has
\beq
  a^{c}_{\m;a'} \equiv D_\m^{ca}\, \tilde\o^{a}_{a'} \,,
\label{iso}
\eeq
where
\beq
  \tilde\o^{a}_{a'}
  = {\d^a_{a'}\over g}\, { (x- x_0)^2 \over (x- x_0)^2 + \r^2 }
  = {\d^a_{a'}\over g} \left( 1 - { \r^2 \over (x- x_0)^2 + \r^2 } \right) \,.
\label{oiso}
\eeq
The last term in parenthesis generates a proper gauge
transformation, whereas the first term generates a {\it global}
SU(2) transformation.
During the quantization procedure described in
Appendix~B.2 one introduces global SU(2) collective coordinates $\z_{a'}$
alongside with the translation and dilatation ones.
As follows from \seeq{oiso}, the isospin modes obey
$a_{\m;a'} = a_{\m,a'} - D_\m\, \o_{a'}$
where $g\, \o^a_{a'} = \d^a_{a'}\,\r^2/((x- x_0)^2 + \r^2)$,
in agreement with the general formula \seeq{covder}. Then,
in gauge invariant correlation functions the isospin collective coordinates
factor out, yielding a group-volume factor~\cite{thooft}.

We next turn to the gaugino. Because of the chiral nature of
instanton amplitudes it is often more natural to use a Weyl notation,
where the Majorana-field gaugino is split as
$\l_{\rm majorana} \to \l_\a, \bar\l_{\dt\a}$.
Similarly, we will distinguish between the SUSY variations $\ds_\a$ and
$\bar\ds_{\dt\a}$ (both of which are accounted for in \seeq{susvar}).
In SU(2)  SYM there are two pairs of gaugino zero modes.
In normalized form they are given by
\bqry
  \l^{SS}(\beta)^c_\a
  & = & {\sqrt{2}\r^2\over\p}{\s^c_{\a\b}\over ((x-x_0)^2+\r^2)^2}\,,
\quad\quad \b =1,2, \\
  \l^{SC}(\dt\a)^c_\a
  & = & {\r\over\p}{i\bar\h_{c\m\n} (x-x_0)_\m (\s_\n)_{\a\dt\a}
  \over ((x-x_0)^2+\r^2)^2}
\,,\quad\quad \dt\a =1,2.
\label{zm}
\eqry
In terms of Weyl fields,
the instanton-sector fermion propagator~\cite{BCCL} obeys
\beq
\left( \begin{array}{cc}
  0  &  \bar\s_\m D_\m \\
  \s_\m D_\m  &  0
\end{array}\right)
\left( \begin{array}{cc}
  0  &  \svev{\bar\l(x)\, \l(y)}_0 \\
  \svev{\l(x)\, \bar\l(y)}_0  &  0
\end{array}\right)
=
\d^4(x-y) -
\left( \begin{array}{cc}
  0  &  0 \\
  0  &  P_F(x,y)
\end{array}\right) \,,
\eeq
where $P_F$ is the projector on the (left-handed) fermionic zero modes,
and $\vev{\cdots}_0$ refers to the tree-level instanton propagators.
In the more compact Majorana notation the same equation reads
\beq
  L_F\, G_F(x,y) = \d^4(x,y) - P_F(x,y) \,.
\label{GF0}
\eeq
The transverse vector-field propagator~\cite{BCCL} obeys
\beq
  L_B\, G_B(x,y) = \d^4(x,y) - P_B(x,y) - P^\parallel(x,y) \,,
\label{GB0}
\eeq
where $P_B$ projects on the eight bosonic zero modes (cf.\ \seeq{PB})
and $P^\parallel$ is the longitudinal projector\seneq{longP}.
(In SYM $\O \to D_\m$;
also, $P_B \, L_B = P^\parallel\, L_B = 0$, cf.\ \seeq{elim}.)

Last we give the semi-classical measure
($d^n\z\, \exp(-\scl-W_1)$ of \seeq{vevz}),
which reads
\beq
  2^{10}\p^6 g^{-8} \LYM^6\,  d\r \,\r^3  \, d^4x_0\,,
\label{msr}
\eeq
where $\LYM$ is the one-loop renormalization-group invariant scale.
In SYM the functional determinants, cf.\ \seeq{semiclgg},
cancel each other exactly. The matrix $C_0$ (\seeq{C0}) is diagonal.
The entries related to the translation
and the dilatation modes have mass dimension zero, while those
related to the three isospin zero modes have dimension minus two.
The $\r^3$ factor in \seeq{msr} thus arises from
the isospin entries of ${\rm det}^\half C_0$.

In (potentially) singular cases,
the integration over the translation collective coordinates $x_0^\m$
should always be done before the $\r$-integration. This guarantees
translation invariance. Also, it implies that the integral
of the total $x_0$-derivative in \seeq{dOint} vanishes.
The isospin collective coordinates too do not give rise to any
Hilbert-space surface term, because group-integration is compact.

%%%%%%%%%%%%
%\newpage
\sbsect{3.1}{Super Yang-Mills}

In \seeq{dOint}, a non-zero result may arise in the instanton sector
(only) from the $\r$-integral of the total $\r$-derivative.
In this subsection we compute the anomalous Ward identity
\beq
  \svev{\ds_\a \,\, \co_{\m}^{\,i\b}(0) \,\, \l\l(y)}
  = - { 2^{12.5} 3 \p^2 \LYM^6 \over 7 g^7}\,
  {\bar\h_{i\m\t} [\s_\n \, \bar\s_\t]_\a^{\,\b}  \, y_\n \over  (y^2)^3} \,.
\label{anml}
\eeq
Here $\ds_\a$ acts on everything to its right,  $\l\l=\l^a_\a \, \l^{a\a}$ and
\beq
  \co_{\m}^{\,i\b}
  = \left(
  \hat{D}_\m^{eb}(\e_{bcd}\, \l^{c\a} \, \s^{i\g}_\a \, \l^d_\g) \right)
  \left( (\hat{D}^2)^{ef} \, \l^{f\b} \right)\,,
\label{O}
\eeq
where
\beq
  \hat{D}_\m = \partial_\m -ig T^c A_\m^c = D_\m - ig T^c \a_\m^c \,,
\label{fullder}
\eeq
and $\s^i = {1\over 4}\, \bar\h_{i\m\n}\s_{\m\n}$.
Using \seeq{dOint}, the \lhs of \seeq{anml}
is expressed as a limit
\bqry
  \hspace{-12mm}
  \svev{\ds_\a \,\, \co_{\m}^{\,i\b}(0) \,\, \l\l(y)}
& = & 2^{10}\p^6 g^{-8} \LYM^6 \, \int d\r\,
  {\partial\over \partial\r} \,
  \left\{
  \r^3 \, \svev{(\ds_\a\,\r) \,\, \co_{\m}^{\,i\b}(0) \,\, \l\l(y)}_{\!\r}
  \right\}
\NON
  \phantom{ \mbox{\huge[ }}
& = &  - 2^{10}\p^6 g^{-8} \LYM^6 \, \lim_{\r\to 0} \, \r^3 \,
    \svev{(\ds_\a\,\r) \,\, \co_{\m}^{\,i\b}(0) \,\, \l\l(y)}_{\!\r} \,.
\label{lim}
\eqry
For $\r^2 \gg y^2$ the $\r$-integrand behaves like $y/\r^7$
(see subsection~4.1 below).
The integral is infra-red convergent, and the $\r=\infty$
boundary point drops out from \seeq{lim}.

\seEq{lim} says that the anomalous term can be expressed solely
in terms of an amplitude of a zero-size instanton.
This mathematical statement has one misleading aspect.
Namely, the {\it dominant contribution to the $\r$-integral} which leads
to \seeq{anml} {\it comes from finite-size instantons} with $\r^2 \sim y^2$.
\seEq{anml} is {\it insensitive} to the contributions of
vanishingly-small instantons. If we restrict the
integration to $\bar\r \le \r < \infty$ for some $0 < \bar\r^2 \ll y^2$,
the relative change in the result will be $O(\bar\r^2/y^2)$.
The limit $\bar\r \to 0 $ is smooth.
We return to these observations in Sec.~4.

With \seeq{lim} at hand the calculation is straightforward.
We begin with $\ds_\a \r$ (cf.\ \seeq{dzeta}).
To leading order one has
\beq
  \ds_\a\,\r
  = {g^2 \over 16\p^2}
  \int d^4x\, (\partial a^{c}_\m / \partial\r) \, \ds_\a A^c_\m \,,
\label{drho}
\eeq
where $\ds_\a A^c_\m = (i/\sqrt{2})(\s_\m)_{\a \dt\a}\, \bar\l^{c \dt\a}$,
and
\beq
  {16\p^2 \over g^2} = \int d^4x\, (\partial a^{c}_\m / \partial\r)^2 \,,
\label{Crr}
\eeq
is the $\r\r$-entry of $C_0$. Note that $\ds_\a \r = O(g)$
since $\partial a^{c}_\m / \partial\r = O(1/g)$.

As can be seen from the above equations,
$\ds_\a \r$ contains a $\bar\l^{\dt\a}(x)$ field.
To arrive at \seeq{anml},
one gaugino field from $\l\l(y)$ is contracted with $\bar\l^{\dt\a}(x)$
to form a propagator $G_F(y,x)$ (cf.\ \seeq{GF0}),
and the second one is saturated by a $\l^{SC}(\dt\a)$ mode.
(If $\bar\l^{\dt\a}$ is contracted with one of the $\l$'s
in $\co_{\m}^{\,i\b}(0)$, one obtains a contribution
that vanishes in the limit $\r\to 0$ due to an extra
damping factor of $\r^2/y^2$.)
Now, because of the operator $\co_{\m}^{\,i\b}$,
the integration over the instanton position is dominated by $x_0 \sim \r$.
The integral in \seeq{drho}, in turn, is dominated by $x \sim y$.
In the calculation of the $\r \to 0$ boundary term,
this allows us to replace the (singular gauge) fermion propagator
$G_F(y,x)$ by a {\it free} propagator.
We find
\beq
  \svev{\l\l(y)\, \ds_\a\r}_{\!\r,\, x_0 \sim \r ; \dt\a}
   = {3\r^2\, g \over 2^{2.5} i \p^3}\,
   {(\s_\n)_{\a\dt\a}\, y_\n \over (y^2)^3}\,,
\label{prop}
\eeq
where we have indicated which (superconformal)
Grassmann amplitude has been integrated over.

The other three zero modes saturate the operator $\co_{\m}^{\,i\b}$.
The $x_0$-integration yields
\beq
  \int d^4 x_0 \, \svev{\co_{\m}^{\,i\b}(0)}_{\r,x_0;\dt\b}
  = { 2^5 i \over 7 \p \r^5}\,
  \bar\h_{i\m\t}\, \e_{\dt\b \dt\g}\, \bar\s_\t^{\dt\g \b} \,.
\label{x0int}
\eeq
(At this order,
only the background covariant derivative $D_\m$ contributes, cf.\ \seeq{O}.)
We now put together \seeqs{prop} and\seneq{x0int}, include
a symmetry factor $\e^{\dt\a \dt\b}$, and substitute
everything on the \rhs of \seeq{lim}. The result is \seeq{anml}.

%%%%%%%%%%%%
%\newpage
\sbsect{3.2}{Diagrams}

In this subsection we describe in detail how \seeq{anml}
works at the level of Feynman diagrams.
This provides a concrete leading-order example of the general
diagrammatic identities of Appendix~C.
The renormalization of the loop diagrams encountered in the derivation
will be discussed in the next subsection.

Taking a slightly different course from Appendix~C,
our starting point is the observation that, in SYM,
\beq
  \int d^4x\, \partial_\mu \svev{\hat{S}_\mu(x)\, \co}_{\r,x_0} = 0 \,,
\label{dmuSmu}
\eeq
for any (multi)local operator $\co$. Here, $\co$ will
be taken to be the bi-local operator in \seeq{anml}.
In the generic notation of Appendix~A,
SUSY invariance of the action can be expressed as the
{\it off-shell identity}
\beq
  \partial_\mu S_\mu =  \Big( \hb^\dagger \lvec{L}_B + \cs_B \Big) \G\hJ
  + \overline{\ds \J}\, \Big( \,\rvec{\! L}_{\! F} \hJ + \cs_F \Big) \,,
\label{susEoM}
\eeq
where
\beq
  \cs_F = {\partial \cl_F^{\rm int} \over \partial \hJ } \,, \qquad
  \cs_B = {\partial \cl_B^{\rm int} \over \partial \hb }
          + {\partial \cl_F^{\rm int} \over \partial \hb } \,,
\label{SFSB}
\eeq
and the interaction lagrangians are defined in \seeqs{LF} and\seneq{LB}.
The expressions inside the parenthesis in \seeq{susEoM} are recognized as the
(bosonic and fermionic)  equations of motion.
Below we will repeatedly use \seeqs{dmuSmu} and\seneq{susEoM}.
(We have used the notation $\partial_\m S_\m$ in the above off-shell relation
to distinguish it from the operator $\partial_\m \hat{S}_\m$;
the latter is discussed in subsection~3.5 below.)

The first diagrams that contribute to $\svev{\ds_\a\co}$ are $O(1/g^2)$
in comparison to the \rhs of \seeq{anml}. At this order, $\svev{\ds_\a\co}$
vanishes for rather trivial reasons, as we now explain.
For any SUSY theory, the leading-order terms in \seeq{susEoM} are
\beq
  \partial_\mu S_\mu^{(0)} = \cs_B^{(1)} \G\hJ
  + \overline{\ds \J}{}^{(0)}\, \rvec{\! L}_{\! F} \hJ  \,,
\label{dS0}
\eeq
where \seeqs{SFSB} and\seneq{LB} were used. We have expanded
\beq
  \ds \J = \ds \J^{(0)} + \ds \J^{(1)} + \ds \J^{(2)} \,,
\label{expandl}
\eeq
and the superscript counts how many quantum fields
occur in each term. In SYM, \seeq{dS0} reduces to
\beq
  \partial_\mu S_{\mu\a}^{(0)}
  = \ds_\a \l_\b^{(0)}\, \Sl{D}_{\!\b\dt\b}\, \bar\l^{\dt\b}
  = i/(2\sqrt{2})\, \s_{\a\b}^{\m\n} f_{\m\n}\,
  \Sl{D}_{\!\b\dt\b}\, \bar\l^{\dt\b}\,,
\label{dS0YM}
\eeq
Like \seeq{susEoM}, \seeq{dS0YM} is an off-shell
relation in which only the {\it classical} equation of motion is used.
The fermion propagator emanating from $\partial_\mu S_\mu^{(0)}$
ends on any one of the five $\l$'s in the operator $\co$.
The other four $\l$'s are saturated by zero modes.
We next eliminate this fermionic propagator using \seeq{GF0}.
Only the delta-function from \seeq{GF0} contributes to $\svev{\ds_\a\co}$.
The terms with the projector $P_F$ cancel each other by antisymmetry.
These steps are depicted in Fig.~2. (See Fig.~1 for our Feynman rules.)
To this order, the diagrams contributing to $\svev{\ds_\a\co}$
involve the fermionic zero modes
and the classical field, and nothing else.
As in \seeq{dmuSmu}, the vanishing result is obtained {\it before}
the integration over any of the collective coordinates.

We now turn to the more interesting next order, that
corresponding to \seeq{anml}.
First, there is a set of {\it disconnected} diagrams which is trivially zero,
since it consists of the {\it same} diagrams
considered above (Fig.~2) times a sum of bubble diagrams.
(The bubble diagrams (Fig.~3) were discussed in detail in ref.~\cite{MRS}.
Note that, besides the familiar two-loop diagrams,
Fig.~3 contains a one-loop diagram coming from the off-diagonal
terms of the jacobian\seneq{J} or,
equivalently, from the interaction term on the last row of \seeq{fpint}.
This is the only place where that term occurs at this order.)

We next consider Fig.~4. The shown diagrams are related
by interchanging the insertion of
$\partial_\mu S_\mu^{(0)}$ with one of the zero modes.
Thanks to this antisymmetrization, the fermionic projector terms
of \seeq{GF0} cancel out.
After the application of \seeq{GF0}, the second diagram
gives a contribution to $\ds\co$.

In the first diagram of Fig.~4,
the propagator emanating from $\partial_\mu S_\mu^{(0)}$
is attached to a fermionic interaction vertex.
On the \lhs of Fig.~5 we show only the
connected part of this diagram containing
$\partial_\mu S_\mu^{(0)}$. Now, the terms in \seeq{susEoM}
which are linear in both the fermion and the boson fields read
\beq
  \partial_\mu S_\mu^{(1)} = \hb^\dagger \lvec{L}_B\, \G\hJ
  + \overline{\ds \J}{}^{(1)} \rvec{\! L}_{\! F}\, \hJ
  + \overline{\ds \J}{}^{(0)} \cs_F  \,.
\label{dS1}
\eeq
After the application of \seeq{GF0}, the \lhs of Fig.~5
gives rise to an insertion of (minus the integral of) the last term
in \seeq{dS1} (middle Fig.~5). Using \seeq{dmuSmu},
the latter is traded with an insertion of the first two terms
on the \rhs of \seeq{dS1}. Explicitly, this insertion is
\beq
   \Bmath( \G\hJ \Bmath| L_B \Bmath| \hb \Bmath)
  + \Bmath( \ds \J^{(1)} \Bmath| L_F \Bmath| \hJ \Bmath) \,.
\label{genwi1}
\eeq
This step is depicted on the \rhs of Fig.~5.
(\seEq{genwi1} coincides with
\seeq{genwi} up to the replacement $\ds\J \to \ds\J^{(1)}$.)

We now apply \seeqs{GB0} and\seneq{GF0} to the diagram containing
insertion\seneq{genwi1}.
The terms with the fermionic projector cancel by antisymmetry
as before. All other terms are shown in Fig.~6.
The last diagram, which involves a longitudinal projector,
will be discussed later. The first diagram on the \rhs of Fig.~6
is a contribution to $\ds\co$. (This contribution is a tree or a one-loop
diagram, depending on whether the two external legs correspond to
different spacetime points or to the same one.)
The second diagram vanishes by the same Fiertz rearrangement used
in proving the SUSY  invariance of the action.
(Here we are ignoring the need for regularization, see the next
subsection).

The third diagram on the \rhs of Fig.~6
is a contribution to the \rhs of \seeq{anml}.
The bosonic zero mode $b_{;n}$ at the upper-left corner of the triangle,
the $\bar\l^{\dt\a}$ at the same spacetime point
and the thick dashed line representing $C_0^{-1}$, together
form the leading order of $\ds_\a \z_n$. The other bosonic zero
mode is attached to a fermion line emanating from a fermionic zero mode.
Denoting the zero mode by $\c^{F0}$,
this part of the diagram gives the $\z_n$-derivative
$\c^{F0}_{;n} = - G_F \, (L_F)_{;n} \, \c^{F0}$.

We now consider two more diagrams with insertion\seneq{genwi1}.
The third diagram on the \rhs of Fig.~7
gives the $\z_n$-derivative of the fermionic
propagator emanating from $\ds\z_n$.
Fig.~8 contains a one-loop tadpole which is actually zero in SYM.
In general, the third diagram on the \rhs of Fig.~8
gives the $\z_n$-derivative of the
(logarithm of the) functional determinants.

Let us consider separately the boson, fermion and ghost contributions
in the second diagram on the \rhs of Fig.~8. The fermion-loop
contribution is by itself zero after a Fiertz rearrangement.
The ghost-loop diagram will be considered later.
For the sum of the remaining one-(bosonic)-loop diagram, plus
the first diagram on the \rhs of Fig.~7, we use
\beq
  \partial_\mu S_\mu^{(2)}
  = \cs^{(3,B)}_{\sst \tl{I}\tl{J}\tl{K}}(b)\;
  \hb_{\sst \tl{I}} \hb_{\sst \tl{J}} \G_{\sst \tl{K}\tl{L}} \hJ_{\sst \tl{L}}
  + \overline{\ds \J}{}^{(2)} \rvec{\! L}_{\! F}\, \hJ
  + \overline{\ds \J}{}^{(1)} \cs_F  \,.
\label{dS2}
\eeq
This allows us to trade the two bosonic-loop diagrams with a single diagram
containing an insertion of
$\overline{\ds \J}{}^{(2)} \rvec{\! L}_{\! F}\, \hJ$.
Using \seeq{GF0} once more,
we obtain the contribution(s) to $\ds\co$ coming from $\ds \l^{(2)}$
(Fig.~9).

Another type of diagrams with insertion\seneq{genwi1}
is obtained by attaching the bosonic propagator to a quantum gauge field
coming from a covariant derivative (cf.\ \seeq{fullder}).
One obtains two contributions to $\ds\co$ and a contribution to the \rhs
of \seeq{anml}, in which the classical field in the background
covariant derivative is differentiated with respect to $\z_n$ (Fig.~10).

We next attach the bosonic propagator from insertion\seneq{genwi1}
to a vertex coming from the expansion of the jacobian (Fig.~11).
The second term on the \rhs of Fig.~11 gives the $\z_n$-derivative
of the bosonic zero mode contained in $\ds\z_n$ itself.

The third diagram on the \rhs of Fig.~11 gives the $\z_n$-derivative
of $\,\log {\rm det}^\half C_0$.
Particularly relevant for \seeq{anml} are $\r$-derivatives.
As explained earlier, the $\r$-dependence of ${\rm det}\, C_0$
comes from the entries related to the three isospin zero modes.
In the matrix $C_1$ (\seeq{C1}), the isospin-isospin entries
$(C_1)_{a'c'}$ involve the integral of
$g \e_{abc}\, \tilde\o^{b}_{c'}\, a^{c}_{\m,a'}$ (cf.\ \seeq{oiso}).
After taking the product with $\partial a^{a}_\m / \partial\r$,
the result can be written as
$ig\,{\rm tr}\, a_{\m,a'} [\tilde\o_{c'} \,,\, \partial a_\m / \partial\r]$.
This yields the $\r$-derivative of ${\rm det}^\half C_0$, since
\beq
  {\partial \over \partial\r}\, a_{\m,c'}
  =  {\partial \over \partial\r}\, D_\m\, \tilde\o_{c'}
  =  - ig  \left[ {\partial a_\m \over \partial\r} , \tilde\o_{c'} \right]
     + D_\m\, {\partial \tilde\o_{c'} \over \partial\r}\,.
\label{rr}
\eeq
The contribution of the last term is zero after integration by parts
(which is allowed since $\partial \tilde\o_{c'} / \partial\r$ vanishes
rapidly enough at infinity) using $D_\m\, a_{\m,a'} = 0$.
(The above is an example of the commutator formulae of appendix~B.3.)

Last we discuss all diagrams with the longitudinal projector.
Compare the last two diagrams in Fig.~6. These diagrams have
the same topology, but the meaning of the various elements is different.
Recalling the spectral decomposition of the ghost propagator,
the analog of the zero mode $b_{;n}$ is now
$\O^a c^a_p$ where $c^a_p$ is a ghost eigenstate
(see below \seeq{GB0}, and Appendix~B.4).
Similarly, the corresponding inverse eigenvalue (associated with
the ghost propagator)
is the analog of $C_0^{-1}$ (the thick dashed line).
Thus, those elements which constitute
$\ds\z_n$ in the third diagram, correspond in the last diagram to
$\ds \o_p = \bmath(c^a_p \bmath| \ds \o^a \bmath)$ (both to leading order).

Similarly to what we did with the bosonic modes $b_{;n}$,
the insertion of $\O^a c^a_p$ on a fermion line
can be traded with a local gauge transformation with parameter $c^a_p$,
acting on the field(s) at the end(s) of the line.
The sum of the resulting diagrams, where the gauge transformation
acts on the fields of $\co$, vanish by gauge invariance of $\co$.
In addition there are diagrams that vanish thanks to gauge invariance
of the semi-classical jacobian, or of $\ds \o_p$  itself.
One last piece is provided by the ghost-loop diagram that has remained
from the second term on the \rhs of Fig.~8.
It is recognized as a contribution to the gauge transformation
of the product $c^a_p\, \lvec\O{}^{\dagger a}\, \G\hJ$ occurring in
$\ds \o_p$.

The above completes the diagrammatic analysis of \seeq{anml},
{\it except} for counterterm diagrams. These will be discussed
in the next subsection.

%%%%%%%%%%%%
 \newpage
\sbsect{3.3}{One-loop renormalization}

  Several types of loop diagrams occur in the calculation of \seeq{anml},
and the corresponding divergences are renormalized by counterterms.
In this subsection we complete the proof of \seeq{anml},
assuming that renormalized perturbation theory is supersymmetric.
(This statement means that the perturbative S-matrix is supersymmetric,
and that perturbative matrix elements of composite operators fall into
supermultiplets.)
We assume that the counterterms were constructed
using the background field method~\cite{Ab}.
For definiteness we will refer to dimensional
regularization, but the discussion generalizes to any other consistent
regularization as well. As mentioned earlier, the below arguments are
actually sufficient to prove \seeq{dOint} after renormalization,
whenever its \rhs consists of tree diagrams only.

As explained in the previous subsections, the two-loop bubble diagrams
that occur on the \lhs of \seeq{anml} are multiplied by zero.
Thus, we need not concern ourselves here with the corresponding counterterm
diagrams. (There {\it are} delicate points in the renormalization
of the bubble diagrams, see ref.~\cite{MRS}; these will become relevant
at higher orders.)

We next consider the ghost-loop diagram
contained in the second term on the \rhs of Fig.~8, and all the
(loop) diagrams with the longitudinal projector $P^\parallel$.
As discussed in the previous subsection, the sum of these diagrams is
zero by gauge invariance. Since (background) gauge invariance is preserved by
the regularization, this cancelation continues to hold.

The remaining divergences arise from one-loop diagrams
containing boson or fermion propagators, and no ghost propagator.
Renormalization of these diagrams requires us to face two
obstacles. First, dimensional regularization
treats a loop as a single entity. Equations like \seeq{GF},
\seneq{GB0} or\seneq{GF0} do not hold for $d-4 \ne 0$.
We must therefore replace these equations by ones that hold for arbitrary $d$.
The second (related) complication is that SUSY, as expressed by \seeq{susEoM},
is broken by terms proportional to $d-4$.

We now explain how dimensional regularization is applied to an
instanton-sector diagram.
First, repeating $n$ times the recursion relation for
the generic fermion propagator $G_F$, \seeq{GFGvac}, we obtain
\beq
  G_F
  = \left( 1 - P_F^{\rm exact} \right) G_F^{\rm vac}
    \sum_{k=0}^n  \left(- \tilde{V}_F G_F^{\rm vac} \right)^k
    + G_F \left(- \tilde{V}_F G_F^{\rm vac} \right)^{n+1} \,.
\label{rcr}
\eeq
(The superscript ``vac'' denotes free-field quantities,
and $\tilde{V}_F = L_F - L_F^{\rm vac}$. Notice the terms
with the projector $P_F^{\rm exact}$, which ``cut open''
any loop containing $G_F$.)
We now take $n$ (finite and) high enough,
such that any loop containing $L_F$ times the {\it last} term in \seeq{rcr}
will be finite. \seEq{GF} may be applied to the last term,
and the result is
\beq
  L_F\, G_F =
  \left( L_F^{\rm vac} + \tilde{V}_F \right) G_F^{\rm vac}
  \sum_{k=0}^{n-1} \left(- \tilde{V}_F G_F^{\rm vac} \right)^k
  + L_F^{\rm vac} G_F^{\rm vac} \left(- \tilde{V}_F G_F^{\rm vac} \right)^n
  - P_F^{\rm exact} \,.
\label{dne4}
\eeq
We have used (see below \seeq{GFGvac})
\beq
  P_F^{\rm exact} \left(- \tilde{V}_F G_F^{\rm vac} \right)^n
  = P_F^{\rm exact} \,.
\label{Pcrc}
\eeq

\seEqs{rcr} and\seneq{dne4} provide the key for regularization.
Comparing \seeq{dne4} with \seeq{GF},
we see that the term with $P_F^{\rm exact}$ has been
separated out explicitly.
The remaining terms, which replace the delta-function in \seeq{GF}, are
identical to what one would have found in a standard perturbative expansion.
Any (loop) diagram containing (only)
these perturbative-expansion terms is amenable to dimensional
(or any other) regularization.
(If $L_F^{\rm vac}\, G_F^{\rm vac}(x,y) = \d^4(x-y)$ is true in
the regularized theory, then \seeq{dne4} reduces to \seeq{GF}.)

(In SYM, the bosonic and fermionic propagators in the instanton sector
obey \seeqs{GB0} and\seneq{GF0} respectively,
and the above analysis is applicable to both of them.
In the bosonic equivalent of \seeq{dne4}, $P_F^{\rm exact}$
is replaced by $P_B + P^\parallel$.
The ghost propagator contained in the longitudinal projector
(cf.\ \seeq{longP})
obeys the standard recursion relation\seneq{GG0}.
In the most general case $G_B$ is defined by \seeq{GB}.
The last term in this equation
cuts open any loop since (like projectors) it involves a localized source.
As for $\hG_B$, it is convenient (cf.\ Appendix~B.6)
to consider first $\x \ne 0$.
Then $\hG_B$ obeys \seeq{hatGB} and the standard recursion
relation\seneq{GG0}. Using \seeq{GbGgh}, all term involving $1/\x$
may be replaced by expressions that have a smooth $\x \to 0 $ limit.)

Let us now examine, in comparison, the background-field
fermion propagator in the vacuum sector.
This propagator obeys the standard recursion relation\seneq{GG0}
and, therefore, identities analogous
to \seeqs{rcr} and\seneq{dne4} but {\it without} projector terms.
Similarly, the vacuum-sector identities
for the background-field transverse boson propagator
contain the longitudinal projector, but no analog of the projector $P_B$.

When the manipulations based on
\seeq{susEoM} are carried out in the vacuum sector,
all (finite) violations of SUSY that survive the removal
of the regularization can be canceled by an appropriate
({\it non}-supersymmetric) set of counterterms~\cite{pt}.
The cancelation of SUSY violations arising from the loop regularization
continues to hold in the instanton sector.
As a result, whenever $L_B\,G_B$ or $L_F\,G_F$ arise inside
some (amputated) one-loop diagram (from
the application of \seeq{susEoM}) this yields a {\it renormalized}
diagrammatic identity in the instanton sector which differs
from the corresponding one in the vacuum sector
precisely by the {\it projector} terms $P_F^{\rm exact}$ and $P_B$.
As shown in the previous subsection, all diagrams with an insertion
of $P_F^{\rm exact}$ cancel by antisymmetry, while those containing
$P_B$ combine with diagrams that involve the
functional jacobian to form the total $\z_n$-
(and in particular $\r$-) derivatives.

We now explain how the last statement works in practice.
Consider first the loop diagram on the \lhs of Fig.~7,
which originates from the self-energy diagram in Fig.~12(a).
There is a corresponding counterterm diagram, shown in Fig.~12(b).
(As mentioned earlier, the sum of one-loop
tadpoles in Fig.~8 is zero in SYM.)
Now, the \rhs of Figs.~7 and~8 also contains loop diagrams, to which
we have to apply \seeq{dS2}, cf.\ Fig.~9. That equation, which is an expression
of SUSY  of the action, is broken in dimensional regularization
by an amount proportional to $d-4$. When multiplied by
$(d-4)^{-1}$ coming from the loop divergence,
a (finite) explicit breaking of SUSY may result.
This explicit breaking is canceled, however, by an appropriate
{\it non}-supersymmetric set of counterterms.
Consequently, the renormalized equalities represented by
figs.~7, 8 and~9 hold
after adding the counterterm diagram.

A second set of one-loop diagrams can be found in Figs.~4, 5 and~6
(provided both of the external legs go to the same spacetime point).
The divergences in these figures correspond to the composite operators $\l\l$
or $\co_{\m}^{\,i\b}$ (\seeq{O}), or to their SUSY variations.
For example, $\ds (\l\l) = (i/\sqrt{2})\, \s_{\m\n} F_{\m\n}\, \l$.
However, in general
\beq
  \ds [\l\l]^{(1)} \ne (i/\sqrt{2}) [\s_{\m\n} F_{\m\n}\, \l]^{(1)} \,,
\label{ct}
\eeq
where $[\cdots]^{(1)}$ is the corresponding one-loop counterterm.
The \lhs of {\it inequality}\seneq{ct} is shown in Fig.~13(a),
while its \rhs is Fig.~13(b). (The counterterms are normally $O(g^2)$,
but when they involve a classical field they become $O(g)$.)
Again, the counterterms are designed to compensate for any discrepancy
that may have arisen from the loop regularization. With the counterterm
diagrams added, the equalities of Figs.~4, 5 and~6 hold after renormalization.
Finally, the loops of Fig.~10 are treated in a similar manner.
We leave it for the reader to work out the corresponding counterterm diagrams.
This completes the proof of \seeq{anml}.

At higher orders, the renormalization of \seeq{anml}
is more complicated, because one must deal with the divergences
arising from the coupling between the discrete- and the continuous-index
parts of the jacobian. This problem was
addressed in ref.~\cite{MRS}. A related problem is that one must deal with
the divergences of $\ds\r$.  In subsection~3.6 below we
use a renormalization-group argument to determine the
next-order logarithmic corrections to \seeq{anml}.

%%%%%%%%%%%%
%\newpage
\sbsect{3.4}{SUSY theories with matter}

It is easy to generalize \seeq{anml} to SU(2) SQCD.
Let the number of flavors be $N_f$. (By convention, for SU(2)
each flavor corresponds to two chiral supermultiplets in
the fundamental representation. Each (massless) Weyl-fermion field has
one zero mode.) One has
\beq
  \svev{\ds_\a \,\, \co_{\m}^{\,i\b}(0) \,\, \l\l(y) \,\,
  \prod_{k=1}^{N_f} \widetilde{\j_k\j_k}(0) }
  = - { 2^{12.5} 3 \p^2 \LYM^{6-N_f} \over 7 g^7}\,
  {\bar\h_{i\m\t} [\s_\n \, \bar\s_\t]_\a^{\,\b} \, y_\n \over  (y^2)^3} \,,
\label{anmlsqcd}
\eeq
where
\beq
  \widetilde{\j_k\j_k}(q) = \int d^4x\, e^{iqx}\, \j_k\j_k(x) \,.
\label{Ftrans}
\eeq
We have assumed $y^2 m^2 \ll 1$ where $m$ is a generic matter-field mass.
This implies that only the zero modes contribute in \seeq{Ftrans}.
As shown in subsection~4.1 below, both \seeq{anml} and \seeq{anmlsqcd}
cannot be modified by subtractions and, hence, constitute
a supersymmetry anomaly. \seEq{anmlsqcd} reflects a
general feature, namely, the anomalous term may be a polynomial in
some of the external momenta (here the polynomial in $q_\m$
reduces to a constant).

For SU(N), One should replace $\co_{\m}^{\,i\b}$
by an operator whose generic structure is
$\l (D_\m\, \l) (D^2 \l) (\l\l)^{N-2}$. 
The generalization of \seeq{anml} will have on its \rhs 
$c_N\,g^{1-4N}\,\LYM^{3N-N_f}$ times similar spacetime dependent factors. 
We have not worked out the numerical constant $c_N$. 

%%%%%%%%%%%%
\sbsect{3.5}{Local form of the anomaly}

The local version of \seeq{dOint} is obtained by applying the
SUSY variation of the fields only at the point $x=z$.
To this end, one multiplies the \lhs of \seeqs{var} and\seneq{varF}
by $\d^4(x-z)$. One can now repeat the construction of Sec.~2.
An easy way to read the local variations from those of Sec.~2.1
is to associate the  $\d^4(x-z)$ factor with the matrix $\G$.
Whenever $\G\hJ$ occurs inside some integral, it is replaced
by the corresponding integrand at $x=z$.
For the local variations of the collective coordinates
one obtains
\beq
  \ds\z_m(z) = C^{-1}_{mn}
  \left\{ b_{;n}^\dagger(z)\, \G\hJ(z)
  + ig \int d^4x\, b_{;n}^\dagger(x)\, T^a \hb(x)\, \d\o^a(x;z)
  \right\} \,,
\label{dzetax}
\eeq
where
\bqry
  \ds\o^a(x;z) & = & \cf^{-1}_{ab}(x,z)\, \O^{b\dagger}(z)\, \G\hJ(z) \NON
  & & + ig \int d^4y\, \cf^{-1}_{ab}(x,y)\,
  b_{;m}^\dagger(y)\, T^b \hb(y)\;
  C^{-1}_{mn} \; b_{;n}^\dagger(z)\, \G\hJ(z) \,.
\label{domegax}
\eqry
The variation of the measure is zero as before, and the local
form of \seeq{dOint} reads
\beq
  \partial_\m \svev{ \hat{S}_\m(z) \, \co}
  = \mbox{contact terms}
  - \int d^n\z\; {\partial\over \partial\z_n}
  \left(e^{-\scl-W_1-W_2}\,\svev{\ds\z_n(z) \, \co}\subv{\z}\right) \,,
\label{dO}
\eeq
where the contact terms are the expected variations
of the (multi)local operator $\co$.
The last term in \seeq{dO} is, by definition,
a matrix element of the operator $\partial_\m \hat{S}_{\m}(z)$.
Another way of reaching these results is to keep track of the
diagrammatic identities of Sec.~3.2, but now without performing
the integral in \seeq{dmuSmu}.
For the local form of \seeq{anml} we find
\beq
  \partial_\n \svev{ \hat{S}_{\n\a}(z) \,\, \co_{\m}^{\,i\b}(0) \,\, \l\l(y) }
  = \mbox{contact terms}
  - \svev{ \partial_\n \hat{S}_{\n\a}(z) \,\, \co_{\m}^{\,i\b}(0) \,\, \l\l(y)}
  \,,
\label{dS}
\eeq
where the matrix element of $\partial_\n \hat{S}_{\n\a}(z)$ is
\bqry
  \hspace{-12mm}
  \svev{ \partial_\n \hat{S}_{\n\a}(z) \,\, \co_{\m}^{\,i\b}(0) \,\, \l\l(y)}
& = &  - 2^{10}\p^6 g^{-8} \LYM^6 \, \lim_{\r\to 0} \, \r^3 \,
    \svev{(\ds_\a\,\r(z)) \,\, \co_{\m}^{\,i\b}(0) \,\, \l\l(y)}_{\!\r}
\NON
& \hspace{-20mm} = \hspace{20mm} & \hspace{-20mm}
  { 2^{11.5} \LYM^6 \over 7 g^7 }\,
  { \h_{b\r\n}\, z_\n \over (z^2)^2 }\,
  { (y - z)_\k \over ((y-z)^2)^2 }\,
  { \h_{b\l\s}\, y_\s \over (y^2)^2 }\,
  \bar\h_{i\m\t} [\s_\r\, \bar\s_\k\, \s_\l\, \bar\s_\t]_\a^{\,\b} \,.
\label{local}
\eqry
\seEq{local} is consistent with locality of
$\partial_\n \hat{S}_{\n\a}(z)$.
As expected, one recovers the \rhs of \seeq{anml}
by integrating over $z$ in \seeq{local}.

%%%%%%%%%%%%
%\newpage
\sbsect{3.6}{Renormalization group considerations}

At the next-to-leading order, the Hilbert-space surface term on the last
row of \seeq{lim} may pick up logarithmic corrections.
Depending on the power of $\log\r$, the result
of the $\r \to 0$ limit could blow up, remain finite,
or vanish.

In fact, the logarithmic corrections may be either $\log\r$
or $\log y^2$. This is because the limit in \seeq{lim} is not uniform:
while the effective range of the $x_0$-integration scales to
zero with $\r$, the distance scale $y^2$ is kept fixed.
Taking this delicate point into consideration,
we now give a renormalization-group (RG) argument
that the $\log\r$ factors arising from 
$\ds_\a\r$ and $\co_{\m}^{\,i\b}$
should cancel those arising from two-loop bubble diagrams.

Given a renormalization point $\m$, we rewrite \seeq{anml}
to one higher order as
\beq
  g^5(\m) \svev{\ds_\a \,\, \co_{\m}^{\,i\b}(0) \,\, \l\l(y)}
  = - \L_2^6 \, g^2(\m) \,
  { 2^{12.5} 3 \p^2
  \bar\h_{i\m\t} [\s_\n \, \bar\s_\t]_\a^{\,\b} \, y_\n \over 7 (y^2)^3
  }\,
  (1 + {\rm logs}) \,.
\label{anml2}
\eeq
The two-loop RG-invariant scale is
\beq
  \L_2^6 = \m^6 \, g^{-4}(\m) \, \exp(-8\p^2/g^2(\m)) \,.
\label{Ltwo}
\eeq
The reason for including the $g^5(\m)$ factor in \seeq{anml2} is as follows.
First, the operator $\ds_\a\r$ is RG-invariant
because the collective coordinate $\r$ is independent of the renormalization
point $\m$, and the SUSY variation $\d_\a$ respects RG-invariance.
More generally,
the basic building blocks of RG-invariant operators are $gF_{\m\n}$
and $g \l$ (or $g\bar\l$). We have therefore multiplied $\co_{\m}^{\,i\b}(0)$
and $\l\l(y)$ by appropriate powers of $g(\m)$.

We now analyze the expected $\log\r$ corrections,
starting with the contribution
of the two-loop bubble diagrams~\cite{MRS}.
The semi-classical instanton measure\seneq{msr} involves
the product $\L_2^6\, g^{-4}(\m)$. RG invariance requires that,
for size-$\r$ instantons, the logarithms arising from the bubble
diagrams should turn this product into $\L_2^6\, g^{-4}(\r^{-1})$.
Next consider the operator $\ds_\a\r$. As can be seen from \seeqs{drho}
and\seneq{Crr}, the coupling constant explicitly
contained in $\ds_\a\r$ is associated
with the classical field.
The logarithmic corrections to $\ds_\a\r$ should modify
that $g(\m)$ to $g(\r^{-1})$.

Finally, RG invariance of $g^3 \co_{\m}^{\,i\b}$
means that the exponential involving its anomalous dimension,
$\exp[\int_{g(\m)}^{g(\r^{-1})} dg'\, \g_\co(g')/\b(g') ]$,
must be proportional to $g^3(\r^{-1})/g^3(\m)$. In other words,
the renormalization of $\co_{\m}^{\,i\b}$ provides $\log\r$ factors
which are just enough to turn $g^3(\m)$ into $g^3(\r^{-1})$. 
Putting together the expected $\log\r$ factors arising from all sources
we find that they cancel each other.
(Of course, it is important to confirm this conclusion by a direct
next-order calculation.) Moreover, all higher-order
corrections to both $\g_\co(g)$ and $\b(g)$ cannot give rise
to $\log\r$ terms in the solution of the RG equation.

Finally, if the scale $y^2$ is varied, one expects the renormalization
of $\l\l(y)$ to generate $\log y^2$ factors that turn
$g^2(\m)$ into $g^2(|y|^{-1})$. We thus expect the next-order result to be
\beq
  g^5(\m) \svev{\ds_\a \,\, \co_{\m}^{\,i\b}(0) \,\, \l\l(y)}
  = - \L_2^6 \, g^2(|y|^{-1}) \,
  { 2^{12.5} 3 \p^2
  \bar\h_{i\m\t} [\s_\n \, \bar\s_\t]_\a^{\,\b} \, y_\n \over 7 (y^2)^3
  } \,.
\label{anml3}
\eeq

Assume now that $y^6 \LYM^6 \ll 1$.
While SYM is confining, the contributions of sectors with
additional instanton-antiinstanton {\it pairs} should be damped compared
to \seeq{anml} (or \seeq{anml3}) by extra powers of $y^6 \LYM^6$.
What we are calculating is thus the leading-order result in
an expansion in the physical parameter
$y^6 \LYM^6 = \exp(-8\p^2/g^2(y^{-1}))$.

%%%%%%%%%%%%%%%%%%%%%%%%%%%%%%%%%%%%%%%%%%%%%%%%%%%%
 \newpage
\vspace{5ex}
\noindent {\large\bf 4.~~The } $\Bmath{\r \to 0}\;$  {\large\bf limit}
\vspace{3ex}
\secteq{4}

The limit of a vanishing-size instanton, $\r \to 0$,
may be singular in correlation functions that involve operators of
sufficiently high dimension. In this section we investigate several aspects
of this limit. Our main result (subsection~4.1)
is that the anomalous Ward identity
\seeq{anml} cannot be modified by non-perturbative subtractions
and, hence, constitutes a supersymmetry anomaly.
In subsection~4.2 we discuss point splitting, and explain
why it does {\it not} regularize
the operator $\co_{\m}^{\,i\b}$ occurring in \seeq{anml}.
In subsection~4.3 we describe an ambiguity that arises in
a case where point splitting can be used and comment on its
implications. In subsection~4.4 we explain how the anomaly
should arise on the lattice.

%%%%%%%%%%%%
\sbsect{4.1}{Non-perturbative subtractions}

In this subsection we first examine separately the two correlators
$\vev{(\ds_\a\, \co_{\m}^{\,i\b}) \,\, \l\l}$ and
$\vev{\co_{\m}^{\,i\b} \,\, (\ds_\a\, \l\l)}$,
whose difference is \seeq{anml}.
We show that the corresponding $\r$-integrals
are convergent in the limit $\r \to 0$.

For comparison, we next consider another Ward identity where the corresponding
integrals are divergent for $\r \to 0$.
In that case non-perturbative subtractions are needed,
and may in fact be used to recover SUSY.
Having worked out this explicit example,
we list the general properties of non-perturbative subtractions.
It then easily follows that \seeq{anml} {\it cannot} be modified by any
non-perturbative subtraction.

In subsection~3.1 we have computed the correlation function
$\vev{(\ds_\a\,\r) \,\, \co_{\m}^{\,i\b}(0) \,\, \l\l(y)}_{\!\r}$
(\seeq{lim}) in the limit $\r \to 0$. Let us now generalize the
calculation to any $\r$.
The operator product on the \lhs
of \seeq{prop} may be written as
$2 \l^{SC}(y;\dt\a)\, \l^{{\rm ind}}(y;\a)$
where
\beq
  \bar\s_\m^{\dt\b \b} D_\m^{ab} \, \l^{{\rm ind,}\,b}_\b(\a)
  = {g^2\, \bar\s_\m^{\dt\b \a} \over 2^{4.5} \p^2} \,
    {\partial a^{a}_\m \over \partial \r} \,.
\label{indl}
\eeq
In a regular gauge, the solution is
\beq
  \l^{{\rm ind,}\, b}_\b(y;\a)
  = {\r\, g \over 2^{3.5} i \p^2} \,
    { \s^{b\a}_\b \over (y-x_0)^2 + \r^2 } \,.
\label{indlsol}
\eeq
Performing the $x_0$-integration we find
(ignoring irrelevant constants)
\beq
  \r^3 \svev{(\ds_\a\,\r) \,\, \co_{\m}^{\,i\b}(0) \,\, \l\l(y)}_{\!\r}
  \propto
  {\LYM^6\, \bar\h_{i\m\t} [\s_\n\, \bar\s_\t]_\a^{\,\b} \over g^7}\,
  {y_\n \, h(\r^2/y^2)\over (y^2)^3 }  \,,
\label{anyrho}
\eeq
where $h(s)$ can be represented as a Feynman-parameters integral
(see Appendix~E of ref.~\cite{it}) and
\beq
  h(s) = \left\{ \begin{array}{ll}
    1 + h_1\, s + h_2\, s^2 + \cdots, & \quad s \ll 1 \,, \\
    h_{-3}\, s^{-3}  + h_{-4}\, s^{-4} + \cdots, & \quad s \gg 1 \,.
  \end{array} \right. 
\label{rhovery}
\eeq
\seEq{anml} thus involves the integral
\beq
  {y_\n \over (y^2)^3 } 
  \int d\r \, {\partial \over \partial\r}\,  h(\r^2/y^2) \,.
\label{intrho}
\eeq
This integral is convergent, with the main contribution to
it arising from $\r^2 \sim y^2$. For small $\r^2/y^2$, 
the integrand in\seneq{intrho} behaves like $\r y/(y^2)^4$.

We now turn to the correlators
$\vev{(\ds_\a\, \co_{\m}^{\,i\b}(0)) \,\, \l\l(y)}$
and
$\vev{\co_{\m}^{\,i\b}(0) \,\, (\ds_\a\, \l\l(y))}$.
We claim that both the small- and the large-$\r$ behavior
of the corresponding integrals is the same as 
in \seeq{intrho} above.
In the limit $\r \to \infty$ this can be established 
simply on dimensional grounds.
Turning to the $\r \to 0$ limit we first examine
$\vev{(\ds_\a\, \co_{\m}^{\,i\b}(0)) \,\, \l\l(y)}$.
At tree level, if the product of zero modes at the point $y$
is $\l^{SC}(y)\, \l^{SS}(y)$ then the  presence of $y/(y^2)^4$ is
evident and, on dimensional grounds, implies a $\r y/(y^2)^4$ behavior 
of the $\r$-integrand.
Alternatively, if one places both superconformal modes at the point $y$
then the factorized $y$-independent piece,
$\int d^4x_0\, \vev{\ds_\a\, \co_{\m}^{\,i\b}(0)}_{x_0,\r}$, is zero because
the $x_0$-integral is odd (compare \seeq{x0int}).
The first non-zero contribution again goes like $\r y/(y^2)^4$.
At the one-loop level one reaches the same conclusion
by  examining the (small-$\r$ behavior of the) fermion propagator
in the relevant partial wave. 
Having established this (large- and) small-$\r$ behavior of
$\vev{(\ds_\a\, \co_{\m}^{\,i\b}(0)) \,\, \l\l(y)}$,
\seeqs{anyrho} to\seneq{intrho} imply the {\it same}
behavior for the $\r$-integrand in
$\vev{\co_{\m}^{\,i\b}(0) \,\, (\ds_\a\, \l\l(y))}$.
(Reaching this conclusion directly is more difficult:
the correlator involves a diagram with a  {\it vector-boson}
propagator connecting the points 0 and $y$, and in an instanton background
this propagator decreases very slowly~\cite{BCCL};
see, however, ref.~\cite{AM}.)

Before turning to the physical implications of \seeq{anml},
we wish to explain in what way things could be different.
To this end, we consider another anomalous SUSY Ward identity,
which is
\beq
  \svev{\bards_{\dt\a}\,\, \l\l F^3_{ijk}(0)\,\, \s F \l_\a(y)}
  = - \e_{ijk}\, { 2^{22.5} \p^4 \LYM^6 \over g^{11} (y^2)^4}\,
    (\s_\m)_{\a\dt\a}\, y_\m \,,
\label{kon1}
\eeq
where
$\s F \l_\a \equiv \s^{\m\n}_{\a\b} F_{\m\n}^c \l^{c\b}$ and
\beq
  F^3_{ijk} \equiv \e_{abc} F^a_{\m\n} F^b_{\l\r} F^c_{\s\t}
  \bar\h_{i\m\n} \bar\h_{j\l\r} \bar\h_{k\s\t} \,.
\label{f3}
\eeq
The calculation is similar to subsection~3.1, and is in fact simpler.
$\bar\ds_{\dt\a} \r$ involves a $\l$ field,
which is saturated by one of the $\l^{SC}$ modes,
while the other $\l^{SC}$ goes to the operator $\s F \l_\a$.
Note that \seeq{kon1} exhibits a $y^{-7}$ fall-off
which is faster than the $y^{-5}$ fall-off in \seeq{anml}.
This kinematic difference will turn out to play an important role.

We now show that the operator $\l\l F^3_{ijk}$ requires a
{\it non-perturbative subtraction}.
Consider the following correlator
\beq
  \svev{\l\l F^3_{ijk}(0)\,\, \l\l(y)} \sim \e_{ijk}\,
  {\LYM^6\over g^{11} (y^2)^3} \int {d\r\over \r}\,,
\label{div}
\eeq
where on the \rhs we have indicated the small-$\r$ behavior.
The non-perturbative logarithmic divergence at small $\r$
can be handled as follows.
We first restrict the integral to $\bar\r\le\r<\infty$
where $0<\bar\r$. A subtracted operator is defined via
\beq
  [\l\l F^3_{ijk}] = \l\l F^3_{ijk}
  -  \e_{ijk}\, c_3 \, \log(\m\bar\r) \, {\LYM^6\over g^{11}} \,
  \bar\l\bar\l \,,
\label{rnorm}
\eeq
where $c_3$ is a suitable numerical constant.
The subtracted operator $[\l\l F^3_{ijk}]$ yields a finite $\bar\r\to 0$
result when substituted into the \lhs of \seeq{div}.

Like $\l\l F^3_{ijk}$, the operator $\bar\ds_{\dt\a}\,\l\l F^3_{ijk}$
requires a non-perturbative subtraction too. We now define
the renormalized operator as follows
\beq
  [\bards_{\dt\a}\,\l\l F^3_{ijk}] \equiv \bards_{\dt\a}\,[\l\l F^3_{ijk}]
  + \e_{ijk}\, {2^{21}\p^8 \LYM^6 \over 9 g^{11}}\,
  \bards_{\dt\a}\,\bar\l\bar\l \,.
\label{drnorm}
\eeq
We have chosen a manifestly non-supersymmetric subtraction.
The finite, last term in \seeq{drnorm} was chosen to cancel the \rhs
of \seeq{kon1}. Using the vacuum-sector result
\beq
  \svev{\bar\s F \bar\l_{\dt\a}(0)\,\, \s F\l_\a(y)}_{\rm vac}
  = -{36 i \over \p^4 (y^2)^4} (\s_\m)_{\a\dt\a}\, y_\m \,,
\label{FlFl}
\eeq
\seeq{kon1} becomes
\beq
  \svev{ [\bards_{\dt\a}\, \l\l F^3_{ijk}(0)]\,\, \s F\l_\a(y)
  + [\l\l F^3_{ijk}(0)]\,\, \bards_{\dt\a}\, \s F\l_\a(y)}
  = 0 \,.
\label{kon1s}
\eeq
\seEq{kon1s} means that SUSY has been recovered in the limit $\bar\r \to 0$
after performing the above non-perturbative subtractions.
Moreover, by considering additional Ward identities,
one can verify that the renormalized operators respect
the SUSY {\it algebra}. (Recovering the supermultiplet
structure of composite operators by suitable subtractions is a reminiscent
of the so-called Konishi anomaly~\cite{K}.)

There are two lessons from the above example.
First, when non-perturbative subtractions play a role,
there is at least a chance that SUSY will be recovered.
The other lesson has to do with the general properties of
non-perturbative subtractions.
A divergence at $\r \to 0$ always arises from a kinematic
situation where the instanton sits on top of an operator $\co(x)$
of a sufficiently high dimension.
Consider the instanton-sector correlation function
$\vev{\co(x) \prod_k \co_k(y_k)}$ and assume that
a $\r \to 0$ divergence arises only from the operator $\co(x)$.
The zero modes and the propagators
which are functions of $y_1, y_2, \ldots,$ become
$\r$-independent in the limit  $\r \to 0$.
(More generally, they are expandable in
powers of $\r$; this expansion is used when the leading
divergence is stronger than $\log\r$.)

As a result, the dependence on spacetime points of the $\r \to 0$ divergence
must be that of a {\it vacuum-sector} correlation function
$\vev{\co'(x) \prod_k \co_k(y_k)}_{\rm vac}$
involving another operator $\co'(x)$.
The operators $\co(x)$ and $\co'(x)$ must have the same quantum
numbers {\it except} for their chiral charge, where the mismatch
is given by the number of fermionic zero modes. Because of the
explicit $\L_1^6$ factor which appears in
instanton-sector correlation functions,
the dimension of $\co'$ must be smaller than that of $\co$ at least by 6.
(The difference between the dimensions of $\co(x)$ and $\co'(x)$
is 6 (for SU(2)) if the divergence is logarithmic;
if the divergence goes like some inverse power of $\r$,
the difference is 6 plus that power.)

Previously, we showed that there is no small-$\r$ divergence
in the correlation functions
$\vev{(\ds_\a\, \co_{\m}^{\,i\b}(0)) \,\, \l\l(y)}$
and
$\vev{\co_{\m}^{\,i\b}(0) \,\, (\ds_\a\, \l\l(y))}$.
It is now easy to generalize this result, and show that no
instanton-sector correlation function
can have a $\r \to 0$ divergence associated
with the operators $\co_{\m}^{\,i\b}$ or $\ds_\a\co_{\m}^{\,i\b}$.
If such a divergence were to arise in some correlation function,
then from the spacetime dependence of the divergence
one could read off what is the necessary non-perturbative subtraction.
However, there is {\it no} operator that qualifies
as a non-perturbative subtraction for $\co_{\m}^{\,i\b}$
or for $\ds_\a\co_{\m}^{\,i\b}$.
The mass dimension of $\co_{\m}^{\,i\b}$ is 7.5.
Therefore, for the subtraction one would need
an operator whose dimension is $7.5 - 6 = 1.5$.
Evidently, there is no gauge invariant operator with this dimension.
A similar conclusion applies to $\ds_\a\co_{\m}^{\,i\b}$.
Since \seeq{anml} cannot be modified by any non-perturbative subtraction,
there is no way to recover that SUSY Ward identity.
Hence, \seeq{anml} constitutes a {\it supersymmetry anomaly}.

One may wonder how the physical implications of \seeqs{anml} and\seneq{kon1}
can be so different. Apart from the above mentioned kinematic difference
($y^{-7}$ {\it vs}.\ $y^{-5}$ fall-off),
there are two more significant differences between the two
Ward identities. In \seeq{kon1}, only the fermionic zero modes and
the classical field occur on both sides.
As shown in ref.~\cite{itep}, the collective coordinates and the
fermionic zero modes constitute a finite supersymmetric system.
One does not expect that SUSY will be violated when these are the only
relevant degrees of freedom. On the other hand, in \seeq{anml}
the \lhs involves one-loop diagrams which are outside the scope
of the supersymmetric calculus of ref.~\cite{itep},
and the potential for an anomaly exists.

The other qualitative difference between \seeqs{anml} and\seneq{kon1}
is in the corresponding local Ward identities. In the case of \seeq{kon1},
the $\r \to 0$ limit and the integration over the point $z$,
where the SUSY variation is performed, do {\it not} commute.
Ignoring irrelevant constants,
the product $\l^{SC}(z)\,\partial a_\m/\partial\r(z)$
which occurs in $\bar\ds_{\dt\a}\r(z)$,
is $\r^2 z^2/(z^2+\r^2)^4$.
This becomes a delta-function, $\d^4(z)$, in the limit $\r \to 0$.
As a result, the corresponding local matrix element
is zero {\it except} for $z=0$. When the only violation of the naive
Ward identity is of this type, it is natural to associate this effect
with a modification of the composite-operator transformation rule,
and not with a non-zero $\partial_\m \hat{S}_\m(z)$.

In contrast, the local form of \seeq{anml} (namely \seEq{local})
is non-zero for any $z$.
Moreover, after the $z$-integration one recovers \seeq{anml}, hence
the $z$-integration and the $\r \to 0$ limit commute.
This result is {\it incompatible} with $\partial_\m \hat{S}_\m(z)=0$.

%%%%%%%%%%%%
\sbsect{4.2}{On point splitting}

In this subsection we address the following question: what
happens if one attempts to define the operators $\co_{\m}^{\,i\b}$
and $\ds_\a \co_{\m}^{\,i\b}$
via point splitting? Since $\co_{\m}^{\,i\b}$ involves three gaugino fields,
splitting off one of these fields requires the introduction of a parallel
transporter to maintain gauge invariance.
Our finding are: (a) point splitting does {\it not}
regularize these operators but, rather, introduces new divergences;
(b) in the instanton sector,
the $\r \to 0$ anomalous term of \seeq{anml}
is traded with another anomalous term arising from the SUSY variation
of the parallel transporter (whose evaluation is technically much
more complicated).

For our purpose it is enough to consider
the following partial point splitting
\beq
  \co_\m^{\,i\b}(x,\e)
  = \left( \hat{D}_\m^{eb}
    (\e_{bcd}\, \l^{c\a}(x) \, \s^{i\g}_\a \, \l^d_\g(x)) \right)
    U^{ef}(x,x+\e)  \left( (\hat{D}^2)^{fg} \, \l^{g\b}(x+\e) \right)\,,
\label{Oe}
\eeq
where
\beq
  U = U(x,x+\e) = \exp\left( ig \int_x^{x+\e} ds_\m\, A_\m \right) \,.
\label{ptrans}
\eeq
The SUSY variation of the parallel transporter is
\beq
  \ds_\a\, U = ig \int_x^{x+\e} dt_\m\,
  \exp\left( ig \int_x^t ds_\n\, A_\n \right)
  (\ds_\a\, A_\m(t))
  \exp\left( ig \int_t^{x+\e} ds_\n\, A_\n \right) \,.
\label{varptrans}
\eeq
Let us now replace $U$ by $\ds_\a\, U$ in \seeq{Oe}.
The $\bar\l$ field arising from $\ds_\a\, A_\m$ of \seeq{varptrans}
may be contracted with any one of the three $\l$ fields in \seeq{Oe}.
For at least one of them, the contraction involves same-color
fields and, hence, is given in the short-distance limit
by a {\it free} propagator.
This free propagator gives rise to a {\it quadratic divergence}
in the $t \to x$ (or $t \to x+\e$) limit at {\it finite} $\e>0$.

It is well known that point splitting may be used as a (gauge invariant)
regularization in cases such as the chiral anomaly and the Konishi
anomaly~\cite{K}. What is common to these examples is that
the gauge (and gaugino) fields are external, and only
matter fields are integrated over. When one integrates over
the gauge and the gaugino fields too, point splitting ceases to provide
a regularization as demonstrated above.

What happens if, nevertheless, one attempts to use the definition\seneq{Oe}
in the instanton sector with some other method of regularization
(e.g.\ dimensional regularization)? When the three $\l$'s
are not at the same point the $\r$-integrand is less singular,
and behaves qualitatively as
\beq
  \r^3 \svev{(\ds_\a\,\r) \,\, \co_{\m}^{\,i\b}(0,\e) \,\, \l\l(y)}_{\!\r}
  \sim
  {\LYM^6\, \bar\h_{i\m\t} [\s_\n \, \bar\s_\t]_\a^{\,\b} \over g^7}\,
  {y_\n \, h(\r^2/y^2)\over (y^2)^3 } 
  \left({\r^2\over \e^2 + \r^2} \right)^n \,,
\label{anyrhoe}
\eeq
for some $n \ge 2$. The new denominator $(\e^2 + \r^2)^{-n}$ arises
from the (possibly differentiated) wave function of the zero mode
at $x_\m=\e_\m$.
The numerator $\r^{2n}$ corresponds to the same wave function
in the limit $\e^2 \to 0$ where \seeq{anyrho} should be recovered.
Consequently expression\seneq{intrho} is replaced by
\beq
  {y_\n \over (y^2)^3 } 
  \int d\r \, {\partial \over \partial\r}\,
  \left\{ h(\r^2/y^2) \left({\r^2\over \e^2 + \r^2} \right)^{\! n}\, \right\}
  = 0 \,, \qquad \e^2 > 0  \,,
\label{intrhoe}
\eeq
The $\r \to 0$ surface term thus vanishes for $\e^2 > 0$.
In its place, there is now a new anomalous term, the one
arising from the variation of the parallel transporter\seneq{varptrans}.
After factoring out the leading $y/(y^2)^3$
dependence (and handling the new divergences described above!),
one is left with an integral over the
dimensionless variable $\r^2/\e^2$. In this integral,
the parallel tranporters of \seeq{varptrans} (with $A_\m$ taken
to be the classical field) are $O(1)$ and may not be
approximated by any truncation of their
Taylor series. The resulting integrals a very complicated
and we did not pursue this calculation any further.

%%%%%%%%%%%%
%\newpage
\sbsect{4.3}{A non-perturbative ambiguity}

If one uses the fermionic {\it equation of motion},
the operator $\co_{\m}^{\,i\b}$ may be traded with
\beq
  \tilde\co_{\m}^{\,i\b}
  = \left( \hat{D}_\m^{eb}(\e_{bcd}\, \l^{c\a}\, \s^{i\g}_\a\, \l^d_\g) \right)
    \left( \e_{efg}\, \s_{\m\n} F_{\m\n}^f \, \l^{g\b} \right)\,.
\label{tildeO}
\eeq
After contracting the two epsilon-tensors, one can write
$\tilde\co_{\m}^{\,i\b}$ as a sum of products
\beq
  \tilde\co_{\m}^{\,i\b} = \sum_k \co^{(1)}_k\, \co^{(2)}_k \,,
\label{Oi}
\eeq
where the (composite) operators $\co^{(1)}_k$ and $\co^{(2)}_k$
are all gauge invariant. We nay now consider the point splitting
\beq
  \tilde\co_{\m}^{\,i\b}(x,x+\e)
  =  \sum_k \co^{(1)}_k(x)\, \co^{(2)}_k(x+\e) \,,
\label{Oie}
\eeq
which does not require the introduction of parallel transporters.
In the SUSY Ward identity, the $\r \to 0$
surface term will vanish again (cf.\ \seeq{intrhoe}), and moreover
there will be no new anomalous terms since there are no parallel transporters.
Hence
\beq
  \svev{\ds_\a \,\, \tilde\co_{\m}^{\,i\b}(0,\e) \,\, \l\l(y)}
  = 0 \,, \qquad \e^2 > 0 \,.
\label{anmle}
\eeq

What is the significance of this result? Let us re-introduce the
short-distance cutoff $\bar\r$ on the $\r$-integral used in
subsection~4.1. With both $\bar\r$ and $\e$ non-zero, one has
\beq
  \svev{\ds_\a \,\, \tilde\co_{\m}^{\,i\b}(0,\e) \,\, \l\l(y)}
  \sim
  {\LYM^6\, \bar\h_{i\m\t} [\s_\n \bar\s_\t]_\a^{\,\b} \over g^7}\,
  {y_\n \over (y^2)^3 }\, I(\bar\r^2/y^2,\bar\r/\e) \,,
\label{barre}
\eeq
%\newpage \noindent
where
\bqry
  I(\bar\r^2/y^2,\bar\r/\e) & = & 
  \int_{\bar\r}^\infty d\r \, {\partial \over \partial\r}\,
  \left\{ h(\r^2/y^2) \left({\r^2\over \e^2 + \r^2} \right)^{\! n}\, \right\}
\NON
& = &  - 1 +
  \int_{\bar\r}^\infty d\r \, {\partial \over \partial\r}\,
  \left({\r^2\over \e^2 + \r^2} \right)^n 
\label{I}\\
& = & - \left({\bar\r^2\over \e^2 + \bar\r^2} \right)^n \,,
\nonumber
\eqry
and the approximation $\bar\r^2,\e^2 \ll y^2$ was made on the second row.
The result in now {\it ambiguous}. It
depends on the behavior of the ratio $\bar\r/\e$ in the limit.
If we send $\e \to 0$ before $\bar\r \to 0$, \seeq{anml} is recovered.
On the other hand, if we first send  $\bar\r \to 0$ and only later
$\e \to 0$, then the integrand on the second row behaves like
a {\it delta-function}, $\d(\r)$, and the result is zero.
More generally, if we take the limit with some fixed ratio $\bar\r/\e$,
then $\bar\r^2/(\e^2 + \bar\r^2)$
will interpolate smoothly between 0 and 1.

We have not been able to resolve this ambiguity in a completely satisfactory
way, but we have several comments. First, if
this problem could not be resolved, this would mean an ambiguity in
the {\it physical predictions} of the theory.
The latter conclusion seems to us highly unlikely.

We thus assume that a resolution of the ambiguity
should exist. The obvious next question is what effect
does this have on the SUSY  anomaly. Our conclusion is that the SUSY anomaly
exists regardless of what one does about that ambiguity.
We will soon argue that, in fact, the prescription of sending $\e \to 0$
{\it after} $\bar\r \to 0$ violates physical principles.
Nevertheless, suppose momentarily that that prescription was correct.
\seEq{anmle} would then hold, namely the Ward identity
involving $\tilde\co_{\m}^{\,i\b}$ would not be anomalous. However,
$\tilde\co_{\m}^{\,i\b}$ and $\co_{\m}^{\,i\b}$ are related only though
the (ultimately quantum) equation of motion.
The latter is evidently broken by the prescription
leading to \seeq{anmle}. Thus, even if one adopted that prescription,
the anomalous Ward identity with $\co_{\m}^{\,i\b}$, \seeq{anml},
would still stand,
with the conclusion that a SUSY  anomaly exists.

In a renormalizable quantum field theory, contributions from
the cutoff scale are normally associated with some (logarithmic or
power-law) divergence. The divergences are canceled by counterterms,
and the ambiguity in the finite parts of the counterterms
is fixed by renormalization conditions.

In contrast, the ambiguity we encounter here if of a new type.
We find a finite contribution of a varying magnitude, coming
from an infinitesimal neiborhood of the lower end of the $\r$-integral.
(In other words, from instantons whose size
is as small as the short-distance cutoff.)
Moreover, these undetermined short-distance contributions cannot be
attributed to different subtraction schemes, simply because (subsection~4.1)
they cannot be modified by any subtraction!
This state of affairs contradicts the principles
of renormalization and universality. The only way to avoid
this problem is to adopt the (unique) prescription
where these (otherwise undetermined) contributions vanish.
This prescription amounts to sending $\e \to 0$
{\it before} $\bar\r \to 0$ (equivalently $\e/\bar\r \to 0$).
As expected, in this case the Ward identities for $\tilde\co_{\m}^{\,i\b}$
and $\co_{\m}^{\,i\b}$ agree when the cutoff is removed.

%%%%%%%%%%%%
%\newpage
\sbsect{4.4}{SUSY and the lattice}

It is well-known that SUSY is broken by the lattice regularization.
Arguments that SUSY may be recovered in the continuum limit
were put forward in ref.~\cite{CV}. These arguments
are valid in the context of weak-coupling perturbation theory on the lattice.
However, as we now explain, they do not cover {\it non-perturbative}
effects.

In the appropriate momentum (or distance) range,
the SUSY anomaly should arise on the lattice exactly as in our continuum
treatment. The relevant distance range is defined by $a^2/y^2 \to 0$
where $a$ is the lattice cutoff,
and $y^6 \LYM^6 \ll 1$ but finite. This {\it scaling region}, which
would be probed in deep-inelastic scattering, is controlled by a small
coupling constant. The leading dynamical effect in that region
is the logarithmic evolution of the coupling constant.
This is covered by weak-coupling perturbation theory (both in the
continuum and on the lattice).

In the scaling region, {\it non-perturbative} effects
can be studied {\it systematically} on the lattice
just like in the continuum,
because they are controlled by the small parameter 
$\exp(-8\p^2/g^2(|y|^{-1}))$.
What needs to be done analytically is to
repeat the construction of the instanton-sector path integral
given in Appendix~B, but now on the lattice.
While to our knowledge this was never done in any detail,
we stress that there is no conceptual difficulty here.
In fact, the continuum change-of-variables (Appendix~B.2)
is a formal manipulation, because so is the original
path-integral measure in terms of position eigenstates.
In contrast, the corresponding change-of-variables on the lattice
should be completely well defined (albeit technically more involved).

We recall that, in any non-perturbative sector, the
collective coordinates are determined by listing
the (exact or approximate) {\it bosonic} zero modes,
that must be removed from the fluctuations spectrum
to avoid infra-red divergences.
The instanton-sector path integral on the lattice will, obviously,
feature the same set of collective coordinates as in the continuum
treatment. The anomalous SUSY Ward identity, \seeq{anml},
will thus arise also within the above-sketched analytic lattice treatment
of the instanton sector. (Of course, this applies also to the more general
result, \seeq{dOint}.)

It is interesting to re-examine the ambiguity of the previous subsection
in a lattice context. On the lattice the instanton action is
\beq
  S^{\rm inst}_{\rm latt}(\r) \sim {8\p^2 \over g^2(\r^{-1})}
  + {1\over g_0^2}\, f(\r/a) \,,
\label{latt}
\eeq
where $g_0$ is the bare lattice coupling.
The function $f(\r/a)$ accounts
for the lattice-induced change in the instanton action
when its size $\r$ becomes very small.
First, identifying $\e$ of the previous subsections
with the lattice cutoff $a$,
the above advocated $\e/\bar\r \to 0$ prescription means that
the lattice action should be chosen such that
$f(\r/a) \geqx 1$  for $\r/a \sim 1$~\cite{latt}.
This condition implies that the minimal instanton size,
while being vanishingly small in {\it physical} units,
is infinitely large in {\it lattice} units in
the continuum limit. (Typically, $f(\r/a)$ is a polynomial
in $a/\r$. If $f(\r/a) \sim (a/\r)^{2n}$,
unsuppressed instantons will be ones with $\r \ge a\,g_0^{-1/n}$.)

The opposite prescription, namely $\e/\bar\r \to \infty$, corresponds
on the lattice to choosing
$f(\r/a)$ to be {\it negative} for $\r/a \sim 1$.
This means that the probability of finding lattice-size instantons is
{\it enhanced}. In numerical simulations this enhances
lattice artefacts in physical quantities, which is undesirable.
More seriously, this may become a problem of principles as we now explain.

As discussed earlier, the matrix elements
of the point-split operator $\tilde\co_{\m}^{\,i\b}$
generically contain a piece that behaves like a delta-function $\d(\r)$.
Suppose now that lattice-size instantons are not suppressed.
When their Boltzmann weight is multiplied by the effective $\d(\r)$,
a non-zero contribution to the Ward identity 
involving $\tilde\co_{\m}^{\,i\b}$
is obtained {\it even} in the (would-be) continuum limit $g_0 \to 0$.
As before, this contribution depends
on the precise definition of $\tilde\co_{\m}^{\,i\b}$
but, moreover, it now violates {\it Lorentz} invariance. The reason is that
it originates from lattice-size instantons which must be sensitive
to the orientation of the lattice axes.
(As a by-product, the SUSY Ward identity for  $\tilde\co_{\m}^{\,i\b}$ will
not be recovered on the lattice either.)

In fact, the Boltzmann weight of lattice-size instantons must
either diverge or vanish in the limit $g_0 \to 0$.
(It cannot stay finite without ``infinite fine-tuning.'')
We conclude that, as a matter of {\it principle},
when the ambiguity of subsection~4.3 arises,
there will exist a consistent continuum limit
{\it only if}  the lattice action is chosen
such that lattice-size instantons are suppressed.
As in the previous continuum treatment,
the Ward identities for $\co_{\m}^{\,i\b}$
and $\tilde\co_{\m}^{\,i\b}$ will then agree in the continuum limit
of the lattice theory, and will both be anomalous.

%%%%%%%%%%%%%%%%%%%%%%%%%%%%%%%%%%%%%%%%%%%%%%%%%%%%
 \newpage
\vspace{5ex}
\noindent {\large\bf 5.~~Conclusions}
\vspace{3ex}
\secteq{5}

In this paper we have derived a general expression for the anomalous term
in SUSY Ward identities (\seeq{dOint}).
We have analyzed in detail Ward identities in the one-instanton sector
of SU(2) SUSY theories. We have found non-zero anomalous terms,
which moreover cannot be modified by subtractions, both
in SYM (\seeq{anml}) and in SUSY theories with matter (\seeq{anmlsqcd}).

The SUSY anomaly arises from a Hilbert-space surface term,
unlike the chiral anomaly which arises from non-invariance of
the path-integral measure.
In the SUSY case there seem to be no analog
of the familiar anomaly-cancelation mechanism of the chiral anomaly.
Similar anomalous Ward identities should exist with an SU(N) gauge group
(subsection~3.4), and with any matter content.
This would imply that the SUSY anomaly occurs in every
asymptotically-free four-dimensional theory.

Within a fully non-perturbative regularization (such as
a lattice cutoff) the operator $\partial_\m \hat{S}_{\m}$ 
is necessarily non-zero.
In the continuum limit, all matrix elements of
$\partial_\m \hat{S}_{\m}$ vanish to all orders
in perturbation theory but, according to our results,
$\partial_\m \hat{S}_{\m}$ has {\it non-perturbative} 
matrix elements which are not zero. 
As a crucial check, this should be confirmed by 
a calculation which starts directly from the non-zero 
$\partial_\m \hat{S}_{\m}$ of the regularized theory. 

Locality of $\partial_\m \hat{S}_{\m}$ in the regularized
theory implies its locality in the continuum limit.
The local version of the anomalous Ward identity (\seeq{local})
is consistent with this requirement.
In \seeq{local} the correlation between the points $0$ and $z$
must be mediated by a bosonic state. The $z$-dependence,
namely $z_\n/(z^2)^2$, is that of (the derivative of) a massless
boson propagator. However, the apparent single-particle
propagation could also be due to two colinear particles.
Usually the phase space for colinear propagation is zero,
but in the triangle diagram the fermion and the anti-fermion
do in fact become colinear in a special kinematic limit~\cite{FSBY}.
Whether a similar phenomenon takes place in the present case
is another important question. 
Finally, the implications of this new anomaly on the physical spectrum
of SUSY theories should be studied.

\vspace{10ex}
\centerline{\rule{5cm}{.3mm}}

%%%%%%%%%%%%%%%%%%%%%%%%%%%%%%%%%%%%%%%%%%%%%%
 \newpage
\vspace{10ex}
\noindent {\large\bf A.~~Notation}
\secteq{A}
\vspace{3ex}

\noindent For bosons we write:
\beq
  B_{\sst \tl{I}}(x) = b_{\sst \tl{I}}(x) + \hb_{\sst \tl{I}}(x) \,,
\eeq
where $b_{\sst \tl{I}}(x)$ and $\hb_{\sst \tl{I}}$ denote respectively
the classical and quantum parts of each Bose field $B_{\sst \tl{I}}(x)$.
We use a real notation where the generic index
${\st \tl{I}}$ runs over
the gauge field $A_\m$ and over both scalar fields $\F$ and their
complex conjugates $\F^\dagger$. One has
\bqry
  B_{\sst \tl{I}} & \Leftrightarrow & A_\m\,, \,\, \F\,, \,\, \F^\dagger\,,
\NON
  b_{\sst \tl{I}} & \Leftrightarrow &
                            a_\m\,, \,\, \varphi\,, \,\, \varphi^\dagger\,,
\label{bose}\\
  \hb_{\sst \tl{I}} & \Leftrightarrow & \a_\m\,, \,\, \f\,, \,\, \f^\dagger\,,
\nonumber
\eqry
where $\F$ stands for all scalar fields in the theory.
For the fermions we employ
a Majorana-like notation~\cite{CS89} which is valid in
vector-like SUSY theories, such as SQCD. For each pair
of Weyl fermions in complex conjugate representations
$\j^L,\j^L_c$ we define $\j=(\j^L,\bar\j^L_c)$, $\j_c=(\j^L_c,\bar\j^L)$.
We then write
\beq
  \J_{\sst \tl{I}}(x) \Leftrightarrow
  \l\,,  \,\, \j\,, \,\, \j_c\,,
\label{Fermi}
\eeq
where $\l$ is the (Majorana field) gaugino. We will also use the notation
\beq
  Q_{\sst \tl{I}}(x) = b_{\sst \tl{I}}(x) + \hq_{\sst \tl{I}}(x)
\eeq
where $\hq_{\sst \tl{I}}(x)$ stands for any quantum field, boson or fermion, 
and with the understanding that
$b_{\sst \tl{I}}(x)=0$ for fermions
(since there is no classical fermion field).
The notation $\hJ(x)$ will be used as a {\it shorthand}
for the mode expansion of the fermion fields, cf.\ \seeq{qp}.

We introduce a real inner product, defined
for given values $B^{(1)}_{\sst \tl{I}}$ and $B^{(2)}_{\sst \tl{I}}$
of the bosonic fields as
\beq
  \Bmath( B^{(1)}_{\sst \tl{I}} \Bmath| B^{(2)}_{\sst \tl{I}} \Bmath)
  \equiv \int d^4x \,
  \left
  ( A^{(1)}_\m A^{(2)}_\m + \sum
  \left( \F^{(1)\dagger} \F^{(2)} + \F^{(2)\dagger} \F^{(1)} \right)
  \right) \,.
\label{inner}
\eeq
The sum is over all scalar fields.
For fermions the inner product is
\beq
\hspace{-1mm}
  \Bmath( \J^{(1)}_{\sst \tl{I}} \Bmath| \J^{(2)}_{\sst \tl{I}} \Bmath)
  \equiv \int d^4x \,
  \left
  ( \overline\l{}^{(1)}\, \l^{(2)} + \sum
  \left( \overline\j{}^{(1)}\, \j^{(2)}_c
   + \overline\j{}^{(2)}\, \j^{(1)}_c \right)
  \right) \,.
\label{innerF}
\eeq
Here $\overline\l = \l^T C_{\rm conj}$ etc., where $C_{\rm conj}$
is the charge-conjugation matrix.
Finally, the inner product of two adjoint-representation fields
(e.g.\ the last term in \seeq{dzeta}) is simply the integral
of their product.

The SUSY variation of any boson is linear in the fermions.
We write this as follows
\beq
  \ds_\a B_{\sst \tl{I}} = \G_{\a {\sst \tl{I}\tl{J}}} \J_{\sst \tl{J}} \,.
\label{susvar}
\eeq
The index $\a$ runs over all supersymmetries (\ie we do not
distinguish here between $\ds$ and $\bar\ds$).

%%%%%%%%%%%%%%%%%%%%%%%%%%%%%%%%%%%%%%%%%%%%%%%%%%%%%%%%%%%%%%%%%%
%%%%%%%%%%%%%%%%%%%%%%%%%%%%%%%%%%%%%%%%%%%%%%%%%%%%%%%%%%%%%%%%%%
%\newpage
\vspace{5ex}
\noindent {\large\bf B.~~Transverse-field Feynman rules}
\secteq{B}
\vspace{3ex}

In this appendix we derive Feynman rules for any non-perturbative
sector of a weakly-coupled, renormalizable gauge theory.
We restrict the discussion to the background Landau gauge.
Both gauge and scalar classical fields will be considered.
We face two technical problems.
First, we are dealing with a coupled quantization problem, involving
both (discrete) collective coordinates
and (infinitely many) gauge degrees of freedom.
This was discussed in the literature before, see e.g.\ \cite{AR,LY}.
The second complication arises because we do {\it not} assume that
the background field is a solution of the classical field equations.
Many relations used in the quantization
around an exact solution do not apply here.

The below Feynman rules do not involve constraints~\cite{FY}.
As a result there are linear (classical tadpole) terms in the action.
The latter can be treated perturbatively provided
the background field is an approximate solution of the classical
equations. (Verification of the last statement must be done
on a case by case basis. We will discuss the one-instanton sector
of SUSY-Higgs theories elsewhere.)

This Appendix is organized as follows. In subsection~B.1 we define
the background gauge. Quantization in a given non-perturbative
sector is discussed in subsection~B.2. The mode expansion
of the quantum fields, and various commutator formulae,
are given in subsection~B.3. In subsection~B.4 we give
a compact formula for the path integral,
which treats the ordinary collective coordinates
and the infinitely-many gauge degrees of freedom
on a uniform basis. This formula provides the basis for the discussion
in Sec.~2. In subsection~B.5 we defined the ghost
action, and in subsection~B.6 we complete the Feynman rules
by giving explicit formulae for the functional determinants
and the tree-level propagators. Finally, in subsection~B.7 we explain
a complication that arises when one attempts to go to
a general $\x$-gauge.

%%%%%%%%%%%%
\sbsect{B.1}{Background gauge}

We work in the background gauge
whose explicit form will be repeatedly used.
We first record the action of a gauge transformation
parametrized by $\o^a(x)$ on the classical field(s)
\bqry
  \d_\o a^c_\m & = &  - D_\m^{ab} \o^b \,, \NON
  \d_\o \varphi_{\sst A} & = &
  -ig\, T^a_{\sst\!\! AB}\, \varphi_{\sst B}\, \o^a \,,
\qquad
  \d_\o \varphi_{\sst A}^\dagger =
  ig\, \varphi_{\sst B}^\dagger\, T^a_{\sst\!\! BA}\,  \o^a \,,
\label{defOmega}
\eqry
where
\beq
  D_\m = \partial_\m -ig a^a_\m T^a \,,
\label{Dmu}
\eeq
is the background covariant derivative.
In compact notation this reads
\beq
  \d_\o b_{\sst \tl{I}}(x)
  \equiv \O^a_{\sst \tl{I}} (b(x))\, \o^a(x) \,,
\label{Omega}
\eeq
where $\O^a_{\sst \tl{I}}$ is a linear differential operator
that maps the Lie-algebra valued function $\o^a$ into the space
of all (classical) bosonic fields.
For each target Bose field, the action of $\O^a_{\sst \tl{I}}$
can be read off from \seeq{defOmega}.

The quantum fields transform homogeneously.
In particular, $\a^c_\m(x)$ transforms in
the adjoint representation, with $T^a_{bc} = -i f_{abc}$.
We summarize all the transformation rules by
\beq
  \d_\o Q = (\O^a -ig T^a \hq) \o^a \,.
\label{dOQ}
\eeq

\noindent The background gauge is defined by
\beq
  \O^{\dagger a}_{\sst \tl{I}} \hb_{\sst \tl{I}}
  = D_\m^{ab} \a_\m^b
    +ig \sum(\varphi^\dagger T^a \f - \f^\dagger T^a \varphi)
  = 0 \,.
\label{backgg}
\eeq
The classical field depends on the collective coordinates.
In general, the ordinary $\z_n$-derivatives of the classical field
$b_{,n} = \partial b / \partial\z_n$
do not obey the background gauge.  We introduce covariant
$\z_n$-derivatives by making a compensating gauge transformation
\beq
  b_{;n} = b_{,n} + \O^a \o^a_n \,,
\label{covder}
\eeq
where $\o^a_n$ is defined by imposing the background gauge condition(s)
\beq
  \O^{\dagger a}\, b_{;n}=0 \,.
\label{covdergauge}
\eeq
A solution $\o^a_n$ always exists, since
\beq
  L_{\rm gh}^{ab} \equiv \O^{\dagger a}\O^b
  = - (D^2)^{ab} + \sum \varphi^\dagger \{T^a,\, T^b\} \varphi \,,
\label{OdgO}
\eeq
is a positive operator. Under a gauge transformation,
$b_{;n}$ transforms homogeneously.
We define second (or higher) covariant derivatives
of the classical field via \eg
\beq
  b_{;n;m} = b_{;n,m} -ig T^a b_{;n}\, \o^a_m \,.
\label{covder2}
\eeq

%%%%%%%%%%%%
\sbsect{B.2}{Quantization}

Formally, the path integral is defined in terms of
position eigenstates, and has the measure $\cd B(x) \cd\J(x)$.
Quantization in a non-perturbative sector means a change
of variables at the price of introducing a jacobian.
The new variables include the collective coordinates $\z_n$,
the amplitudes of the quantum modes,
and the gauge degrees of freedom $\o^a(x)$ which decouple later.
The change of variables is facilitated by introducing the following
identity into the path integral
\bqry
  1 & = &  \int d^n\z \int \cd\o(x) \,  J
\NON
  & & \times \prod_m\, \delta
  \left[ \Bmath(  {}^{\cu}b_{;m} \Bmath| B^{(\o)} - {}^{\cu}b \Bmath) \right]
\label{FPiden}\\
  & & \times \prod_{x,a}\, \delta
  \left[ \,\, {}^{\cu}\O^{\dagger a}\!
  \left( B^{(\o)}(x) - {}^{\cu}b(x) \right) \right] \,.
\nonumber
\eqry
Here $B^{(\o)}(x)$ denotes a finite gauge transformation
of the bosonic fields generated by $\exp(-ig\o^a(x))$.
The role of the first delta-function
is to remove the infinitesimal variations $b_{;n}$ from
the fluctuations spectrum, and to trade them with collective coordinates.
The second delta-function fixes the gauge, as in the standard
Faddeev-Popov procedure.

A frequently encountered technical problem is that
one obtains ordinary $\z_n$-derivatives in the Jacobian,
and their replacement by covariant $\z_n$-derivatives requires
some justification~\cite{AR}. We overcome this problem
by considering $\o^a_n = \o^a_n(x;\z)$ as a {\it connection} on
the space spanned by the collective coordinates.
We introduce a parallel transporter on this space
\beq
  \cu(x;\z) = \cp \exp\left( -ig \int^\z ds_n \,  \o^a_n\, T^a \right) \,,
\label{cu}
\eeq
where $\cp$ denotes path ordering and
the integration runs along some path from the origin to $\z$.
In \seeq{FPiden} ${}^{\cu}b(x)$ denotes the action of the finite gauge
transformation defined by $\cu$ on $b(x)$, and
${}^{\cu}\O^{\dagger a} = \O^{\dagger a}({}^{\cu}b(x))$.

In the computation of the jacobian $J$,
one must remember that $B(x)$ and $\o^a(x)$ are {\it independent}
of the collective coordinates. The $\z$-dependence enters
through $b=b(x;\z)$ and  $\cu=\cu(x;\z)$. Thanks to the presence of $\cu$
we obtain covariant $\z_n$-derivatives everywhere. One has
\beq
  \cu_{,n} = -ig\, \cu\, T^a \o^a_n(\z) + \mbox{\rm path dependent terms}\,,
\label{dUdz}
\eeq
where the first term on the \rhs comes from the variation of
the end point. By choosing a sequence of paths judicious,
the path dependent terms vanish in the limit~\cite{CS89}.
After computing the $\z_n$-derivatives we set
$B^{(\o)} - {}^{\cu}b = {}^{\cu}\hb$. All the $\cu$ factors, as well as
the explicit $\o^a(x)$-dependence, drop out now.
The generalized Faddeev-Popov jacobian is
\beq
\hspace{-0mm}
    J = \Det \left( \begin{array}{cc}
  - C_{mn} &
  -ig b_{;m}^\dagger(y) T^b \hb(y) \\
  ig b_{;n}^\dagger(x) T^a \hb(x)\quad &
  \O^{a\dagger}(x)\left(\O^b(x)-igT^b\hb(x)\right) \delta^4(x-y)
\end{array} \right)
\label{J}
\eeq
where $C = C_0 + C_1 $ and
\beqabc{B.14}
  (C_0)_{mn} & = & \bmath( b_{;m} \bmath| b_{;n} \bmath) \,,
\label{C0}\\
  (C_1)_{mn} & = & \bmath( b_{;m} \bmath| \hb_{;n} \bmath)
               = - \bmath( b_{;m;n} \bmath| \hb \bmath) \,.
\label{C1}
\eeqabc{14}
\renewcommand{\theequation}{B.\arabic{equation}}\hspace{-4.5mm}
In the derivation of \seeq{J} we have also used \seeq{covdergauge}
and $(\O^a)_{;m} = -ig T^a b_{;m}$. The matrix $C$ is symmetric.
For $C_0$ this is obvious. For $C_1$ this follows using the commutation
relations of the next subsection and \seeq{backgg}.

We may now evaluate the delta-functions in \seeq{FPiden}.
The first delta-function removes the $b_{;n}$ modes
from the bosonic fluctuations, yielding a
factor of ${\rm det}^{-\half}(C_0)$. Similarly, the second (functional)
delta-function removes the longitudinal modes
and yields a factor of ${\rm Det}^{-\half}(L_{\rm gh})$.
The partition function thus reads
\beq
  Z = \int d^n\z\; \cd \hq \,\,
  {J \over {\rm det}^\half(C_0)\,{\rm Det}^\half(L_{\rm gh})}
  \,\, \exp(-S) \,,
\label{Znoghost}
\eeq
where $\cd \hq = \cd\hb\, \cd\hJ$. The quantum bosonic field
obeys the gauge-fixing condition\seneq{backgg} and
\beq
  \bmath( b_{;m} \bmath| \hb \bmath) = 0 \,.
\label{ortho}
\eeq

%%%%%%%%%%%%
%\newpage
\sbsect{B.3}{Mode expansion and commutators}

We now turn to a more detailed discussion of the functional measure.
$\cd\hb\, \cd\hJ$ is defined by
expanding each boson and fermion quantum-field in
eigenmodes of the corresponding small fluctuations operator in the classical
background. Repeating \seeq{qp} one has
\beq
  \hq(x) = \sumint{p} \c_p(x) \hq_p \,,
\label{qpagain}
\eeq
where $\hq_p$ denotes the amplitude of any quantum mode, and
$\c_p(x)$ is the corresponding eigenfunction.
In the case of fermions, the mode expansion includes both
the {\it zero modes} and the continuum modes.
In the case of bosons there are usually only continuum modes.
(However, the formalism can accommodate
discrete bosonic modes too, should any survive after the removal of
the $b_{;n}(x)$ modes.)

Anticipating the need for ghost fields $\h^a(x)$ and $\bar\h^a(x)$
(subsection~B.5 below) we also give their mode expansion
\beq
  \h^a(x) = \sumint{p} c^a_p(x) \h_p
\,, \qquad
  \bar\h^a(x) = \sumint{p} c^a_p(x) \bar\h_p \,,
\label{etap}
\eeq
where the $c^a_p(x)$ are the (real) eigenfunctions
of $L_{\rm gh}^{ab}$.
The {\it longitudinal} bosonic modes which are eliminated by
the background gauge are $\hb_p^\parallel(x) \propto \O^{a} c^a_p(x)$.

The small fluctuations operators
depend on the collective coordinates through the classical
field. Through the eigenfunctions, this implies dependence
of the quantum fields too on the collective coordinates.
The $\z_n$-derivative of a quantum field is
\beq
  \hq_{,n}(x) = \sumint{p} \c_{p,n}(x) \hq_p \,.
\label{dqdz}
\eeq
Covariant $\z_n$-derivative are defined by
\beq
  \hq_{;n} = \hq_{,n} -ig T^a \hq\, \o^a_n \,,
\label{covdqdz}
\eeq
and similarly for the ghost field.

We will also need various commutators. First, the commutator
of two covariant derivatives of the classical field is
\beq
  b_{;m;n} - b_{;n;m} = \O^a \cff^a_{mn} \,,
\label{comutb}
\eeq
where
\beq
  \cff^a_{mn} = \o^c_{m,n} - \o^c_{n,m} + g f_{abc}\, \o^a_n\, \o^b_m \,.
\label{fmn}
\eeq
Including the commutator for any quantum field this reads
(compare \seeq{dOQ})
\beq
  Q_{;m;n} - Q_{;n;m} = (\O^a - igT^a\hq) \cff^a_{mn} \,,
\label{Comutb}
\eeq
in agreement with our interpretation of $\o^a_n$ as the components
of a connection in $\z$-space.

Next we consider the commutator of a covariant $\z_n$-derivative and
a gauge transformation. The local parameter of the gauge transformation
is identified with a ghost eigenstate, $\o^a(x)=c^a_p(x)$
(see the next subsection). Since $c^a_p(x)$ is a function of
the collective coordinates, one has
\beq
   (\d_\o Q)_{;n} - \d_\o (Q_{;n}) = (\O^a - igT^a\hq) \cff^a_{\o n} \,,
\label{Comutmix}
\eeq
where
\beq
  \cff^c_{\o n} = (\o_{;n})^c =  \o^c_{,n} - g f_{abc}\, \o^a \o^b_n\,.
\label{fmix}
\eeq
(The above may also be applied to the improper gauge transformation
that generates the isospin zero modes, see Sec.~3.)
Finally, for the commutator of two gauge transformations
we have the familiar result
\beq
  \d_{\o_2} \d_{\o_1} Q - \d_{\o_1} \d_{\o_2} Q
  = (\O^a - igT^a\hq) \cff^a_{\o_1 \o_2} \,,
\label{Comutgg}
\eeq
with
\beq
  \cff^c_{\o_1 \o_2} = g f_{abc}\, \o_2^a \o_1^b \,.
\label{fgg}
\eeq

%%%%%%%%%%%%
%\newpage
\sbsect{B.4}{Uniform treatment of discrete and gauge
collective coordinates}

Let us expand the local parameter of the gauge transformations $\o^a(x)$
in terms of (its natural basis of) ghosts eigenstates
\beq
  \o^a(x) = \sumint{p} c^a_p(x)\, \o_p \,.
\label{ggpar}
\eeq
Each (unnormalized) longitudinal mode $\O^a c^a_p(x)$ is
the infinitesimal variation of the classical field with respect to the
{\it gauge collective coordinate} $\o_p$. This observation
provides the basis for a uniform treatment of all parts of the
jacobian\seneq{J}. Taking the inner product of the continuous-index terms
with ghosts eigenstates (which amounts to a unitary change of basis),
the jacobian is rewritten as
\beq
    J =  \Det \left( \begin{array}{cc}
  C_{mn}
  &
  -ig \bmath( b_{;m} \bmath| T^b \hb \bmath| c^b_q \bmath)
\\
  -\bmath( (\O^a c^a_p)_{;n} \bmath| \hb \bmath)\quad
  &
  \bmath( c^a_p \bmath|
  \O^{a\dagger} ( \O^b-igT^b\hb )
  \bmath| c^b_q \bmath)
\end{array} \right)
\label{Jperp}
\eeq
Notice that $(c^a_p)_{;n}\, \O^{a\dagger} \hb = 0$ by \seeq{backgg}.
Hence the lower-left entry in \seeq{Jperp} agrees with \seeq{J}.
(We return to this observation in subsection~B.7 below).
The jacobian now takes the compact form
\beq
  J = \Det \, \cc \,,
\label{JperpC}
\eeq
\beq
  \cc_{\sst MN} = (\cc_0)_{\sst MN} + (\cc_1)_{\sst MN} \,,
\label{cc}
\eeq
\beq
  (\cc_0)_{\sst MN} = \Bmath( b_{;{\sst M}} \Bmath| b_{;{\sst N}} \Bmath) \,,
\qquad
  (\cc_1)_{\sst MN} =
                 -\Bmath( b_{;{\sst M};{\sst N}} \Bmath| \hb \Bmath) \,.
\label{cc1}
\eeq
The capital indices ${\st M,N}$ stand for the discrete indices
$m,n$ that label the ordinary collective coordinates,
as well as for the indices $p,q$ of the ghosts eigenstates,
cf.\ \seeq{ggpar} above.
(E.g., for the classical field one has explicitly
$b_{;{\sst M}} = b_{;m}$ for ${\st M} = m$
and $b_{;{\sst M}} = \O^a c^a_p$ for ${\st M}=p$.)
Substituting \seeq{Jperp} into \seeq{Znoghost} we obtain
\beq
  Z = \int d^n\z\; \cd \hq \,\,
  {\Det\, \cc \over {\rm Det}^\half \, \cc_0} \,\, e^{-S} \,.
\label{Zperp}
\eeq
\seEq{Zperp} provides the basis for the general discussion of Sec.~2.

%%%%%%%%%%%%
\sbsect{B.5}{Ghost action}

We complete the definition of the Feynman rules
by giving expressions for the functional determinants
and for the tree-level propagators. In this subsection we discuss
the ghost action. In a diagrammatic expansion, the discrete-
and the continuous-index parts of the jacobian\seneq{J}
must be treated separately. Given a general square matrix
\beq
  A=\left(\begin{array}{cc}
  B_{11} & B_{12} \\ B_{21} & B_{22}
  \end{array}\right)\,,
\nonumber
\eeq
one has $\det A = \det B_{11} \det D$ where
$D = B_{22} - B_{21} B_{11}^{-1} B_{12}$ (if $B_{11}^{-1}$ exists).
Applying this formula to \seeq{J} and
introducing the ghost and the anti-ghost fields $\h(x)$ and $\bar\h(x)$,
the jacobian\seneq{J} can be written as
\beq
  J = \det C\, \int \cd\h \cd\bar\h\, \exp(-S_{\rm fp}) \,,
\label{fpPI}
\eeq
where the ghost action is
\beq
 S_{\rm fp} = \int d^4x d^4y \, \bar\h^a(x)\, \cf^{ab}(x,y)\, \h^b(y) \,,
\label{sfp}
\eeq
\beq
  \cf^{ab}(x,y) = \cf^{ab}_0(x,y) + \cf^{ab}_{\rm int}(x,y) \,.
\label{fp0int}
\eeq
The tree-level and the interaction terms are
\bqry
  \cf_0^{ab}(x,y) & = &  L_{\rm gh}^{ab}(x)\delta^4(x-y)\,,
\label{fp0} \\
  \cf_{\rm int}^{ab}(x,y) & = &
  -ig\O^{a\dagger}(x)T^b\hb(x) \delta^4(x-y) \NON
  & & + g^2 b_{;m}^\dagger(x) T^a \hb(x)\, C^{-1}_{mn}\,
  b_{;n}^\dagger(y) T^b \hb(y)  \,,
\label{fpint}
\eqry
where $L_{\rm gh}^{ab}$ is defined in \seeq{OdgO}.
The ghost propagator $G_{\rm gh}$ is defined by
\beq
  L_{\rm gh}\, G_{\rm gh}(x,y) = \d^4(x-y) \,.
\label{Ggh}
\eeq
Next, the covariant derivatives $b_{;m}$ are $O(1/g)$.
Hence, a systematic expansion for $C^{-1}$ (cf.\ \seeq{fpint})
is obtained by writing
\bqry
   C^{-1} & =  C_0^{-1} -  C_0^{-1} \, C_1 \, C_0^{-1} + \cdots \,.
\label{expndC}
\eqry
This may be represented in terms of discrete-ghost
propagator and vertices, see Fig.~1. The Feynman rules depicted
in Fig.~1 also generate the expansion of $\det (C/C_0)$.
The partition function is now
\beq
  Z = \int d^n\z\; \cd \hq \cd\h \cd\bar\h\,\,
  {\det C \over {\rm det}^\half(C_0)\,{\rm Det}^\half(L_{\rm gh})}\,\,
  \exp(-S-S_{\rm fp}) \,.
\label{Zfull}
\eeq

%%%%%%%%%%%%
%\newpage
\sbsect{B.6}{Propagators, functional determinants and interactions}

The last step is to expand the bosonic fields in the action
around the classical background.
One obtains terms containing any number of quantum fields
between zero (the classical action) and four.
The interaction lagrangians are
\beq
  \cl^{\rm int}_F = \half\, \cs^{(3,F)}_{\sst \tl{I}\tl{J}\tl{K}}\,
  \hJ_{\sst \tl{I}}  \hb_{\sst \tl{J}}  \hJ_{\sst \tl{K}} \,,
\label{LF}
\eeq
and
\beq
  \cl^{\rm int}_B(b,\hb) = \cs^{(1)}_{\sst \tl{I}}(b)\; \hb_{\sst \tl{I}}
  + {1\over 6}\, \cs^{(3,B)}_{\sst \tl{I}\tl{J}\tl{K}}(b)\;
  \hb_{\sst \tl{I}}  \hb_{\sst \tl{J}}  \hb_{\sst \tl{K}}
  + {1\over 24}\, \cs^{(4)}_{\sst \tl{I}\tl{J}\tl{K}\tl{L}}\,
  \hb_{\sst \tl{I}}  \hb_{\sst \tl{J}}  \hb_{\sst \tl{K}} \hb_{\sst \tl{L}} \,.
\label{LB}
\eeq
As explained earlier, the tadpole (linear) terms involving
$\cs^{(1)}_{\sst \tl{I}}(b) = \d \cl_B / \d b_{\sst \tl{I}}$ arise because
we are not assuming that the background field is an exact
solution of the classical equations.

The bilinear terms serve to define the tree-level propagators
and the functional determinants for bosons and fermions.
The semi-classical measure (including the ghosts contribution) is
\beq
  \exp(-W_1) = {\rm det}^\half (C_0)\; {\rm Det}^\half (L_{\rm gh})\;
               \Det^{-\half} (L_B^\perp)\; \Det^{\half} (L_{\rm F})\,.
\label{semiclgg}
\eeq
The one-half power of the fermionic determinant is due to our
Majorana-like notation (Appendix~A). As usual, exact zero modes
are excluded from the fermionic determinant.
Notice that, unlike in a general $\x$-gauge,
we have here the ghosts determinant to a one-half power.
This is compensated by the absence of longitudinal-mode
contributions in the transverse bosonic determinant.

%%%%%%%%%%%%
\vspace{2ex}
\noindent {\bf Bosons.} \hspace{1ex}
We now turn to a detailed discussion of the propagators and
the functional determinants, starting with the bosons.
We focus on the propagator for those (gauge and scalar) fields
that have a classical value.
(For bosonic fields that do not have a classical part,
the below derivation reduces to the textbook case,
and the propagator is the inverse of the differential operator
that occurs in the bilinear action.) Let the quadratic bosonic action be
\beq
  S_B^{(2)}(b,\hb) \equiv \half \bmath( \hb \bmath| L_B \bmath| \hb \bmath) \,.
\label{bilin}
\eeq
The second-order differential operator $L_B=L_B(b)$ depends on
the classical field.
Our task is to construct a bosonic propagator $G_B$
which is the inverse of $L_B$ on the {\it transverse} subspace.
The latter is the complement of the subspace spanned by the $b_{;n}$
and by the infinitely-many longitudinal modes.
It is convenient to consider first a general $\x$-gauge,
and then take the limit $\x\to 0$. This means that we first project
out the finite-dimensional space spanned by the $b_{;n}$ for any $\x$,
and in the end remove the longitudinal modes by sending $\x\to 0$.

We begin by considering the differential operator
\beq
  L_B^\x = L_B + {1\over\x}\, \O^a\O^{\dagger a} \,,
\label{Lxi}
\eeq
where the last term is the contribution of the gauge-fixing action for
$\x \ne 0$. When the background field is not a classical solution,
$L_B^\x$ has no (exact) zero modes.
Therefore it has an inverse $\hG_B$ such that
\beq
  L_B^\x\, \hG_B(x,y) = \d^4(x-y) \,.
\label{hatGB}
\eeq
We will now express $G_B$ in terms of $\hG_B$ and $b_{;n}$.
(The below construction is similar to that of Levine and Yaffe~\cite{LY},
see also ref.~\cite{AM}.) We consider the bosonic
gaussian integration in the presence of an external source $K(x)$
at fixed values of the collective coordinates $\z_n$.
We introduce a bosonic field $\hb'(x)$ which obeys the orthogonality
conditions\seneq{ortho} but {\it not} the background gauge\seneq{backgg},
as well as a completely unconstrained field $\hb''(x)$.
We let
\beq
  \tilde{b}_{m}= (C_0^{-\half})_{mn}\, b_{;n}
\label{bnnorml}
\eeq
denote the normalized covariant derivatives of the classical field. Now
\bqry
  Z_\x(K) & \equiv & \int \cd\hb'\,
  \exp\left[ -\half \bmath( \hb' \bmath| L^\x_B \bmath| \hb' \bmath)
  + \bmath( K \bmath| \hb' \bmath) \right]
\NON
           & = & \int \cd\hb''\,
  \prod_n \delta\left[ \bmath( \tilde{b}_{n} \bmath| \hb'' \bmath) \right]\,
  \exp\left[ -\half \bmath( \hb'' \bmath| L^\x_B \bmath| \hb'' \bmath)
  + \bmath( K \bmath| \hb'' \bmath)\right]
\label{Z0} \\
           & = & \int \cd\hb''\,
  \prod_n \int d\a_n
  \exp\left[ -\half \bmath( \hb'' \bmath| L^\x_B \bmath| \hb'' \bmath)
  + \bmath( K + i\sum_n \a_n \tilde{b}_{n} \bmath| \hb'' \bmath) \right] \,,
\nonumber
\eqry
where, on the last row, we have introduced an
integration over auxiliary variables $\a_n$ to enforce the constraint.
It is straightforward to perform the gaussian integration,
first over $\hb'(x)$, and then over $\a_n$.
The result is
\beq
  Z_\x(K) = {\rm Det}^{-\half}\, (L_B^{\x\perp})\,
  \exp\left[ \half \bmath( K \bmath| G_B \bmath| K \bmath) \right] \,.
\eeq
The bosonic determinant is
\beq
  \Det\, (L_B^{\x\perp}) \equiv {\rm Det}^{-1} (G_B)
  = \Det\, (L_B^\x) \; \det (\tL^{-1}) \,,
\label{detB}
\eeq
where
\beq
  \tL^{-1}_{mn} =
  \bmath( \tilde{b}_{m} \bmath| \hG_B \bmath| \tilde{b}_{n} \bmath) \,.
\label{lmbda}
\eeq
The bosonic determinant in \seeq{semiclgg} is the $\x \to 0$ limit
of \seeq{detB}.
(${\rm Det}(L_B^\x)$ contains an infinite power of the gauge-fixing
parameter $\x$; this power, which is formally equal to the number
of longitudinal modes, cancels against a matching infinite power
upon dividing by the free vacuum-sector determinant.)

The normalization factor $C_0^{-\half}$ in \seeq{bnnorml}
cancel out in the expression for the bosonic propagator, which reads
\beq
  G_B = \hG_B - \hG_B \, \L \, \hG_B \,,
\label{GB}
\eeq
\beq
  \L =
  \sum_{mn} \bmath| b_{;m} \bmath)\, \L_{mn} \,
  \bmath( b_{;n} \bmath| \,,
\label{Lamun}
\eeq
\beq
  (\L^{-1})_{mn} =
  \bmath( b_{;m} \bmath| \hG_B \bmath| b_{;n} \bmath) \,.
\label{Lun}
\eeq
At this stage the bosonic propagator obeys the orthogonality relation
$G_B\bmath| b_{;n} \bmath) = 0$ and the completeness relation
\beq
  L_B^\x\, G_B(x,y) - \int d^4z\,  P_B(x,z) L_B^\x\, G_B(z,y)
  = \d^4(x-y) - P_B(x,y) \,,
\label{Bcomplt}
\eeq
where
\bqry
  P_B(x,y)
& = & \sum_n \tilde{b}_{n}(x)\, \tilde{b}^\dagger_{n}(y)
\NON
& = & \sum_{mn} b_{;n}(x)\, (C_0^{-1})_{mn} \, b^\dagger_{;n}(y) \,,
\label{PB}
\eqry
is the projector on the space spanned by $b_{;n}$.

The last step is to take the Landau-gauge limit $\x \to 0$.
A very useful identity is
\beq
  G_{\rm gh} \O^\dagger L_B \hG_B
  = G_{\rm gh} \lvec\O{}^\dagger - {1\over \x}\, \O^\dagger \hG_B \,,
\label{GbGgh}
\eeq
which relates the longitudinal part of $\hG_B$
to the ghost propagator. \seEq{GbGgh} is derived
by multiplying \seeq{hatGB} on the left by $G_{\rm gh} \lvec\O{}^\dagger$
and integrating by parts.
(If the classical field is an exact solution, the longitudinal
modes are zero modes of the bosonic small fluctuations operator
and the \lhs of \seeq{GbGgh} is zero;
\seeq{GbGgh} holds in fact for $G_B$ too,
as one can check using \seeq{GB} and $\O^\dagger \L = 0$.)
Eliminating the gauge-fixing term from \seeq{hatGB} using \seeq{GbGgh},
we rewrite the former as
\beq
  L_B\, \hG_B(x,y) - \int d^4z\, P^\parallel(x,z) L_B\, \hG_B(z,y)
  = \d^4(x-y) - P^\parallel(x,y) \,,
\label{hatBcomp}
\eeq
where
\beq
  P^\parallel(x,y) = \O \, G_{\rm gh}(x,y)\, \lvec\O{}^\dagger \,,
\label{longP}
\eeq
is the longitudinal projector. \seEq{hatBcomp} holds for any $\x$,
and has a smooth $\x \to 0$ limit. Finally, substituting \seeq{hatBcomp}
into \seeq{GB} we find
\bqry
  L_B\, G_B & = & 1 - \L\, \hG_B - P^\parallel(1 - L_B\, G_B) \NON
            & = & 1 - (P_B + P^\parallel)(1 - L_B\, G_B) \,.
\label{elim}
\eqry
The last expression exhibits again that
the covariant derivatives $b_{;n}$ and the longitudinal modes
have a similar role.
This result is used in the diagrammatic identities of Appendix~C.2.
We comment that \seeqs{Bcomplt}, \seneq{hatBcomp}
and\seneq{elim} all have the generic form
$(1-P) L\, G = 1 - P$, which is the completeness relation for
a constrained propagator obeying $P\, G = G\, P = 0$.

%%%%%%%%%%%%
\vspace{2ex}
\noindent {\bf Fermions.} \hspace{1ex}
In Majorana-like notation (cf.\ \seeq{innerF}) the bilinear fermion action is
\beq
  S_F^{(2)}(b,\hJ) \equiv \half \bmath( \hJ \bmath| L_F \bmath| \hJ \bmath) \,.
\label{bilinF}
\eeq
The euclidean Dirac operator $L_F$ can be chosen to be hermitian~\cite{CS89}.

In a generic SUSY theory that contains both explicit mass terms
for matter fields and a scalar VEV, the fermion spectrum contains no exact
zero modes, and $L_F$ has an inverse $\hG_F$ such that
$L_F\, \hG_F(x,y) = \d^4(x-y)$.
Alternatively, if there are no explicit mass terms, $L_F$ anti-commutes
with the generator of some (typically anomalous) $R$-symmetry
even in the presence of a Higgs field.
In a sector with a single classical object (one unit of topological charge)
the classical field is spherically symmetric
(or can be chosen to be so, in case it is not an exact solution).
This implies the existence of a conserved index for {\it each}
value of the total angular momentum. In this case
the fermion spectrum contains exact zero modes. One has
\beq
  L_F\, G_F(x,y) = \d^4(x-y) - P_F^{\rm \, exact}(x,y) \,,
\label{GF}
\eeq
where
\beq
  P_F^{\rm \, exact}(x,y) = \sum_i \c_i^{F0}(x) \, \c_i^{F0\dagger}(y) \,,
\label{PFexact}
\eeq
is the projector on the exact fermionic zero modes $\c_i^{F0}$.

In general, the fermion propagator $G_F$ in the instanton
sector contains a contribution from {\it approximate} zero modes too.
The latter correspond to those zero modes of the exact instanton solution
which couple through the explicit mass terms and/or the Higgs field.
For practical calculations, it may be convenient to separate out
the approximate zero modes. To this end one needs fermionic propagator
and determinant which are constrained
to the complement of the subspace spanned by both exact and
approximate zero modes. These are defined via the same algebraic
construction used above for bosons.

For the general diagrammatic proof of SUSY Ward identities (Appendix~C)
it is convenient to use the propagator $G_F$.
The projector on the exact fermionic zero modes $P_F^{\rm \, exact}$ arises
at intermediate steps in the derivation.
However, the sum of all terms involving $P_F^{\rm \, exact}$
is zero due to anti-symmetry.

%%%%%%%%%%%%
\vspace{2ex}
\noindent {\bf Recursion relations.} \hspace{1ex}
A key role in the renormalization of (SUSY) Ward identities
is played by recursion relations between the {\it non-perturbative}
and {\it free} propagators. Let $H = H_0 + V$ be a Schr\"odinger operator.
Assume that both $H_0$ and $H$ have no zero modes, and let $G_0$ and
$G$ be the corresponding Green functions.
One has the textbook relation
\beq
  G = (1 - G\, V) G_0 \,,
\label{GG0}
\eeq
which may be iterated a finite number of times.
In our case, \seeq{GG0} may be applied to the ghost propagator (\seeq{Ggh}) and
to the unconstrained bosonic propagator $\hG_B$ (\seeq{hatGB})
for $\x \ne 0$.

The Born series is obtained by iterating \seeq{GG0} infinitely many times.
However, the series do not converge if $H_0$ and $H$ have
different zero-modes spectra. Thus,  a generalization
of \seeq{GG0} is needed for the fermion propagator $G_F$
which obeys \seeq{GF}.  We first write
\beq
  L_F = L_F^{\rm vac} + \tilde{V}_F \,,
\label{LFLvac}
\eeq
where $L_F^{\rm vac}$ stands for the free (vacuum-sector) Dirac operator(s)
for all fields participating in the exact zero modes.
Its inverse, the free fermion propagator, is denoted $G_F^{\rm vac}$.
We now claim that
\beq
  G_F = (1-P_F^{\rm exact} - G_F \tilde{V}_F) G_F^{\rm vac} \,.
\label{GFGvac}
\eeq
Let $\D$ denote the difference between the \rhs and the \lhs of
the above equation. One has
$\D (\lvec{L}_F - \tl{V}_F) = \D \lvec{L}{}_F^{\rm vac} = 0$.
This implies $\D=0$ since $L_F^{\rm vac}$ has no zero modes.
One can also verify directly that the \rhs of \seeq{GFGvac}
is a right-inverse:
$L_F\D = P_F^{\rm exact}(1+\tilde{V}_F\, G_F^{\rm vac}) =
P_F^{\rm exact} L_F\, G_F^{\rm vac} = 0 $.
The role of the above recursion relations
in the renormalization procedure is discussed in subsection~3.3.

%%%%%%%%%%%%
%\newpage
\sbsect{B.7}{The case $\bmath{\xi \ne 0}$}

As can be seen from the previous discussion,
it is natural to work with a transverse bosonic field.
Since \seeq{dOint} deals with gauge-invariant operators,
it should hold for any $\xi$, and one may want to check
this explicitly. However, there is a technical difficulty
that has prevented us from generalizing the Feynman rules
to $\x \ne 0$.

As explained below \seeq{Jperp}, one must use the
gauge condition\seneq{backgg} in order to prove that \seeqs{J}
and\seneq{Jperp} define the same jacobian.
For $\x \ne 0$, however, the constraint $\O^\dagger \hb = 0$
is not respected. Therefore, \seeqs{J}
and\seneq{Jperp} do {\it not} define the same jacobian.
We did not attempt to resolve the question of what {\it is}
the correct jacobian for $\x \ne 0$. (The gauge-fixing constraint
is used also in the calculation of $\ds\o^a$, cf.\ \seeq{domega};
therefore it is likely that $\ds\o^a$ too will pick up terms
proportional to $\O^\dagger \hb$ for $\x \ne 0$.
In the calculation of \seeq{anml} it is in fact possible to work
with any $\x$-gauge, since the above subtlety is relevant only
at the next order.)

%%%%%%%%%%%%%%%%%%%%%%%%%%%%%%%%%%%%%%%%%%%%%%%%%%%%%%%%%%%%%%%%%%%%%%%
%%%%%%%%%%%%%%%%%%%%%%%%%%%%%%%%%%%%%%%%%%%%%%%%%%%%%%%%%%%%%%%%%%%%%%%
\vspace{5ex}
%\newpage
\noindent {\large\bf C.~~General diagrammatic proof of SUSY
Ward identities}
\vspace{3ex}
\secteq{C}

In this appendix we give a general diagrammatic proof
of \seeq{dOint}. The underlying algebraic structure
is similar to Sec.~2. However, the bilinear part of the action
plays a special role, being the starting point for any diagrammatic
expansion. As a result things are technically
more involved.

In more detail, besides algebraic manipulations, what entered
the results of Sec.~2 is the physical boundary conditions of the
quantum fields. In a path-integral context we had to invoke a
finite-volume cutoff in order to impose these boundary conditions.
However, the propagators automatically obey correct
boundary conditions in infinite volume.
Therefore a diagrammatic expansion should
be consistent with the vanishing of $\ds S$ and $\ds\m$
without having to resort to a finite-volume cutoff.

%%%%%%%%%%%%
\sbsect{C.1}{No gauge fields}

In this subsection we prove \seeq{dOint} in the physically
less interesting case of
(a non-perturbative sector of) a SUSY theory with no gauge fields.
This simplifies matters in a number of ways.
First, there is no need to introduce a ghost {\it field}.
In addition we now deal with ordinary (in place of covariant) derivatives
with respect to the collective coordinates, which {\it commute}.
Expectation values are given by \seeq{vevz}, which results after
performing the integration over the quantum fields in \seeq{Zperp}
with $\cc \to C$ (cf.\ eqs.~(B.14) and\seneq{cc}).
The semi-classical measure (\seeq{semiclgg}) simplifies to
\beq
  \exp(-W_1) = {\rm det}^\half\, (C_0)\;
               \Det^{-\half} (L_B^\perp)\; \Det^{\half} (L_{\rm F})\,.
\label{semicl}
\eeq
The interaction terms arising from the action are given by \seeqs{LF}
and\seneq{LB}. Additional vertices arise from the expansion
of $\det (C/C_0)$, cf.\ \seeq{expndC}.

The basic strategy is to construct separately Ward identities
for the various terms in the transformation laws\seneq{bosvar}
and\seneq{fervar} of the quantum amplitudes.
The final Ward identity is obtained by piecing together
the individual Ward identities after suitable arithmetical manipulations.
We begin by considering all diagrams that define the expectation value
of some (multi)local operator $\co$, times an insertion of
\beq
  \bmath( \ds Q \bmath|  L \bmath| \hq \bmath)
  = \bmath( \G\hJ \bmath| L_B \bmath| \hb \bmath)
  + \bmath( \ds \J \bmath| L_F \bmath| \hJ \bmath) \,,
\label{genwi}
\eeq
In each diagram  we act with $L_B$ ($L_F$)
on the propagator whose one leg corresponds to $\hb$ ($\hJ$) of \seeq{genwi}.
Using \seeqs{GB} and\seneq{GF}
we obtain the diagrammatic identity
\beq
  \bmath( \ds Q \bmath|  L \bmath| \hq \bmath) \co
  \;\eqv\;
  \hmath( \ds \J \hmath| 1-P_F^{\rm \, exact} \hmath|
  {\partial\co \over \partial \hJ} - \cs_F\, \co \hmath)
  + \Bmath( \G\hJ \Bmath| 1 - \L \hG_B \Bmath| X \Bmath) \,.
\label{WIsus}
\eeq
Here
\beq
  X = {\partial\co \over \partial \hb} +
  \left(
  \tr\, {\partial C_1 \over \partial \hb}\,C^{-1} - \cs_B
  \right) \co \,,
\label{X}
\eeq
and $\cs_B$ and $\cs_F$ are defined in \seeq{SFSB}.
The $\eqv$ sign denotes equality under the expectation value defined
by integrating over the quantum fields only.
(This is an equality between {\it disconnected} diagrams,
which correspond to  $\exp(-W_2)\,\svev{\cdots}_{\!\z}$ in \seeq{vevz}.)

As in Sec.~3, all terms with $P_F^{\rm \, exact}$ in \seeq{WIsus}
cancel each other by antisymmetry. Using SUSY invariance of the action,
as expressed in \seeq{susEoM}, we find
\beq
  \ds \co \;\eqv\;
  \Bmath( \G\hJ \Bmath| \L \hG_B \Bmath| X \Bmath)
- \hmath( \G\hJ \hmath|
  \co \; \tr\, {\partial C_1 \over \partial \hb}\,C^{-1}
  \hmath)  \,.
\label{WIss}
\eeq
\seEq{WIss} says that
the fixed-$\z$ expectation value of $\ds\co$
does not vanish because of terms arising from the
jacobian in \seeq{Zperp}, or from the constraints
obeyed by the bosonic field (\seeq{ortho}), or from both.
Our task will be to simplify the \rhs of \seeq{WIss}.

The hint comes from \seeqs{bosvar} and\seneq{fervar}.
These equations show that, besides the expected piece arising
from $\ds Q$, the variation of a quantum amplitude
contains additional pieces needed to compensate for
the $\z$-dependence of the field.
We thus proceed by writing down a diagrammatic identity for the variation
$\d_1 \hq(x) \equiv \ds\z_n\, \hq_{,n}(x)$ (cf.\ \seeq{dqdz}).
To this end, we consider the new insertion
\beq
  \half\ds\z_n \left\{
  \nord \bmath( \hq_{,n} \bmath|  L \bmath| \hq \bmath) \nord
  + \nord \bmath( \hq \bmath|  L \bmath| \hq_{,n} \bmath) \nord
  \right\} \,,
\label{normal}
\eeq
where $\ds\z_n$ is given by \seeq{dcoll} with $\cc \to C$.
The normal ordering prescription means that the $\hq$ and $\hq_{,n}$
must not be contracted with each other.
We momentarily introduce a finite volume cutoff, which
allows us to drop total derivatives of currents
involving $\hq_{,n}$ as in Sec.~2.
As before, we let $L$ act on the propagator attached to the $\hq$-leg
of the insertion, and the resulting
terms with $P_F^{\rm \, exact}$ cancel out.
We obtain the diagrammatic identity
\bqry
  0 & \!\eqv\! &  \ds\z_n \left\{
  \half\,\nord \bmath( \hq \bmath|  L_{,n} \bmath| \hq \bmath) \nord
  - \nord S^{(2)}_{,n} \nord
  \right\} \co
\NON & &
  +\, \ds\z_n   \hmath( \hJ_{,n} \hmath|
  {\partial\co \over \partial \hJ} - \cs_F\, \co \hmath)
  +  \hmath( \hJ_{,n} \hmath|
  {\partial (\ds\z_n) \over \partial \hJ}  \hmath) \co
\\ \label{WIqn} & &
  +\, \ds\z_n \Bmath( \hb_{,n} \Bmath| 1 - \L \hG_B \Bmath| X \Bmath)
  +  \hmath( \hb_{,n} \hmath|  1 - \L \hG_B \hmath|
  {\partial (\ds\z_n) \over \partial \hb}  \hmath) \co\,.
\nonumber
\eqry
In a finite volume one has
\beq
  S^{(2)}_{,n}
  = \half \Bmath( \hat{q} \Bmath| L \Bmath| \hat{q} \Bmath)_{\!,n}
  = \half \sum_p \l_{p,n}\; \hq_p^2 \,,
\label{ds2dz}
\eeq
where the $\z$-derivatives act on the (discrete) eigenvalues
(compare \seeq{S2modes}). Also recall that each propagator
has the mode expansion
\beq
  G(x,y) = \sum_p \c_p(x)\, \l_p^{-1} \, \c_p^\dagger(y) \,.
\label{propexp}
\eeq
%\newpage \noindent %%%%%%%%%%%%%%%%%%%%%%%%%%%%%%%%%%%%%%%
This allows us to rewrite \seeq{WIqn} as
\bqry
  0 & \!\eqv\! &  \half\;
  \nord \bmath( \hq \bmath|  L_{,n} \bmath| \hq \bmath) \nord\; \ds\z_n\; \co
\NON & &
  - \hmath( \hb_{,n} \hmath| \L \hG_B \hmath|
  \ds\z_n\, X + {\partial (\ds\z_n) \over \partial \hb}\,  \co \hmath)
\label{WIqn2}\\
& & + \exp(S^{\rm int})\,
  \nord
  \hmath( {\partial \over \partial \hq} \hmath|  G_{,n} \hmath|
  {\partial \over \partial \hq} \hmath)
  \nord\;
  \exp(-S^{\rm int})\, \ds\z_n\, \co
\NON
& & +\, \ds\z_n   \hmath( \sum_i \c^{F0}_{i,n}\, \hq^{F0}_i \hmath|
  {\partial\co \over \partial \hJ} - \cs_F\, \co \hmath)
  +  \hmath( \sum_i \c^{F0}_{i,n}\, \hq^{F0}_i \hmath|
  {\partial (\ds\z_n) \over \partial \hJ}  \hmath) \co\,.
\nonumber
\eqry
\seEq{WIqn2} holds in the infinite-volume limit.
On the third row, $S^{\rm int}$ contains all the interaction terms (that arise
from the expansion of both the action and the jacobian).
The functional operator on that row differentiates
(the product of) all {\it propagators} in
a given diagram with respect to $\z_n$.
On the last row, $\c^{F0}_i$ denote the exact fermionic zero modes
(see \seeq{PFexact})
and $\hq^{F0}_i$ are the corresponding Grassmann amplitudes.
The role of the last row is to differentiate all (exact)
fermionic zero modes with respect to $\z_n$.

It may be helpful to examine the content of identity\seneq{WIqn2} in
a simpler case, where it is applied
to a scalar field that does not have a classical part.
The second and forth rows are then missing, and the identity reduces to
the familiar relation
\beq
  G_{,n}(x,y) + \int d^4z\, G(x,z)\, L_{,n}\, G(z,y) = 0 \,,
\label{diffprop}
\eeq
which holds in any volume. As for
the terms on the second and forth rows of \seeq{WIqn2}, 
all of them involve localized sources
which guarantee a smooth infinite volume limit.
(At coinciding points, $x=y$, \seeq{diffprop} may however be violated
by the ultra-violet cutoff (see subsection~3.3).
If angular momentum cutoff is employed, the partial-wave
propagators obey \seeq{diffprop} even for $x=y$.
In that case the \lhs of \seeq{diffprop} is formally $\d_{,n}(0)$,
and the spectral trace in \seeq{dmcom} may be regarded as its integral.)

Consider now the sum of \seeqs{WIss} and\seneq{WIqn2}.
In order to arrive at \seeq{dOint} we need,
besides $\z$-derivatives of the propagators,
also $\z$-derivatives of all elements of the Feynman rules
that depend explicitly (not through the eigenmodes) on the classical field.
The latter must come from
the \rhs of \seeq{WIss} plus the first two rows on the \rhs of \seeq{WIqn2}.
Verifying that this is indeed the case is essentially
an arithmetical task. The reader who is not interested in these
algebraic details is advised to skip to subsection~C.2,
where we explain how we deal with the extra complications of a gauge theory.

We start with the terms involving $X$ in
\seeqs{WIss} and\seneq{WIqn2}.
Writing $\L \hG_B = (\L \hG_B - P_B) + P_B$ (cf.\ \seeq{PB}) we get
\beq
  \Bmath( \G\hJ - \ds\z_n\, \hb_{,n} \Bmath| \L \hG_B \Bmath| X \Bmath)
  = \ds\z_n\, \Bmath( b_{,n} + d_n \Bmath| X \Bmath) \,,
\label{Xsprd}
\eeq
where
\beq
  d_m = \left( (C_0)_{ml} \L_{ln} - \d_{mn} \right) b_{,n}
\label{adn}
\eeq
and $\L_{ln}$ is defined in \seeq{Lun}.
We now consider the \rhs of \seeq{Xsprd} term by term using
\seeqs{LF}, \seneq{LB}, \seneq{SFSB} and\seneq{X}.
First, $b_{,n}\, \partial\co/\partial\hb = b_{,n}\, \partial\co/\partial b$
since $\co=\co(b+\hb,\hJ)$.
This gives the $\z$-derivative of the classical part of $\co$.
Next
$\bmath( b_{{\sst \tl{I}},n} \bmath| \cs^{(1)}_{\sst \tl{I}}(b) \bmath)
= b_{,n}\, \partial S^{(0)} / \partial b$
gives the $\z$-derivative of the classical action, and
$\bmath( b_{{\sst \tl{I}},n} \bmath|
\cs^{(4)}_{\sst \tl{I}\tl{J}\tl{K}\tl{L}}\,
\hb_{\sst \tl{J}} \hb_{\sst \tl{K}} \hb_{\sst \tl{L}} \bmath)$
gives $b_{,n}\, \partial/\partial b$ acting on
$\cs^{(3,B)}_{\sst \tl{I}\tl{J}\tl{K}}(b)\,
\hb_{\sst \tl{I}} \hb_{\sst \tl{J}} \hb_{\sst \tl{K}}$.

Next consider those terms in $\cs_B$ arising from \seeq{LF} and from
the remaining (the middle) term in \seeq{LB}. Combined with the first term
on the \rhs of \seeq{WIqn2} we find
\beq
  \nord \bmath( \hq \bmath|  L_{,n} \bmath| \hq \bmath) \nord
  - \bmath( b_{{\sst \tl{J}},n} \bmath|
  \cs^{(3)}_{\sst \tl{I}\tl{J}\tl{K}}(b)
  \hq_{\sst \tl{I}} \hq_{\sst \tl{K}} \bmath)
  \;\eqv\;
  -  {\rm Tr}\, (-)^F\, L_{,n}\, G \,.
\label{ddetdn}
\eeq
We used $L_{{\sst \tl{I}\tl{K}},n} = b_{{\sst \tl{J}},n}\,
  \cs^{(3)}_{\sst \tl{I}\tl{J}\tl{K}}(b)$.
Including a one-half factor,
this gives the logarithmic $\z$-derivative of the functional determinants,
{\it except} for those bosonic fields that have a classical part,
see below.

We next turn to the $d_n$ term in \seeq{Xsprd}. In the limit
where the classical field is an exact solution, $d_n$ vanishes.
(This can be worked out using Appendix~B.6.)
In our Feynman rules, not having an exact classical solution
is reflected in the tadpole terms
and in the peculiar formulae for the bosonic propagator and determinant.
We will see that the $d_n$-dependent terms are indeed related
to those elements of the Feynman rules. First, similarly to
\seeq{Xsprd} we rewrite the second term on the second row
of \seeq{WIqn2} as $\co$ times
\bqry
  - \Bmath( \hb_{,n} \Bmath| \L \hG_B \Bmath|
  \partial (\ds\z_n)/\partial\hb \Bmath)
& = & - (C_0^{-1})_{ml} (C_1)_{ln}
  \Bmath( b_{,m} \Bmath| \partial (\ds\z_n)/\partial\hb \Bmath)
\NON
& &  + \left(\d_{mn} - (C_0^{-1})_{ml} C_{ln}\right)
  \Bmath( d_m \Bmath| \partial (\ds\z_n)/\partial\hb \Bmath)  \,.
\label{znsprd}
\eqry
The $d_n$-dependent term in \seeq{Xsprd} plus the first term on
the second row of \seeq{znsprd} give
\bqry
\hspace{-16mm}
  \hmath( d_n \hmath|
  \ds\z_n\, X + {\partial (\ds\z_n) \over \partial \hb}\,  \co \hmath)
& = & - \hmath( b_{,n} \, L_B\, G_B \hmath|
  \ds\z_n\, X + {\partial (\ds\z_n) \over \partial \hb}\,  \co \hmath)
\NON
& \eqv & -  \Bmath( b_{,n}\, \partial \cs^{(1)}_{\sst \tl{I}}/\partial b
            \Bmath| \hb_{\sst \tl{I}} \Bmath) \ds\z_n\, \co\,.
\label{dninsrt}
\eqry
This is the $\z$-derivative of the classical source
$\cs^{(1)}_{\sst \tl{I}}(b)$.
The remaining $d_n$-dependent term in
\seeq{znsprd} plus the \rhs of \seeq{ddetdn} (for those fields
that have a classical part) give
\beq
  - \half\, \ds\z_n\, {\rm Tr}\, L_{,n}\, G
  -  (C_0^{-1})_{ml} C_{ln}
  \Bmath( d_m \Bmath| \partial (\ds\z_n)/\partial\hb \Bmath)
= \ds\z_n\, {\partial\over\partial \z_n}\,
  \log \Det^{-\half} (L_B^\perp) \,.
\label{ddetperp}
\eeq
(Note that the definition of the bosonic determinant\seneq{detB}
involves the normalized modes of \seeq{bnnorml}.)

We are now left with three terms: the last term on the \rhs of \seeq{WIss},
$\co$ times the first term on the \rhs of \seeq{znsprd}, and
\beq
  \ds\z_n\, \co\,
  \hmath( b_{,n} \hmath| {\partial (C_1)_{kl} \over \partial \hb}
  \hmath) C^{-1}_{lk}
  = - \ds\z_n\, \co\,
  \Bmath( b_{,n} \Bmath| b_{,k,l} \Bmath)\, C^{-1}_{lk}
\,,
\label{Xleftover}
\eeq
which is the last term that has remained from the \rhs of \seeq{Xsprd}.
In order to complete the derivation of \seeq{dOint}
we must show that the sum of these three terms is equal to $\co$ times
\beq
  \Bmath( b_{,n} \Bmath| \partial(\ds\z_n)/\partial b \Bmath)
  + \ds\z_n\, C^{-1}_{lk}
  \Bmath( b_{,n} \Bmath| \partial C_{kl} / \partial b \Bmath)
  - \half\, \ds\z_n\, (C_0^{-1})_{lk}
  \Bmath( b_{,n} \Bmath| \partial (C_0)_{kl} / \partial b \Bmath) \,.
\label{msrterms}
\eeq
This expression amounts to $b_{,n}\, \partial/\partial b$
acting on (the logarithm of) the jacobian and on $\ds\z_n$ itself.
The needed result follows after a few algebraic manipulations using
the commutativity of ordinary $\z_n$-derivatives.
This completes the diagrammatic proof of \seeq{dOint}
for a SUSY  theory that contains only scalar and fermion
fields, and no gauge fields.

%%%%%%%%%%%%%%%%%%%%%%%%%%%%%%%%%%%%%%%%%%%%%%%%%%%%
%\newpage
\sbsect{C.2}{Gauge theories}

Before we turn to the full-fledged diagrammatic proof of \seeq{dOint}
in a gauge theory, there is one more exercise that we can do.
In the discussion of the previous subsection, let us make the
(formal) replacement $C \to \cc$, $m,n \to {\st M,N}$ etc (cf.\ Appendix~B.4).
In other words, we consider both the discrete and the continuous
collective coordinates. Accordingly, we relax the assumption
that $\z_{\sst N}$-derivatives commute. Repeating the entire derivation
we now arrive at \seeq{dOint}, but with one extra term which
is left after the algebraic manipulations.
This term is ($\co$ times) the commutator from \seeq{dmcom}.

The careful diagrammatic treatment of the boson and the fermion fields
has confirmed the vanishing of the corresponding spectral traces
in \seeq{dmcom}. This is simply because no such terms
are left in the final diagrammatic answer of subsection~C.1.
Now, as discussed in Sec.~2, the commutator is really a spectral
trace over the ghost field. As long as we treat the continuous-index
part of the jacobian formally, we cannot escape the appearance
of this commutator. The full gauge-theory
Feynman rules are needed in order to confirm the vanishing
of the ghost-field spectral trace.

The general strategy will be the same as in Appendix~C.1.
We begin by establishing a number of key diagrammatic identities
which correspond to various pieces in the transformation
rules of the quantum amplitudes\seneq{bosvar} and\seneq{fervar}.
The rest amounts to very lengthy arithmetical manipulations that we omit.
The gauge-theory partition function is defined by \seeq{Zfull}.
The semi-classical measure is given by \seeq{semiclgg}.
The interaction lagrangians arising from the action are \seeqs{LF}
and\seneq{LB}. The ghosts lagrangian as well as the
vertices arising from the discrete part of the
jacobian are given in Appendix~B.5.

The first diagrammatic identity that we need is a generalization
of \seeq{WIss}. Again we consider insertion\seneq{genwi}.
Performing the same steps as in Appendix~C.1 and using \seeq{elim} we obtain
\bqry
  \ds \co
& \eqv &
  \hmath( \G\hJ \hmath| (P_B + P^\parallel) (1 - L_B\, G_B) \hmath| X
    - {\partial S_{\rm fp}^{\rm int} \over \partial \hb}\, \co \hmath) \NON
& & - \hmath( \G\hJ \hmath|
  \left\{ \tr\, {\partial C_1 \over \partial \hb}\,C^{-1}
    - {\partial S_{\rm fp}^{\rm int} \over \partial \hb}\right\} \co
  \hmath) \,,
\label{WIssgg}
\eqry
where $X$ is defined in \seeq{X} and the ghost interaction action
$S_{\rm fp}^{\rm int}$ can be read off from \seeqs{sfp} and\seneq{fpint}.
On top of the \rhs of \seeq{WIss}, we see here the contributions of
the ghost action.

The second diagrammatic identity corresponds to the variation
$\d_1^{\rm cov} \hq(x) \equiv \ds\z_n\, \hq_{;n}(x)$. Following the same
steps as in Appendix~C.1 we obtain
\bqry
  0 & \!\eqv\! &  \half\;
  \nord \Bmath( \hq \Bmath|  L_{;n} \Bmath| \hq \Bmath) \nord  \ds\z_n\; \co
\label{WIqngg}\\
& & - \hmath( \hb_{;n} \hmath| (P_B + P^\parallel) (1 - L_B\, G_B) \hmath|
  \ds\z_n \hmatn(
      X  - {\partial S_{\rm fp}^{\rm int} \over \partial \hb}\, \co
  \hmatn)
+ \, {\partial (\ds\z_n) \over \partial \hb}\,
 \co \hmath)
\NON
& & + \exp(S^{\rm int})\,
  \nord \hmath( {\partial \over \partial \hq} \hmath|  G_{;n} \hmath|
  {\partial \over \partial \hq} \hmath) \nord\;
  \exp(-S^{\rm int})\, \ds\z_n\, \co
\NON
& & +\, \ds\z_n   \hmath( \sum_i \c^{F0}_{i;n}\, \hq^{F0}_i \hmath|
  {\partial\co \over \partial \hJ} - \cs_F\, \co \hmath)
  +  \hmath( \sum_i \c^{F0}_{i;n}\, \hq^{F0}_i \hmath|
  {\partial (\ds\z_n) \over \partial \hJ}  \hmath) \co\,.
\nonumber
\eqry
As with \seeq{WIqn2}, the functional operator on the third row
differentiates all bosonic and fermionic propagators with
respect to $\z_n$, while the last row does this for
the exact fermionic zero modes. For the $\z$-derivative of the ghost
propagator we may use (the covariant version of)
the simpler relation \seeq{diffprop}.

The last two identities correspond to the variation
$\d_2\, \hq(x) \equiv -ig\, \ds\o^a(x)\, T^a\hq(x)$.
This variation looks like a gauge transformation of the quantum fields,
except that we have replaced  the local parameter $\o^a(x)$
by its SUSY variation $\ds\o^a(x)$, see \seeq{domega}.
For bosons the identity reads
\bqry
  0 & \!\eqv\! &  {ig\over 2}\;
  \nord \Bmath( \ds\o^a\, T^a\hb \Bmath|  L_B \Bmath| \hb \Bmath) \nord \; \co
\NON & &
  -\; ig\, \hmath( \ds\o^a\, T^a\hb \hmath|
   1 - (P_B + P^\parallel) (1 - L_B\, G_B) \hmath|
      X  - {\partial S_{\rm fp}^{\rm int} \over \partial \hb}\, \co \hmath)
\label{WIbosgg}\\
& & -\; ig  \int d^4x\, d^4y\, {\partial (\ds\o^a(x)) \over \partial \hb(y)}\,
    T^a \hb(x)
    \left[ \d^4(x-y) - P_B(x,y) - P^\parallel(x,y) \phantom{\int}
    \right.
\NON & &
  + \left. \int d^4z\, \left(P_B(x,z) + P^\parallel(x,z)\right)
    L_B\, G_B(z,y)
    \right] \co \,.
\nonumber
\eqry
and for fermions
\bqry
  0 & \!\eqv\! &  {ig\over 2}\;
  \nord \Bmath( \ds\o^a\, T^a\hJ \Bmath|  L_F \Bmath| \hJ \Bmath) \nord\; \co
  -\; ig\, \hmath( \ds\o^a\, T^a\hJ \hmath|
   {\partial \co \over \partial \hJ} - \cs_F\, \co \hmath)
\NON
& & -\; ig  \int d^4x\;  {\partial (\ds\o^a(x)) \over \partial \hJ(x)}\;
    T^a \hJ(x)\; \co \,.
\label{WIfergg}
\eqry

With the above diagrammatic identities at hand,
\seeq{dOint} follows after lengthy arithmetical manipulations
which will not be repeated here. Apart from similar steps to those
of Appendix~C.1,  the commutator formulae
(Appendix~B.3) play an important role.
In addition we use gauge invariance of the theory in general,
and of the operator $\co$ and the functional determinants in particular.
(The detailed discussion of Sec.~3.2 provides a leading-order
explanation on how \seeq{dOint} works in a gauge theory.)

%%%%%%%%%%%%%%%%%%%%%%%%%%%%%
\vspace{5ex}
%\newpage
\noindent {\large\bf D.~~Translation invariance}
\secteq{D}
\vspace{3ex}

It is instructive to see how translation invariance is proved
using the general formalism of Sec.~2.
Let us define $\d_\m Q = Q_{;\m}$. Using \seeq{var}
one easily finds that the translation collective coordinates $x_0^\n$
transform according to $\d_\m\, x_0^\n = -\d_\m^\n$.
Moreover, all other collective and quantum variables are invariant.
The corresponding Ward identity (cf.\ \seeq{WI}) for a gauge-invariant
local operator reads
\beq
  \svev{\co_{,\m}} = - \int d^4 x_0\,
  {\partial\over \partial x_0^\m} \svev{\co}_{x_0} = 0 \,.
\label{transinv}
\eeq
We see that the Hilbert-space surface term coincides with a spacetime
surface term, which vanishes thanks to locality (and gauge invariance)
of $\co$.

Let us now examine the local Ward identity involving
the conserved energy-momentum tensor $T_{\m\n}$ (compare subsection~3.5).
Define $\d_{z\m} Q(x) = \d^4(x-z) Q_{;\m}(x)$. Now other collective
coordinates besides $x_0^\m$ transform as well.
In particular, in the one-instanton sector the scale collective
coordinate $\r$ transforms to leading order as
\beq
  \d_{\m}\r(z) = {g^2 \over 16\p^2}\,
  {\partial a_\n^c(z)\over \partial \r}\, f_{\n\m}^c(z) \,.
\label{localtrans}
\eeq
Using the explicit expressions for the instanton field
it easily follows that $\d_{\m}\r(z)$ is a total $z$-derivative,
whose integral is zero. Suppose now that, for a given operator $\co$,
the $\r \to 0$ surface term involving $\d_{\m}\r(z)$ is finite.
Being a total spacetime derivative,
this term can be absorbed into a redefinition of the matrix element
of $T_{\m\n}(z)$. (An additional total spacetime derivative which is
absorbed into the matrix element of $T_{\m\n}(z)$
comes from the local variation of the measure.)

\vspace{5ex}
\centerline{\rule{5cm}{.3mm}}

%%%%%%%%%%%%%%%%%%%%%%%%%%%%%%%%%%%%%%%%%%%

%\newpage

\newpage

%>>>>>>>>>>>>>>>>%
% FIGURE    1    %
%>>>>>>>>>>>>>>>>%
\begin{figure}[hbt]
\centerline{
\epsfxsize=13.0cm
%\vspace*{0.5cm}
\epsfbox{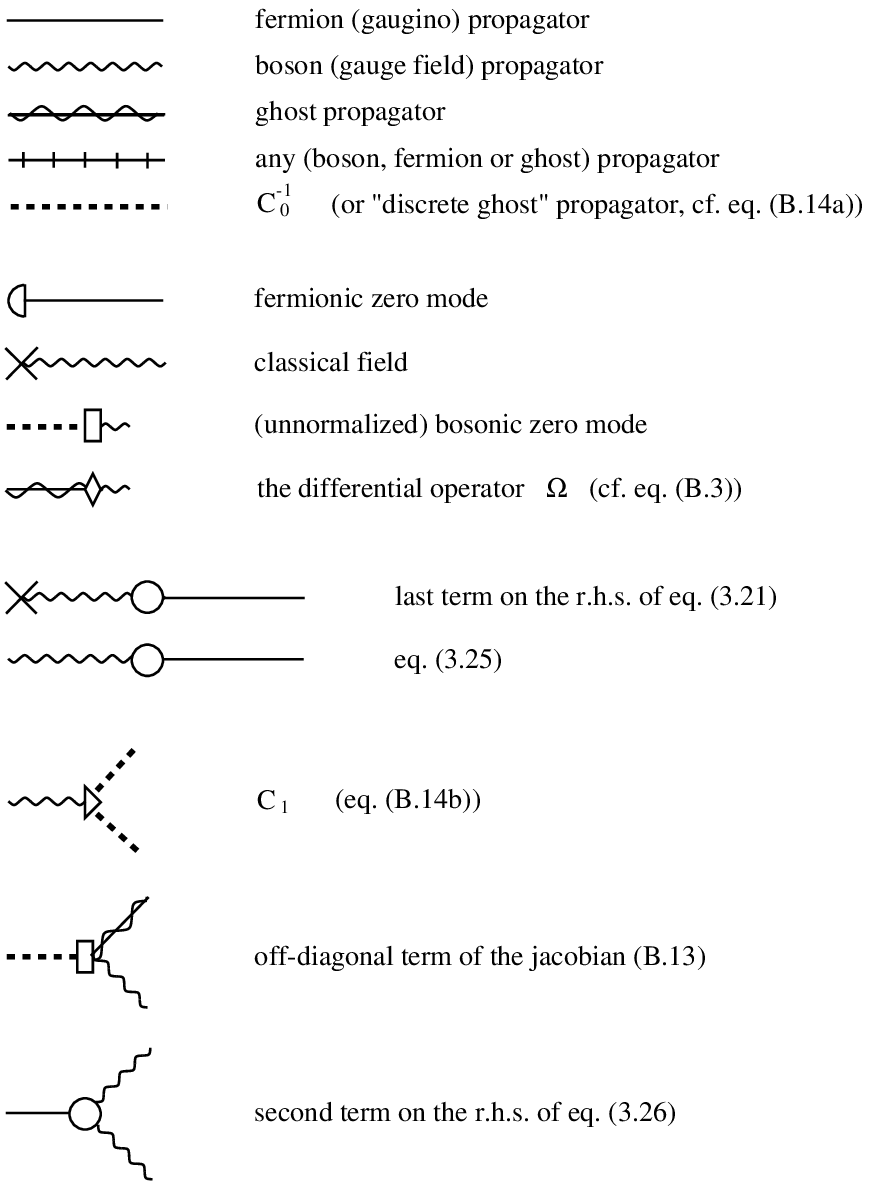}
}
\vspace*{0.8cm}
\caption{ \noindent {  Feynman rules.
}}
\label{fig1}
\end{figure}
%<<<<<<<<<<<<<<<<%
% FIGend    1    %
%<<<<<<<<<<<<<<<<%

%>>>>>>>>>>>>>>>>%
% FIGURE    2    %
%>>>>>>>>>>>>>>>>%
\begin{figure}[hbt]
\centerline{
\epsfysize=4.0cm
%\vspace*{0.5cm}
\epsfbox{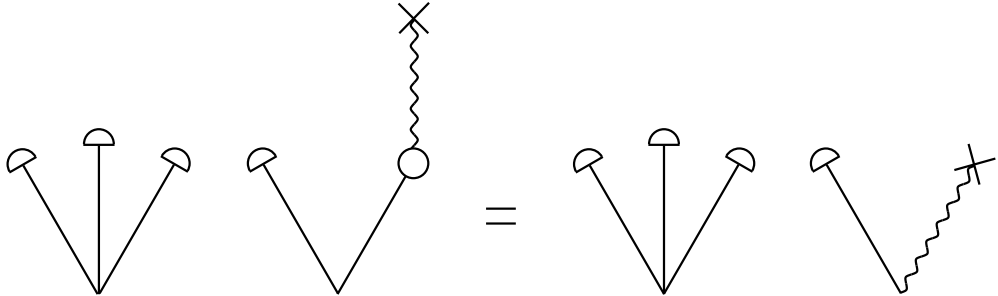}
}
\vspace*{0.8cm}
\caption{ \noindent {  Trivially vanishing leading order.
The three fermionic zero modes coming together saturate the
operator $\co_{\m}^{\,i\b}$. The other two external legs correspond
(on the l.h.s.) to the operator $\l\l$ or (on the r.h.s.) to its leading-order
SUSY variation. In all figures,
the indicated equalities hold after antisymmetrization (see fig.~4).
}}
\label{fig2}
\end{figure}
%<<<<<<<<<<<<<<<<%
% FIGend    2    %
%<<<<<<<<<<<<<<<<%

%>>>>>>>>>>>>>>>>%
% FIGURE    3    %
%>>>>>>>>>>>>>>>>%
\begin{figure}[hbt]
\centerline{
\epsfysize=4.0cm
%\vspace*{0.5cm}
\epsfbox{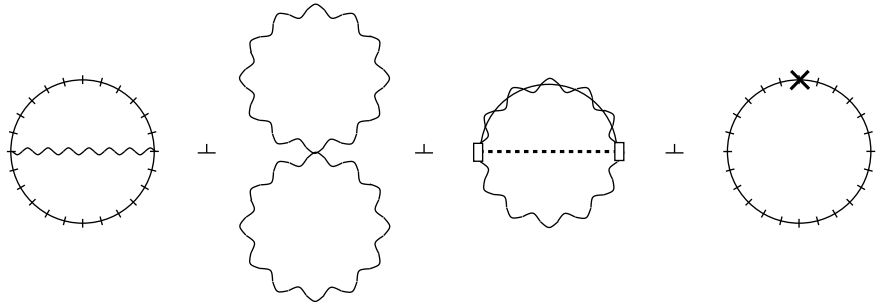}
}
\vspace*{0.8cm}
\caption{ \noindent { Bubble diagrams. The first and last
diagrams contain a sum over boson, fermion and ghost contributions.
The third diagram involves
the off-diagonal terms of the jacobian \seeq{J}. The last diagram
involves (one-loop) self-energy counterterms.
}}
\label{fig3}
\end{figure}
%<<<<<<<<<<<<<<<<%
% FIGend    3    %
%<<<<<<<<<<<<<<<<%

%>>>>>>>>>>>>>>>>%
% FIGURE    4    %
%>>>>>>>>>>>>>>>>%
\begin{figure}[hbt]
\centerline{
\epsfysize=6.0cm
%\vspace*{0.5cm}
\epsfbox{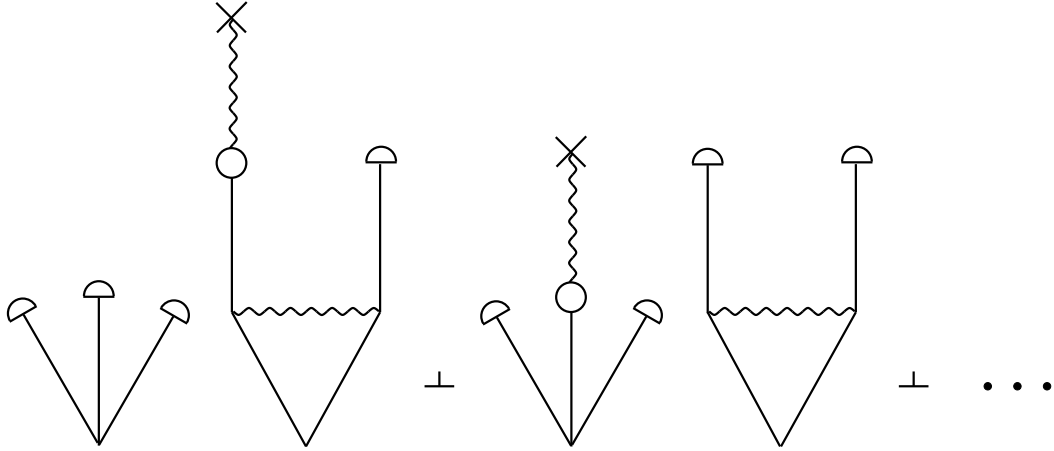}
}
\vspace*{0.8cm}
\caption{ \noindent { Antisymmetrization. The fermionic projectors
(\seeq{GF} or \seeq{GF0}) cancel each other after antisymmetrizing
with respect to the zero modes and the insertion of
$\partial_\m S_\m^{(0)}$.
}}
\label{fig4}
\end{figure}
%<<<<<<<<<<<<<<<<%
% FIGend    4    %
%<<<<<<<<<<<<<<<<%

%>>>>>>>>>>>>>>>>%
% FIGURE    5    %
%>>>>>>>>>>>>>>>>%
\begin{figure}[hbt]
\centerline{
\epsfysize=6.0cm
%\vspace*{0.5cm}
\epsfbox{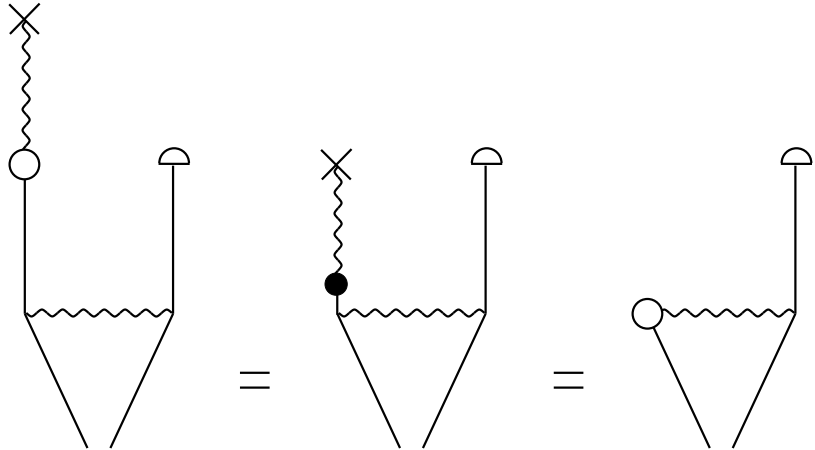}
}
\vspace*{0.8cm}
\caption{ \noindent {  A diagram where $\partial_\m S_\m^{(0)}$
is connected to a fermionic interaction vertex.
Here and in the following figures, those external legs which are simply
saturated by fermionic zero modes are not shown.
As before, the equalities hold after antisymmetrization (including
those zero modes which are not shown here, see Fig.~4).
The filled circle (here in the middle diagram) indicates that a bosonic field
is attached to a fermionic leg of a vertex.
The index-matching differential operators (or matrices)
can be read off from the appropriate terms in \seeq{expandl}.
On the \rhs this is traded with an insertion of
\seeq{genwi1}. These diagrams contain
a loop (the almost-closed triangle) provided the two external legs
are at the same spacetime point.
}}
\label{fig5}
\end{figure}
%<<<<<<<<<<<<<<<<%
% FIGend    5    %
%<<<<<<<<<<<<<<<<%

%>>>>>>>>>>>>>>>>%
% FIGURE    6    %
%>>>>>>>>>>>>>>>>%
\begin{figure}[hbt]
\centerline{
\epsfysize=4.0cm
%\hspace*{-0.2cm}
\epsfbox{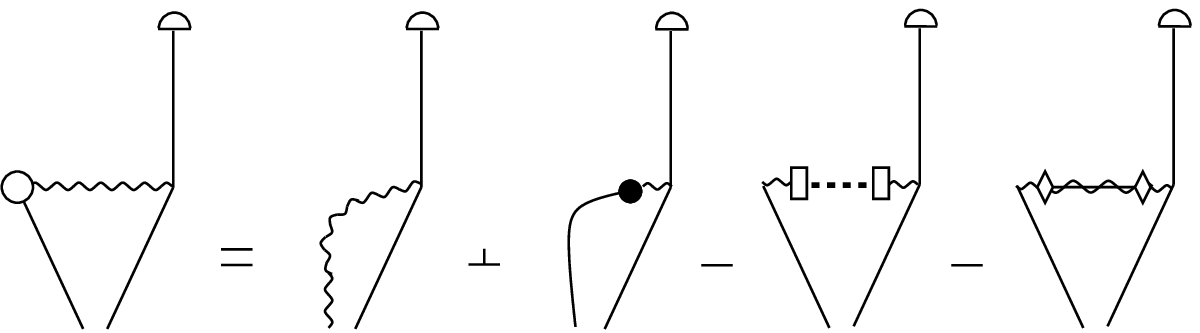}
}
\vspace*{0.8cm}
\caption{ \noindent {The third diagram on the \rhs is
a first contribution to the $\r$-derivative of
$\vev{(\ds_\a\,\r) \,\, \co_{\m}^{\,i\b}(0) \,\, \l\l(y)}_{\!\r}$.
In that diagram, the two rectangles
connected by the think dashed line constitute the bosonic
projector $P_B$, cf.\ \seeqs{GB0} and\seneq{PB}.
The first diagram on the \rhs is a contribution to $\ds\co$,
while the second diagram vanishes by a Fiertz
rearrangement (if the need for regularization is ignored).
The last diagram involves the longitudinal projector\seneq{longP},
and together with other diagrams
add up to zero by gauge invariance, see text.
}}
\label{fig6}
\end{figure}
%<<<<<<<<<<<<<<<<%
% FIGend    6    %
%<<<<<<<<<<<<<<<<%

%>>>>>>>>>>>>>>>>%
% FIGURE    7    %
%>>>>>>>>>>>>>>>>%
\begin{figure}[hbt]
\centerline{
\epsfysize=4.0cm
%\vspace*{0.5cm}
\epsfbox{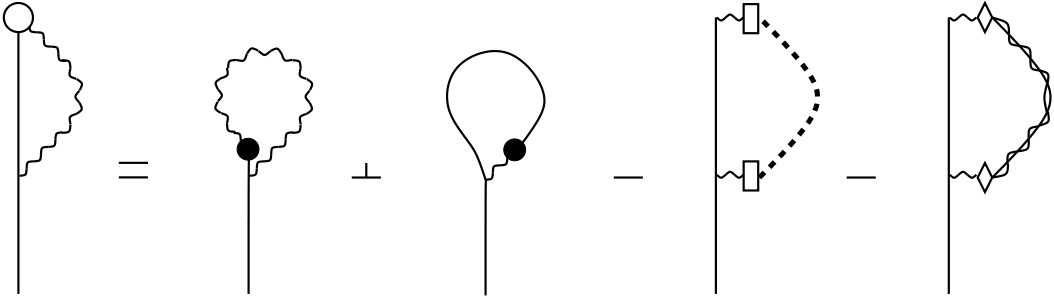}
}
\vspace*{0.8cm}
\caption{ \noindent {Another diagram with an insertion of
\seeq{genwi1}. Again, the second
diagram on the \rhs vanishes after a Fiertz rearrangement,
the third diagram is a contribution to the $\r$-derivative of
$\vev{(\ds_\a\,\r) \,\, \co_{\m}^{\,i\b}(0) \,\, \l\l(y)}_{\!\r}$,
and the last diagram cancels against other diagrams by gauge invariance.
}}
\label{fig7}
\end{figure}
%<<<<<<<<<<<<<<<<%
% FIGend    7    %
%<<<<<<<<<<<<<<<<%

%>>>>>>>>>>>>>>>>%
% FIGURE    8    %
%>>>>>>>>>>>>>>>>%
\begin{figure}[hbt]
\centerline{
\epsfysize=6.0cm
%\vspace*{0.5cm}
\epsfbox{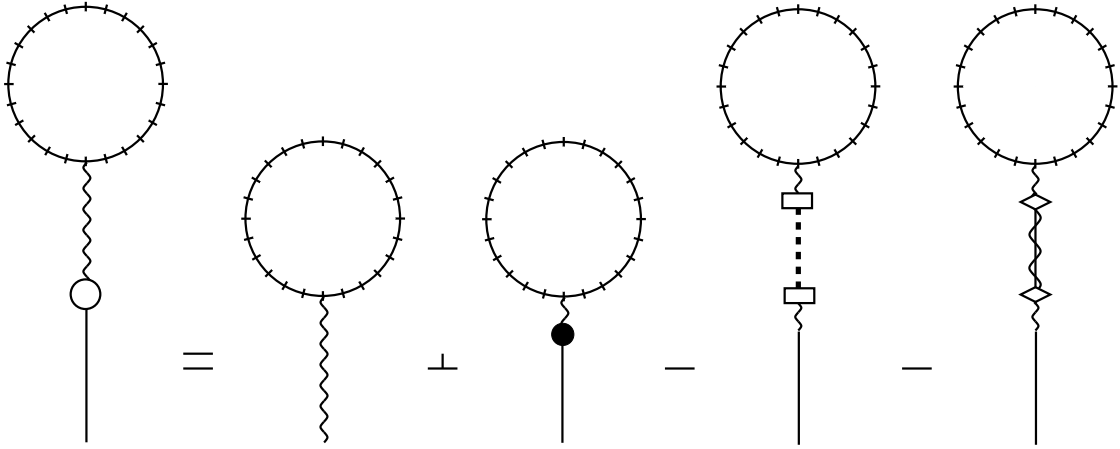}
}
\vspace*{0.8cm}
\caption{ %\noindent
One-loop tadpole diagrams. The loop stands for
the sum of boson, fermion and ghost contributions.
The first diagram on the \rhs is a contribution to $\ds\co$.
In general, the third diagram gives the $\z_n$-derivative of the
logarithm of the functional determinants.
In SYM the functional determinants cancel each other exactly,
and accordingly the one-loop tadpole is zero.
   In the second diagram on the r.h.s.,  it is convenient
for our purpose to consider the contribution of each field
separately. The fermion loop is by itself zero after a Fiertz
rearrangement. The boson loop appears in Fig.~9.
The ghost loop cancels against a sum of diagrams with
the longitudinal projector, see text.
}
\label{fig8}
\end{figure}
%<<<<<<<<<<<<<<<<%
% FIGend    8    %
%<<<<<<<<<<<<<<<<%

%>>>>>>>>>>>>>>>>%
% FIGURE    9    %
%>>>>>>>>>>>>>>>>%
\begin{figure}[hbt]
\centerline{
\epsfysize=4.5cm
%\vspace*{0.5cm}
\epsfbox{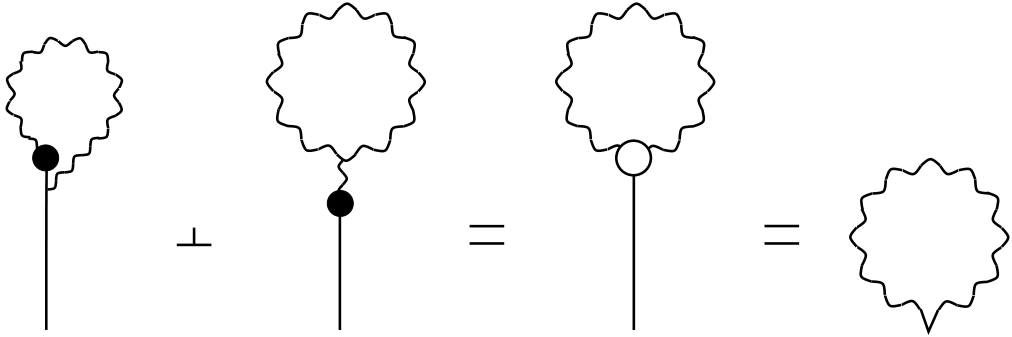}
}
\vspace*{0.8cm}
\caption{ The diagram on the \rhs is  $\ds\l^{(2)}$ (which makes
a contribution to $\ds\co$). The first diagram on the \lhs is from
Fig.~7, and the second one from Fig.~8.
}
\label{fig9}
\end{figure}
%<<<<<<<<<<<<<<<<%
% FIGend    9    %
%<<<<<<<<<<<<<<<<%

%>>>>>>>>>>>>>>>>%
% FIGURE   10    %
%>>>>>>>>>>>>>>>>%
\begin{figure}[hbt]
\centerline{
\epsfysize=2.0cm
%\vspace*{0.5cm}
\epsfbox{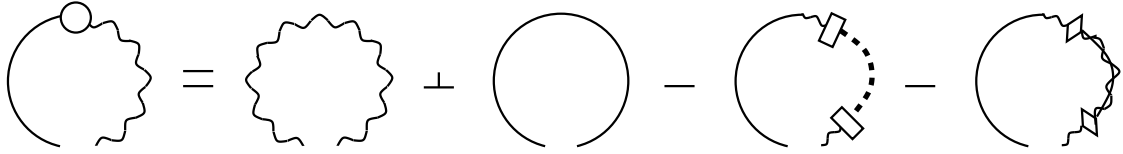}
}
\vspace*{0.8cm}
\caption{ The third diagram gives the $\r$-derivative of
(the classical field contained in) a covariant derivative $\hat{D}_\m$.
The first two diagrams contribute to $\ds\co$.
}
\label{fig10}
\vspace*{2cm}
\end{figure}
%<<<<<<<<<<<<<<<<%
% FIGend    10   %
%<<<<<<<<<<<<<<<<%

%>>>>>>>>>>>>>>>>%
% FIGURE   11    %
%>>>>>>>>>>>>>>>>%
\begin{figure}[hbt]
\centerline{
\epsfysize=6cm
%\vspace*{5cm}
\epsfbox{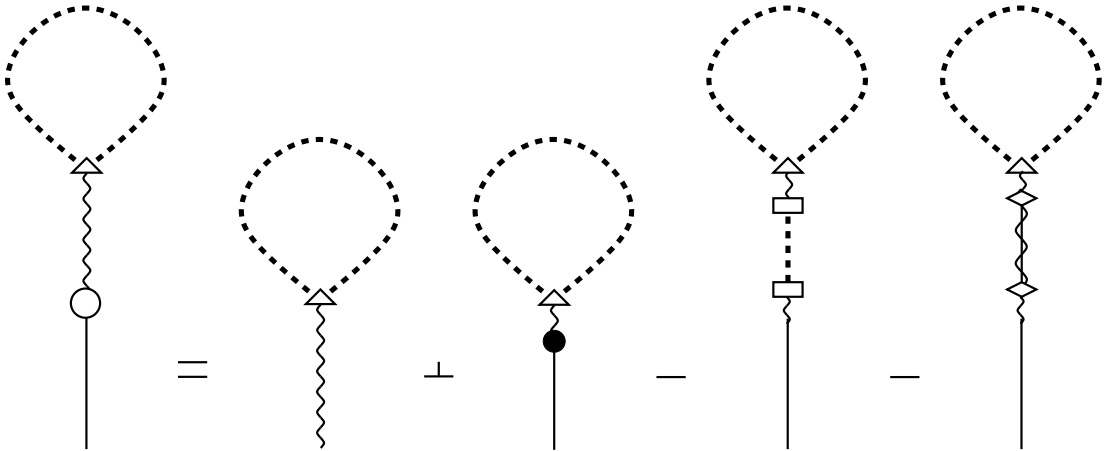}
}
\vspace*{0.8cm}
\caption{Diagrams containing the matrix $C_1$ (\seeq{C1}) from
the discrete-index part of the jacobian.
The first diagram on the \rhs is a contribution to $\ds\co$. The second
diagram gives the $\r$-derivative of the bosonic
zero mode contained in $\ds\r$ itself.
The third diagram gives the $\r$-derivative of
$\log {\rm det}^\half C_0$ (see text).
}
\label{fig11}
\end{figure}
%<<<<<<<<<<<<<<<<%
% FIGend    11   %
%<<<<<<<<<<<<<<<<%

%>>>>>>>>>>>>>>>>%
% FIGURE   12    %
%>>>>>>>>>>>>>>>>%
\begin{figure}[thb]
\centerline{
\epsfxsize=9cm
%\vspace*{0.5cm}
\epsfbox{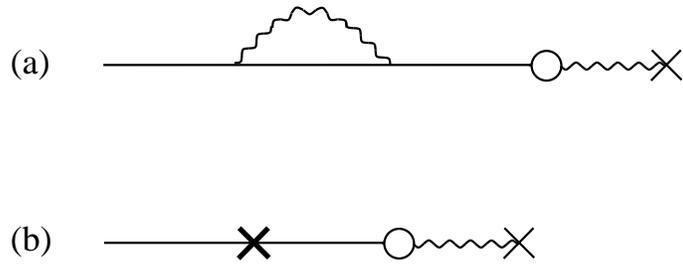}
}
\vspace*{2.0cm}
\caption{(a) A fermion self-energy diagram (cf.\ Fig.~7).
(b) The corresponding counterterm diagram.
}
\label{fig12}
\end{figure}
%<<<<<<<<<<<<<<<<%
% FIGend    12   %
%<<<<<<<<<<<<<<<<%

%>>>>>>>>>>>>>>>>%
% FIGURE   13    %
%>>>>>>>>>>>>>>>>%
\begin{figure}[hbt]
\centerline{
\epsfxsize=9cm
%\vspace*{0.5cm}
\epsfbox{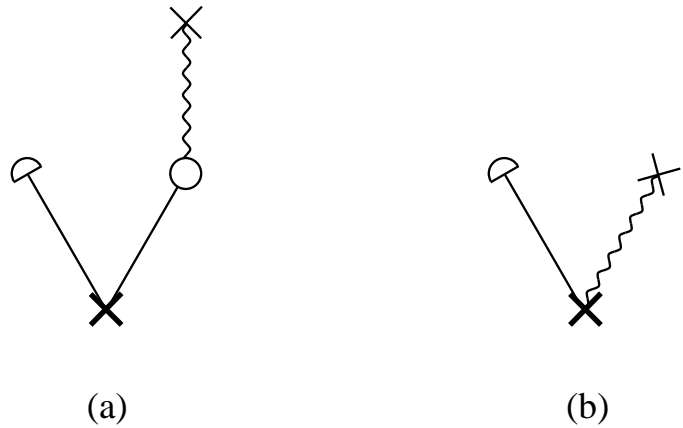}
}
\vspace*{0.8cm}
\caption{(a) A diagram containing a counterterm for the operator $\l\l$.
(b) A counterterm for the operator
$\ds(\l\l) = (i/\sqrt{2})\,\s_{\m\n} F_{\m\n}\,\l$.
The corresponding one-loop diagrams are in Figs.~4, 5 and~6.
}
\label{fig13}
\end{figure}
%<<<<<<<<<<<<<<<<%
% FIGend    13   %
%<<<<<<<<<<<<<<<<%

\end{document}